\documentclass[11pt]{article}
\usepackage[utf8]{inputenc}
\usepackage[T1]{fontenc}
\usepackage{amsmath,amssymb,mathtools,mathrsfs,bm}
\usepackage[margin=1in]{geometry}
\usepackage{graphicx}
\usepackage{caption}
\usepackage{array}             % >{...} column prefixes, for ragged p-columns
\usepackage{float}             % [H] --- place a table exactly where it is called
\usepackage{afterpage}
\usepackage{placeins}
\usepackage{longtable}         % Table E.1, the claims registry, runs over several pages
\usepackage{booktabs}          % plain rule tables, approved by the author
\usepackage[table]{xcolor}     % the shaded crossover row of Table D.1, as the Word master shades it
\definecolor{crossshade}{HTML}{F2F2F2}
\usepackage{cite}
\usepackage{titlesec}
\usepackage[colorlinks=true,allcolors=blue,hypertexnames=false]{hyperref}
\usepackage{orcidlink}
\usepackage{tocloft}
\newcommand{\vv}[1]{\vec{#1}}
\newcommand{\Mscr}{\vv{\mathscr{M}}}
\renewcommand{\thesection}{\arabic{section}}
\renewcommand{\thesubsection}{\thesection.\arabic{subsection}}
\newcommand{\hsection}[2]{\setcounter{section}{\numexpr#1-1}\section{#2}}
\newcommand{\hsubsection}[2]{\setcounter{subsection}{\numexpr#1-1}\subsection{#2}}
\titleformat{\section}{\normalfont\Large\bfseries}{\thesection.}{0.6em}{}
\titleformat{\subsection}{\normalfont\large\bfseries}{\thesubsection}{0.6em}{}

\makeatletter
\newcommand{\fwdeq}[2]{\def\@currentlabel{#2}\label{#1}}
\makeatother
\allowdisplaybreaks

\title{\vspace{-2em}\textbf{Free Precession Revisited:}\\[2pt]
\large A Vector-Only Analysis of Gyroscopic Dynamic Equilibrium}
\author{Dan I. Hariton\,\orcidlink{0009-0005-9168-9425}\\[2pt]
\normalsize Independent researcher, California, USA\\[2pt]
\normalsize \texttt{dan.hariton@comcast.net}\\[2pt]
\normalsize Open Researcher and Contributor ID (ORCID): 0009-0005-9168-9425}
\date{\normalsize August 2026 \ $\cdot$\ revised November 2026 \ $\cdot$\ arXiv category: physics.class-ph}

\newcommand{\TABLEF}{%
\begin{tabular}{crrrc}
\toprule
$\theta$ (degrees) & $M\vv{R}\times\vv{g}$ (N$\cdot$m) & $\Mscr_A$ (N$\cdot$m) & $M\vv{R}\times\vv{g} + \Mscr_A$ (N$\cdot$m) & Torque about $A$ \\
\midrule
$43^\circ$ & $-1.336716786$ & $+1.327396557$ & $-0.009320229$ & DOWN \\[2pt]
\midrule
$44^\circ$ & $-1.361530406$ & $+1.356740891$ & $-0.004789516$ & DOWN \\[2pt]
\midrule
$\mathbf{45^\circ}$ & $\mathbf{-1.385929291}$ & $\mathbf{+1.385929291}$ & $\mathbf{0.000000000}$ & \textbf{EQUALIZED} \\[2pt]
\midrule
$46^\circ$ & $-1.409906009$ & $+1.414953024$ & $+0.005047016$ & UP \\[2pt]
\midrule
$47^\circ$ & $-1.433453255$ & $+1.443803093$ & $+0.010349837$ & UP \\
\bottomrule
\end{tabular}}

\newcommand{\TABLEONE}{%
\renewcommand{\arraystretch}{2.4}%
\setlength{\tabcolsep}{4pt}%
\begin{tabular}{lccc}
\toprule
Solid & $J_0$ (kg$\cdot$m$^2$) & $J_R$ (kg$\cdot$m$^2$) & $J_0 - J_R$ (kg$\cdot$m$^2$) \\
\midrule
Hollow sphere $(D,E)$ & $\tfrac{3}{20}M\dfrac{D^5-E^5}{D^3-E^3}$ & $J_0/3$ & $\tfrac{2}{3}J_0$ \\[2pt]
\midrule
Torus, circular $(D,E)$ & $\dfrac{M(D^2+E^2)}{8}$ & $\dfrac{M(D-E)^2}{64}$ & $\dfrac{M[\,8(D^2+E^2)-(D-E)^2]}{64}$ \\[2pt]
\midrule
Torus, rectangular $(D,E,H)$ & $M\!\left[\dfrac{D^2+E^2}{8}+\dfrac{H^2}{12}\right]$ & $\dfrac{MH^2}{12}$ & $\dfrac{M(D^2+E^2)}{8}$ \\[2pt]
\midrule
\multicolumn{4}{c}{\textit{Classical limits --- recovered from the three bodies above ($E=0$, and $H\to0$ for the disk)}} \\[2pt]
\midrule
Solid sphere $(E=0)$ & $\tfrac{3}{20}MD^2$ & $\dfrac{MD^2}{20}$ & $\dfrac{MD^2}{10}$ \\[2pt]
\midrule
Cylinder / rod $(E=0)$ & $M\!\left(\dfrac{D^2}{8}+\dfrac{H^2}{12}\right)$ & $\dfrac{MH^2}{12}$ & $\dfrac{MD^2}{8}$ \\[2pt]
\midrule
Thin disk $(E=0,\ H\to0)$ & $\dfrac{MD^2}{8}$ & $0$ & $\dfrac{MD^2}{8}$ \\
\bottomrule
\end{tabular}}

\newcommand{\TABLEDONE}{%
\setlength{\tabcolsep}{4pt}%
\begin{tabular}{ccccc}
\toprule
$H/D$ &
\shortstack{$I_\parallel$\\($\times 10^{-4}$ kg$\cdot$m$^{2}$)} &
\shortstack{$I_\perp$\\($\times 10^{-4}$ kg$\cdot$m$^{2}$)} &
\shortstack{$\tfrac{1}{2}(3J_R - J_0)$\\($\times 10^{-4}$ kg$\cdot$m$^{2}$)} &
\shortstack{$\Omega$ exact, (39, which is\\this paper's steady-\\precession relation)\\(rad/s)} \\
\midrule
$0.000$ & $12.500$ & $6.250$   & $-6.250$   & $2.374957$ \\[2pt]
\midrule
$0.200$ & $12.500$ & $6.583$   & $-5.917$   & $2.375219$ \\[2pt]
\midrule
$0.400$ & $12.500$ & $7.583$   & $-4.917$   & $2.376007$ \\[2pt]
\midrule
$0.600$ & $12.500$ & $9.250$   & $-3.250$   & $2.377322$ \\[2pt]
\midrule
$0.800$ & $12.500$ & $11.583$  & $-0.917$   & $2.379168$ \\[2pt]
\midrule
\rowcolor{crossshade}$0.866$ & $12.500$ & $12.500$ & $0.000$ & $2.379895$ \\[2pt]
\midrule
$0.900$ & $12.500$ & $13.000$  & $+0.500$   & $2.380292$ \\[2pt]
\midrule
$1.000$ & $12.500$ & $14.583$  & $+2.083$   & $2.381551$ \\[2pt]
\midrule
$1.200$ & $12.500$ & $18.250$  & $+5.750$   & $2.384475$ \\[2pt]
\midrule
$1.500$ & $12.500$ & $25.000$  & $+12.500$  & $2.389898$ \\[2pt]
\midrule
$2.000$ & $12.500$ & $39.583$  & $+27.083$  & $2.401785$ \\[2pt]
\midrule
$3.000$ & $12.500$ & $81.250$  & $+68.750$  & $2.437117$ \\[2pt]
\midrule
$5.000$ & $12.500$ & $214.583$ & $+202.083$ & $2.566557$ \\
\bottomrule
\end{tabular}}

\newcommand{\TABLEDTWO}{%
\setlength{\tabcolsep}{4pt}%
\begin{longtable}{>{\raggedright\arraybackslash}p{0.28\textwidth}>{\raggedright\arraybackslash}p{0.25\textwidth}>{\raggedright\arraybackslash}p{0.39\textwidth}}
\toprule
Item tested & Against & Result \\
\midrule
\endfirsthead
\toprule
Item tested & Against & Result \\
\midrule
\endhead
\endfoot
\bottomrule
\caption{Table~D.2 --- Every check performed for this appendix, with the reference
tested against and the outcome.}
\endlastfoot
Algebraic continuity of $I_\parallel$, $I_\perp$ across all $H/D$ & SymPy + direct algebra & PASS --- both are smooth polynomials; no pole, no branch, no case split \\[2pt]
\midrule
Crossover located at $H = D\sqrt{3}/2$ & Direct algebra on Table~2 & PASS --- $\tfrac{1}{2}(3J_R-J_0) = \tfrac{1}{2}M(H^{2}/6 - D^{2}/8)$ vanishes exactly there \\[2pt]
\midrule
Thin-disk limit $H \to 0$ & SymPy limit & PASS --- $I_\parallel \to \tfrac{1}{2}MR_{\mathrm{DISK}}^{2}$, $I_\perp \to \tfrac{1}{4}MR_{\mathrm{DISK}}^{2}$, both exact \\[2pt]
\midrule
Thin-rod limit $D \to 0$ & SymPy limit & PASS --- $I_\parallel \to 0$, $I_\perp \to ML_{\mathrm{ROD}}^{2}/12$, both exact \\[2pt]
\midrule
Numeric $\Omega$ through the crossover & Direct evaluation of (39, which is this paper's steady-precession relation) & PASS --- smooth and monotone; no kink or jump at $H = D\sqrt{3}/2$ \\[2pt]
\midrule
Steady precession, thin disk compared at $\theta = 90^\circ$ (2 sources; OpenStax states no tilt) & OpenStax~\cite{ref29}; Jensen and Poling~\cite{ref30} & PASS --- $3.12$ rad/s, exact to OpenStax's precision; $2.01$ rad/s against Jensen and Poling's $\approx 2.0$, to the precision they quote \\[2pt]
\midrule
Steady precession, general $\theta$, both branches & Bolina~\cite{ref27}, arXiv:physics/0005025 & PASS --- symbolically identical, both roots (SymPy difference $=0$) \\[2pt]
\midrule
Steady precession, general $\theta$, both branches; the worked thin disk; the three parameter tables & Slade~\cite{ref33} & PASS --- symbolically identical (SymPy difference $=0$); both inertia constants reproduced exactly; roots $4.645916070$ and $19.369667379$ rad/s; the other entries of its three parameter tables recomputed from its own formulas, as D.5 states \\[2pt]
\midrule
Minimum-spin condition, (D9, which is the discriminant of this paper's \eqref{e:43}, written as a quadratic in $\Omega$ in the paper's symbols $I_\parallel$ and $I_\perp + MR^2$) against (D8, which is the textbooks' minimum-spin condition, written in their symbols $I_1, I_3$ and $l$, Thornton and Marion writing $h$ for $l$, and Goldstein writing it with the strict inequality) & Goldstein \S5.7~\cite{ref2}; Thornton \& Marion \S11.11~\cite{ref4} & PASS --- identical discriminant; at zero discriminant, with $\omega_0 + \Omega \cos \theta$ held fixed, the two branches merge into a double root \\[2pt]
\midrule
Location of the degenerate geometry, $K = 0$ & Direct algebra on (D11, which defines $K = MR^2 + \tfrac{1}{2}(3J_R - J_0)$, the quantity whose zero removes the $\Omega ^2$ term) & PASS --- $MR^{2} = \tfrac{1}{2}(J_0 - 3J_R)$ exactly; $R = D/4$ for a thin disk of any mass, and unreachable for the isotropic sphere, where $J_R = J_0/3$ identically \\[2pt]
\midrule
The degeneracy is distinct from the crossover & Direct algebra on (D11, which defines $K = MR^2 + \tfrac{1}{2}(3J_R - J_0)$, the quantity whose zero removes the $\Omega ^2$ term) & PASS --- at the crossover $K = MR^{2}$, strictly positive for every $R > 0$, so the two zeros can never coincide \\[2pt]
\midrule
Table~D.1's sweep is clear of the degeneracy & Direct evaluation & PASS --- $K$ runs from $0.002975$ to $0.023808$ kg$\cdot$m$^{2}$ across the whole range of heights, never approaching zero \\[2pt]
\midrule
Exact steady precession + finite-cylinder continuity & Brand, \S\S193--196, 214, 217--218~\cite{ref37} (named in the reference list) & \textbf{PASS} --- Brand's exact quadratic maps identically to (39, which is this paper's steady-precession relation). Brand gives finite-cylinder inertia formulas; the continuous sweep is demonstrated here, in Table~D.1. \\[2pt]
\midrule
Exact steady precession, both branches, and the degenerate case $A = C$ & Klein and Sommerfeld, Vol.~I \S III.6~\cite{ref38}, pp.~177--179 (named in the reference list) & \textbf{PASS} --- their equation~(3), the exact steady-precession condition, maps identically to (39, which is this paper's steady-precession relation); their equation~(4), the two roots in closed form, returns 3.753732625419 and 18.460682065373 rad/s for the sphere of Appendix~F, the rates printed there; their case $A = C$ is the condition $K = 0$ of \S D.7. \\
\end{longtable}}

\newcommand{\TABLEDTHREE}{%
\setlength{\tabcolsep}{4pt}%
\begin{tabular}{>{\raggedright\arraybackslash}p{0.24\textwidth}c>{\raggedright\arraybackslash}p{0.44\textwidth}>{\raggedright\arraybackslash}p{0.17\textwidth}}
\toprule
Source & Ref. & What it was consulted for & Validation datum \\
\midrule
OpenStax, University Physics Vol.~1, \S11.4 & \cite{ref29} & Worked numeric example: thin disk, slow-precession formula, no tilt stated & Yes --- D.5, Source~1 \\[2pt]
\midrule
Jensen and Poling, Teaching Rotational Physics with Bivectors & \cite{ref30} & Worked numeric example: horizontal-shaft thin disk at $\theta = 90^\circ$ & Yes --- D.5, Source~2 \\[2pt]
\midrule
Bolina, The Precessing Top & \cite{ref27} & Exact two-branch precession expression, valid at every non-vertical tilt except the horizontal shaft ($\cos\theta = 0$); one of the four load-bearing external checks & Yes --- D.5, Source~3 \\[2pt]
\midrule
Goldstein, Poole and Safko, Classical Mechanics, \S5.7 & \cite{ref2} & Canonical minimum-spin condition for the heavy symmetric top & Yes --- D.6 \\[2pt]
\midrule
Thornton and Marion, Classical Dynamics, \S11.11 & \cite{ref4} & The same canonical condition, independently stated & Yes --- D.6 \\[2pt]
\midrule
MIT OpenCourseWare 8.09, Classical Mechanics III, Assignment~3 & \cite{ref32} & Standard reading assignment mapping the heavy symmetric top onto Goldstein Ch.~4--5 and Thornton and Marion Ch.~11 & No --- no number offered \\[2pt]
\midrule
MIT OpenCourseWare 16.07, Dynamics, Lecture~30 & \cite{ref24} & Exact steady-precession balance, stated before the fast-spin approximation is introduced; also its Euler-angle treatment, consulted for cross-reference & Yes --- \S F.9 and Table~F.4 \\[2pt]
\midrule
Agashe, PHYS 601, Maryland & \cite{ref26} & Uniform-precession condition and its discriminant & Yes --- \S F.9 and Table~F.4 \\[2pt]
\midrule
Slade, Classical Symmetric Top in a Gravitational Field & \cite{ref33} & Exact two-branch precession expression, and a thin-disk-on-a-stem worked example exercising both mass integrals & Yes --- D.5, Source~4 \\[2pt]
\midrule
Brand, \emph{Vectorial Mechanics} (1930) & \cite{ref37} & Exact vectorial steady-precession equation; both branches; exact horizontal case; finite-cylinder inertia and transfer theorem & Yes --- D.5, Source~5 \\[2pt]
\midrule
Klein and Sommerfeld, \emph{The Theory of the Top} (1897--1910) & \cite{ref38} & Exact steady-precession condition and its two roots in closed form, both branches named (Vol.~I \S III.6, pp.~177--178); the existence condition at a given spin; the degenerate case $A = C$ (pp.~178--179) & Yes --- D.5, Source~6 \\
\bottomrule
\end{tabular}}

\newcommand{\TABLEE}{%
\setlength{\tabcolsep}{3pt}%
\begin{longtable}{c>{\raggedright\arraybackslash}p{0.28\textwidth}>{\raggedright\arraybackslash}p{0.105\textwidth}>{\raggedright\arraybackslash}p{0.335\textwidth}>{\raggedright\arraybackslash}p{0.175\textwidth}}
\toprule
\# & Claim (assertion the paper makes about itself) & Equations & Support basis --- (i) shown in paper / (ii) checked / (iii) search-bounded / (iv) open & Verdict \\
\midrule
\endfirsthead
\toprule
\# & Claim (assertion the paper makes about itself) & Equations & Support basis --- (i) shown in paper / (ii) checked / (iii) search-bounded / (iv) open & Verdict \\
\midrule
\endhead
\endfoot
\bottomrule
\caption{Table~E.1 --- Registry of the manuscript's principal claims, with supporting
equations, verification or source comparison, and support-basis code.}
\endlastfoot
1 & \S\ref{sec:7.6}: ``no approximation is made anywhere in the chain from~\eqref{e:32} to~\eqref{e:40}'' & \eqref{e:32}--\eqref{e:40} & (i) chain derived stepwise in \S\ref{sec:6}--\S\ref{sec:7}; (ii) every link re-derived symbolically --- checks~\#24 (\eqref{e:28}$\to$\eqref{e:35}), \#25 (\eqref{e:35}$\to$\eqref{e:38}), \#26--\#27 (\eqref{e:38}$\to$\eqref{e:39}), \#28 (\eqref{e:39}$\to$\eqref{e:40}); each difference simplifies to exactly zero & \textbf{SUPPORTED} \\[2pt]
\midrule
2 & C10 / \S\ref{sec:7.5}: the horizontal-shaft relation~\eqref{e:40} holds exactly, not as a fast-spin approximation; the textbook rate is recovered only afterward as a limiting root & \eqref{e:39}, \eqref{e:40} & (i) stated and derived \S\ref{sec:7.5}; (ii) check~\#28 (exact at $\theta = 90^\circ$); Appendix~D.5 Bolina identity (both roots exact) & \textbf{SUPPORTED} \\[2pt]
\midrule
3 & \S\ref{sec:8.3}: ``informationally equivalent to the classical description on their common ground'' & \eqref{e:39}, \eqref{e:43} & (i) dictionary stated \S\ref{sec:8.3}; (ii) check~\#31 (\eqref{e:43} with the dictionary is~\eqref{e:39} term for term); Appendix~D.5/D.6 (Bolina symbolic identity; Goldstein/Thornton \& Marion discriminant identical) & \textbf{SUPPORTED} \\[2pt]
\midrule
4 & \S\ref{sec:1} and \S\ref{sec:8.3} (closing sentence): derived without the classical axial and transverse moments $I_\parallel$ and $I_\perp$ (\S\ref{sec:8.3}), tensor or matrix formalism or Euler angles, and --- up to the governing torque-vector equation~\eqref{e:35} --- without unit vectors or component decomposition (\S\ref{sec:1}) & the derivation and its exact symbolic solution (\S\ref{sec:1}) & (i) textual property of the manuscript, confirmed by direct symbol audit: no tensor, matrix or Euler-angle symbol is used in the derivation --- the inertia tensor $\mathbf{I}$ and the identity $1\!\!1$ are named in \S\ref{sec:1} and \S\ref{sec:4.2} only as a comparison, and Euler-angle rates appear only in the quoted source relation~\eqref{e:D7}; neither unit vectors nor component decomposition is used in the derivation up to~\eqref{e:35} --- the three unit vectors of the trace identity~\eqref{e:12}, stated in \S\ref{sec:4.1}, enter only in the closure of \S\ref{sec:7.4} --- and the scalar reduction of \S\ref{sec:7.5} and the numerical verification of Appendix~F use them, and say so & \textbf{SUPPORTED} \\[2pt]
\midrule
5 & C1: $J_0$ is a primary inertia descriptor, defined directly by~\eqref{e:1}, never assembled from axial or planar moments & \eqref{e:1} & (i) definitional; (ii) all Table~2 $J_0$ values re-derived directly from $\int(\vv{\xi}\cdot\vv{\xi})dm$ --- checks~\#10, \#12, \#15, \#17 --- with no axial-moment intermediary anywhere in the derivation & \textbf{SUPPORTED} \\[2pt]
\midrule
6 & C2: one mass-vector operator $\vv{F}$ carries all directional inertia information, replacing moments and products of inertia & \eqref{e:2}, \eqref{e:12}, \eqref{e:37} & (ii) check~\#22 ($\vv{F}$'s full axial/transverse eigen-structure follows from $J_0$, $J_R$ alone), numerical check~\#23 (direct volume integration of $\vv{F}(\vv{v})$ for the cylinder, by numerical quadrature, equals~\eqref{e:37} for four arbitrary $\vv{v}$), numerical check~\#9 (trace identity, on a point cloud) & \textbf{SUPPORTED} \\[2pt]
\midrule
7 & C3: internal bookkeeping closes through Propositions~1--3, with no external tensor identities invoked & \eqref{e:16}, \eqref{e:17}, \eqref{e:18} & (i) proofs shown \S\ref{sec:4.3}; (ii) numerical checks~\#8 (Prop~1), \#5 (Prop~2), \#6 (Prop~3); textual half confirmed by the symbol audit (no tensor identities cited) & \textbf{SUPPORTED} \\[2pt]
\midrule
8 & C4: entire dynamics in inertial material derivatives; $\partial/\partial t$ never an independent physical input; no rotating frame carries dynamics, the chosen frame of the transport rule~\eqref{e:6} serving only bookkeeping; no force used in the derivation, only torques & \eqref{e:6}, \eqref{e:7} & (i) confirmed by direct sweep of the present text: the one frame named besides the inertial one, the chosen frame of~\eqref{e:6} in \S\ref{sec:3}, serves only the bookkeeping of the transport rule, and no rotating observer carries any dynamics; (ii) checks~\#1--\#3 establish $\partial/\partial t$ as the derived bookkeeping object of the transport relation, exactly as this paper defines it & \textbf{SUPPORTED} \\[2pt]
\midrule
9 & C5: the balance~\eqref{e:32} follows from Newton's law per mass element, taken as a torque about the pivot --- internal torques canceling pairwise, pivot contributing no torque --- with no angular-momentum primitive & \eqref{e:20}--\eqref{e:25}, \eqref{e:32} & (i) derivation shown \S\ref{sec:5}--\S\ref{sec:7.1}; (ii) numerical check~\#4 verifies the assembly~\eqref{e:23}$\to$\eqref{e:25} on random particle data for an arbitrary deforming closed system. The pairwise cancellation itself is Newton's third law --- a stated physical axiom, not a testable identity & \textbf{SUPPORTED} \\[2pt]
\midrule
10 & C6: inertia about the center of mass enters the torque exclusively through $J_0\vv{\varepsilon} - \vv{F}(\vv{\varepsilon})$ and $\vv{F}(\vv{\omega}) \times \vv{\omega}$; no thin-body idealization needed anywhere & \eqref{e:19}, \eqref{e:28} & (ii) numerical check~\#7 (engine-identity integrand, on a point cloud); checks~\#10--\#17 (the three general bodies and their classical limits resolved by the same pair --- no thin-body limit invoked) & \textbf{SUPPORTED} \\[2pt]
\midrule
11 & C7: the dynamics culminates in one unified vector equation of motion~\eqref{e:35}; each term is physically legible and can be visualized by the corkscrew mnemonic, and the governing vectors are never replaced by coordinate or principal-axis components before the final scalar reduction & ~\eqref{e:35} & (i) structural property of~\eqref{e:35}--\eqref{e:39}:~\eqref{e:35} and~\eqref{e:38} remain intrinsic torque-vector balances; \S\ref{sec:7.5} takes the scalar relation only after all terms are shown to have one common physical direction; (ii) check~\#24 verifies~\eqref{e:35} exactly. `Physically legible' is an editorial quality judgment, not a separate mathematical proposition & \textbf{SUPPORTED} \\[2pt]
\midrule
12 & C8: the two-route consistency theorem~\eqref{e:31} validates the assembly of the inertia torque internally and forecloses the multiple-counting error & \eqref{e:29}, \eqref{e:30}, \eqref{e:31} & (i) theorem shown \S\ref{sec:6.2}, \eqref{e:28} constructed as the sum of rows~\eqref{e:29}+\eqref{e:30}; (ii) numerical check~\#4 verifies the assembly~\eqref{e:23}$\to$\eqref{e:25} on an arbitrarily deforming closed body, and check~\#24 exercises the summed rows against the independent steady-precession route & \textbf{SUPPORTED} \\[2pt]
\midrule
13 & C9: the closing scalar $J_R$ is produced by the formulation itself, as the operator's own response to the shaft vector & \eqref{e:36} & (ii) numerical check~\#23: the direct $\vv{F}(\vv{v})$ integral for the cylinder, by numerical quadrature, returns the axial response $J_R$ to below $10^{-9}$ --- the scalar emerges from the operator, it is not inserted & \textbf{SUPPORTED} \\[2pt]
\midrule
14 & C11: the single quadratic~\eqref{e:39} contains both classical branches, with the parallel-axis $MR^{2}$ arising automatically from the transport term & \eqref{e:39} & (ii) check~\#27 (\eqref{e:39} emerges from~\eqref{e:38} with the $MR^{2}$ term produced by the $M(\vv{\Omega}\cdot\vv{R})(\vv{R}\times\vv{\Omega})$ transport contribution); Appendix~D.5 (both roots match Bolina exactly), D.6 (double-root merger reproduced) & \textbf{SUPPORTED} \\[2pt]
\midrule
15 & C12: isotropic-body collapse~\eqref{e:41}, lamina closure~\eqref{e:42}, and the $H = D\sqrt{3}/2$ crossover are read off the scalar pair by inspection, with no eigen-analysis; continuity and external validation established in Appendix~D & \eqref{e:41}, \eqref{e:42}, Table~2 & (ii) checks~\#29 (isotropic body), \#30 (lamina), \#20--\#21 (crossover coefficient and location), \#18 (isotropy: $J_R = J_0/3$); Appendix~D.2--D.4 (algebraic + numeric continuity, Table~D.1 sweep) & \textbf{SUPPORTED} \\[2pt]
\midrule
16 & C13: the balance, Brand's theorem of moment of momentum for any system of particles (\S183~\cite{ref37} and \S184~\cite{ref37}), holds exactly beyond rigidity --- $M\vv{R}\times\vv{g} = M\vv{R}\times(d^{2}\vv{R}/dt^{2}) + d\vv{S}/dt$ --- with rigidity confined to the closed-form solution & \eqref{e:25}, \eqref{e:32}, \eqref{e:18} & (ii) numerical check~\#4 (assembly holds on an arbitrarily deforming closed body) and numerical check~\#6 (internal torque integral $= d\vv{S}/dt$, by a central difference) & \textbf{SUPPORTED} \\[2pt]
\midrule
17 & C14: one uniform two-integral recipe resolves every solid geometry identically & Table~2, \S\ref{sec:8.1} & (ii) checks~\#10--\#17: all Table~2 entries for the three bodies and their limits re-derived from raw volume integrals by the identical procedure; check~\#19 numeric quadrature cross-check to below $10^{-10}$ & \textbf{SUPPORTED} \\[2pt]
\midrule
18 & C15: a single algebra --- mass integrals and intrinsic cross products, with dot products secondary, appearing inside the two mass integrals and as reductions of cross products --- runs unbroken from the first kinematic statement to the exact torque-vector balance, and each torque contribution retains a traceable cross-product sense until the final scalar reduction, which occurs only after a common physical direction is established & ~\eqref{e:1}--\eqref{e:43} & (i) structural property of the manuscript confirmed by direct symbol audit: no coordinate-component representation replaces the governing vectors in the derivation; dot products are intrinsic scalar products;~\eqref{e:35} and~\eqref{e:38} remain vector equations and~\eqref{e:39} is introduced only after the common direction is established & \textbf{SUPPORTED} \\[2pt]
\midrule
19 & C16: the formulation's boundary conditions are stated as results --- assumptions enumerated (\S\ref{sec:3.1}), existence of the steady solution precisely delimited (\S\ref{sec:7.6}, with the domain of~\eqref{e:39} stated where~\eqref{e:39} is introduced) & \S\ref{sec:3.1}, \S\ref{sec:7.6}, \eqref{e:39} & (i) shown in the paper; (ii) the existence delimitation is the discriminant of \eqref{e:43}, verified identical to the canonical minimum-spin condition in Appendix~D.6 (Goldstein \S 5.7~\cite{ref2}; Thornton \& Marion \S 11.11~\cite{ref4}); at a fixed $\omega_0$ it is the quantity under the square root of \eqref{e:D12}, the existence condition of the note in \S D.6; double-root merger reproduced numerically & \textbf{SUPPORTED} \\[2pt]
\midrule
20 & \S\ref{sec:8.2}: for the finite cylinder the coefficient $\tfrac{1}{2}(3J_R - J_0)$ changes sign at $H = D\sqrt{3}/2$, produced automatically by the two standalone scalars with no eigen-analysis & Table~2, \eqref{e:38} & (ii) checks~\#20--\#21; Appendix~D.2 (transversal zero of a smooth quadratic --- sign change of a continuous function, not a discontinuity) & \textbf{SUPPORTED} \\[2pt]
\midrule
21 & \S\ref{sec:7.6}: the physical limits are exactly those of \S\ref{sec:3.1} (fixed frictionless pivot, uniform gravity, closed constant mass, no other external actions) & \S\ref{sec:3.1}, \S\ref{sec:7.6} & (i) internal consistency of the two statements verified by direct comparison; every symbolic check was run under precisely these premises and no others & \textbf{SUPPORTED} \\[2pt]
\midrule
22 & Appendix~D.8: the continuous disk-to-rod sweep is not externally validated, although the two-scalar machinery is checked at individual geometries, including ones where $J_R$ is not negligible; no published source presents a finite-height cylinder carried across the transition & ~\eqref{e:38}--\eqref{e:39}, Appendix~D & (iv) OPEN as an external-source gap, and this paper says so itself in D.8 item~2. Internal support is complete --- checks~\#10--\#17, \#20--\#21, numerical check~\#23 and Table~D.1. Brand~\cite{ref37} narrows the gap without closing it: his \S\S193, 194 and 196~\cite{ref37} put the axial and the finite-height transverse inertia constants in print, so the ingredients for a continuous principal-moment calculation exist in the literature. What remains missing is a published finite-H comparison, which the searched literature does not contain & \textbf{OPEN (documented)} \\[2pt]
\midrule
23 & Appendix~F: at the equalization tilt the gravity torque and the assembled inertia torque are equal in magnitude and diametrically opposed, so their resultant is the zero vector, and displacing the tilt breaks the equalization in a restoring direction on both sides & \eqref{e:32}, \eqref{e:35}, \eqref{e:39} & (i) shown in Appendix~F; (ii) both torques computed independently and agreeing to $10^{-16}$ N$\cdot$m, with the five-tilt sweep of Table~F.2 and the external checks of Table~F.4 (Bolina~\cite{ref27}, MIT~\cite{ref24}, Agashe~\cite{ref26}, Goldstein~\cite{ref2} / Thornton and Marion~\cite{ref4}) & \textbf{SUPPORTED} \\[2pt]
\midrule
24 & \S\ref{sec:1}--\S\ref{sec:2} / \S\ref{sec:9.2}: in the searched accessible record, no prior source was found that combines the COM-defined $J_0$ and arbitrary-input $\vv{F}$ operator, operator-generated $J_R$, persistent intrinsic three-dimensional vector representation without principal-axis/component reduction, scalar reduction only after the torque-vector balance has one common direction, and continuous term-by-term cross-product traceability through the exact steady-precession balance & ~\eqref{e:1}--\eqref{e:2},~\eqref{e:35}--\eqref{e:39}; \S\ref{sec:2} & (iii) search-bounded literature finding. Brand~\cite{ref37} and Klein and Sommerfeld~\cite{ref38} were checked at the place given for each feature in the ancillary comparison file, and both were explicitly distinguished: same classical physics and end equation, but a working representation carried in components --- on Brand's principal axes, and in Klein and Sommerfeld's components of the impulse and of its change. Milne~\cite{ref8}, Synge and Griffith~\cite{ref9}, Rutherford~\cite{ref35}, Chorlton~\cite{ref36}, Butikov~\cite{ref11}, Hestenes~\cite{ref12}, and the contemporary sources of \S\ref{sec:2} were compared for the stated features. This is not a universal proof that no undiscovered source exists. & \textbf{SUPPORTED (search-bounded)} \\[2pt]
\midrule
25 & Appendix~D / C12--C14: the present finite-cylinder disk-to-rod sweep is actually carried by the same $J_0(H)$, $J_R(H)$ definitions, operator closure, and exact precession relation throughout, without changing the solver interface & Table~2, \eqref{e:38}--\eqref{e:39}, Appendix~D & (i) D.2--D.4 prove and numerically demonstrate the unchanged two-integral path across the entire sweep; (ii) Brand~\cite{ref37} independently validates the exact precession equation and the finite-cylinder inertia ingredients, and Klein and Sommerfeld~\cite{ref38} the exact precession equation; neither is cited as publishing the sweep or the same continuous geometry-to-dynamics methodology & \textbf{SUPPORTED} \\[2pt]
\midrule
26 & \S\ref{sec:1} / C4, C5, C7, C17: the interpretation --- free precession read as a dynamic equilibrium of torque vectors about the pivot, held from the mass element to the governing equation, with the constant tilt read as torque equalization and the restoring imbalance computed on either side of it; the balance itself, and its equalization at the steady tilt, are shared with Brand and with Klein and Sommerfeld, the start from Newton's law for each mass element with Brand, and the inertia turning built from mass-element accelerations with Klein and Sommerfeld's second derivation & \eqref{e:20}, \eqref{e:25}, \eqref{e:32}, \eqref{e:35}, \eqref{e:39} & (i) shown in \S\ref{sec:5}--\S\ref{sec:7}: each mass element enters as the infinitesimal inertia torque~\eqref{e:20}, and the balance~\eqref{e:32} and the unified equation~\eqref{e:35} are complete before the scalar~\eqref{e:39}; the equalization is computed in Appendix~F (Table~F.2); (ii) the ancillary comparison file, each feature checked against the books at the place given: rows~8 and 9 shared with Brand and with Klein and Sommerfeld, row~6 with Brand and row~7 with Klein and Sommerfeld, and rows~4, 5, 12 and 13 this paper's alone among the three works & \textbf{SUPPORTED} \\
\end{longtable}}

\newcommand{\TABLEFONE}{%
\setlength{\tabcolsep}{6pt}%
\begin{tabular}{>{\raggedright\arraybackslash}p{0.46\textwidth}lc}
\toprule
Term of equation~(35, which is this paper's unified vector equation of motion for steady precession), written as it stands on the side opposite the gravity torque $M \vv{R} \times \vv{g}$ & $X$-component & Unit \\
\midrule
$M(\vv{\Omega}\cdot\vv{R})(\vv{R}\times\vv{\Omega})$ & $+\,M\Omega^{2}R^{2}\sin\theta\cos\theta$ & N$\cdot$m \\[2pt]
\midrule
$J_0(\vv{\Omega}\times\vv{\omega}_0)$ & $-\,J_0\Omega\omega_0\sin\theta$ & N$\cdot$m \\[2pt]
\midrule
$-\,\vv{F}(\vv{\Omega}\times\vv{\omega}_0)$ & $+\,J_R\Omega\omega_0\sin\theta$ & N$\cdot$m \\[2pt]
\midrule
$\vv{F}(\vv{\Omega})\times\vv{\Omega}$ & $0$\quad (parallel vectors) & N$\cdot$m \\[2pt]
\midrule
$\vv{F}(\vv{\Omega})\times\vv{\omega}_0$ & $-\,J_R\Omega\omega_0\sin\theta$ & N$\cdot$m \\[2pt]
\midrule
$\vv{F}(\vv{\omega}_0)\times\vv{\Omega}$ & $+\,J_R\Omega\omega_0\sin\theta$ & N$\cdot$m \\
\bottomrule
\end{tabular}}

\newcommand{\TABLEFTHREE}{%
\setlength{\tabcolsep}{6pt}%
\begin{tabular}{crrc}
\toprule
$\theta$ (degrees) & $|M\vv{R}\times\vv{g}|$ (N$\cdot$m) & $|\Mscr_A|$ (N$\cdot$m) & ratio (dimensionless) \\
\midrule
$43^\circ$ & $1.336716786$ & $1.327396557$ & $0.9930275$ \\[2pt]
\midrule
$44^\circ$ & $1.361530406$ & $1.356740891$ & $0.9964823$ \\[2pt]
\midrule
$\mathbf{45^\circ}$ & $\mathbf{1.385929291}$ & $\mathbf{1.385929291}$ & $\mathbf{1.0000000}$ \\[2pt]
\midrule
$46^\circ$ & $1.409906009$ & $1.414953024$ & $1.0035797$ \\[2pt]
\midrule
$47^\circ$ & $1.433453255$ & $1.443803093$ & $1.0072202$ \\
\bottomrule
\end{tabular}}

\newcommand{\TABLEFFOUR}{%
\setlength{\tabcolsep}{3pt}%
\begin{tabular}{>{\raggedright\arraybackslash}p{0.28\textwidth}>{\raggedright\arraybackslash}p{0.28\textwidth}>{\raggedright\arraybackslash}p{0.21\textwidth}c}
\toprule
Source or method & What was tested & Agreement & Verdict \\
\midrule
Brand~\cite{ref37}, \emph{Vectorial Mechanics} (1930) & both equalization rates, exact, every non-vertical tilt except the horizontal shaft & exact --- the same quadratic & PASS \\[2pt]
\midrule
Bolina~\cite{ref27}, arXiv:physics/0005025 & both equalization rates, exact, every non-vertical tilt except the horizontal shaft & $1.8\times10^{-15}$ rad/s & PASS \\[2pt]
\midrule
MIT OpenCourseWare 16.07, Lecture~L30~\cite{ref24} & exact steady-precession balance & machine zero & PASS \\[2pt]
\midrule
Goldstein \S5.7~\cite{ref2}; Thornton \& Marion \S11.11~\cite{ref4} & minimum-spin condition (D8, which is the textbooks' minimum-spin condition, written in their symbols $I_1, I_3$ and $l$, Thornton and Marion writing $h$ for $l$, and Goldstein writing it with the strict inequality), evaluated at this configuration's $\omega_0 + \Omega \cos \theta$ & satisfied: 159.73392 rad/s against 123.47154 rad/s & PASS \\[2pt]
\midrule
Agashe, PHYS 601, Maryland~\cite{ref26} & uniform-precession condition, discriminant & $6.7\times10^{-16}$ N$\cdot$m & PASS \\[2pt]
\midrule
Elementary cross product & gravity torque, all five tilts & $\le 3.4\times10^{-10}$ N$\cdot$m & PASS \\[2pt]
\midrule
Classical $\vv{L}$ and $-\,d\vv{L}/dt = -\,\vv{\Omega}\times\vv{L}$ & inertia torque, all five tilts & $\le 4.8\times10^{-10}$ N$\cdot$m & PASS \\
\bottomrule
\end{tabular}}

\newcommand{\TABLESYM}{%
\small
\renewcommand{\arraystretch}{0.99}%
\setlength{\tabcolsep}{6pt}%
\begin{tabular}{@{}p{0.15\textwidth}p{0.79\textwidth}@{}}
\toprule
\multicolumn{2}{@{}l}{\textit{Geometry and the body}} \\
\midrule
$M$ & total mass of the body, constant \\
$A$ & the fixed pivot, at the shaft end \\
$C$ & the center of mass \\
$P$ & a generic material point of the body \\
$dm$ & the mass element at $P$ \\
$\vv{R}$ & $= \vv{AC}$, the pivot-to-COM vector, directed along the shaft \\
$\vv{\xi}$ & $= \vv{CP}$, the COM-relative material vector of $P$ \\
$\vv{r}$ & $= \vv{AP} = \vv{R} + \vv{\xi}$, the position of $P$ from $A$ \\
$\theta$ & tilt angle of the shaft $\vv{R}$ from the upward vertical \\
$D$, $E$, $H$ & outer and inner diameter, and axial height, of the swept solids \\
\midrule
\multicolumn{2}{@{}l}{\textit{Kinematics}} \\
\midrule
$\vv{\omega}$ & angular velocity of the chosen frame of the transport rule~\eqref{e:6}, \S\ref{sec:3.3}; for the rigid body, from~\eqref{e:14} on, the total angular velocity of the body \\
$\vv{\omega}_0$ & shaft spin, of constant magnitude, carried around by the precession \\
$\vv{\Omega}$ & precession about the vertical through $A$ \\
$\vv{\varepsilon}$ & $= d\vv{\omega}/dt$, the angular acceleration of the chosen frame of~\eqref{e:6}, and of the body from~\eqref{e:14} on; for steady precession, $\vv{\varepsilon} = \vv{\Omega} \times \vv{\omega}_0$ by (33, which is the composition $\vv{\omega} = \vv{\Omega} + \vv{\omega}_0$ of the steady motion and its time derivative) \\
$d/dt$ & inertial time derivative, following the same material element \\
$\partial/\partial t$ & companion derivative defined by the transport rule~\eqref{e:6}, never an independent physical input \\
\midrule
\multicolumn{2}{@{}l}{\textit{Inertia --- the whole apparatus of the method}} \\
\midrule
$J_0$ & $= \int(\vv{\xi}\cdot\vv{\xi})\,dm$, the polar second moment about $C$; non-negative, frame-invariant, additive \\
$\vv{F}$ & the mass-vector operator of~\eqref{e:2}; linear, additive, and satisfying the trace identity~\eqref{e:12} \\
$\vv{v}$ & the argument of $\vv{F}$ in~\eqref{e:2}: any vector the operator acts on; the velocity of a material point is written $\vv{v}_P$ (\S\ref{sec:3}) \\
$J_R$ & $= \vv{R}\cdot\vv{F}(\vv{R})/(\vv{R}\cdot\vv{R})$, the closure scalar the operator produces on its own \\
$J_\perp$ & the response of $\vv{F}$ to every vector across the shaft of the axisymmetric body, $\tfrac{1}{2}(J_0 - J_R)$, \S\ref{sec:7.4}; not the classical transverse moment $I_\perp = \tfrac{1}{2}(J_0 + J_R)$ \\
$\vv{S}$ & the internal spin integral~\eqref{e:13} \\
$K$ & the combination $I_\perp + MR^{2} - I_\parallel$ appearing in the solved relation \\
$I_\parallel$, $I_\perp$ & classical axial and transverse moments --- verification vocabulary only, \S\ref{sec:8.3} and Appendix~D, and in Appendix~F, where the textbooks' $I_3$ is $I_\parallel$ and their $I_1$, taken about the fixed point, is $I_\perp + MR^{2}$ \\
\midrule
\multicolumn{2}{@{}l}{\textit{External quantities and torque}} \\
\midrule
$\vv{g}$ & uniform gravitational field \\
$\Mscr_A$ & the total inertia torque about $A$ \\
$d\Mscr_P$ & $= \vv{r}_P\times(-\,\vv{a}_P\,dm)$, the infinitesimal inertia torque of the element at $P$ about $A$ \\
\midrule
\multicolumn{2}{@{}l}{\textit{Abbreviations}} \\
\midrule
COM & center of mass \\
LHS & left-hand side \\
rpm & revolutions per minute \\
\bottomrule
\end{tabular}%
}

\begin{document}
\maketitle
\pdfbookmark[1]{Abstract}{abstract}%
\addtocontents{toc}{\protect\contentsline{subsection}{Abstract}{\thepage}{abstract.1}}%
\begin{abstract}
\noindent This paper presents a self-contained, vector-only formulation of gyroscope
precession about a fixed pivot --- gravity-driven, not the torque-free Euler--Poinsot
case. The derivation of its governing equation uses vector integrals and cross products of intrinsic vectors, with no component decomposition; no tensor or dyadic formalism, matrix or Euler angle enters. Inertia about the center of mass enters through two mass integrals alone: the polar second moment of mass $J_0$ (isotropic) and the vector operator $\vv{F}$ (the rest, including all non-isotropic inertia).

\medskip\noindent Using material time derivatives referenced to the inertial observer,
the torque about the pivot is assembled once from the complete acceleration of each
mass element, giving an exact, unified vector equation for steady precession. The
balance of the gravity torque against this inertia torque reproduces the classical
steady-precession relation of the heavy symmetric top exactly at non-vertical tilts, on
both the slow and fast branches, with the horizontal-shaft case exact rather than
approximate. Closed forms are given for the disk, sphere, and circular and rectangular
toruses. The constant tilt angle of dynamic equilibrium is explained as torque
equalization: the two torques about the pivot balance exactly at that tilt and
unequally on either side of it, treated throughout as a geometric function of the tilt
under constant precession and self-spin rates, and never as a function of time. Informationally equivalent to the classical description, the work contributes, in order, an interpretation (the classical balance read in torque vectors throughout) and the vector method as its tool. The exact relation predates this work: Klein and Sommerfeld stated it in 1897, Brand in 1930. Begun by the author in 1975 and re-verified symbolically, the work is intended for undergraduate readers, with every step of the derivation stated.
\end{abstract}


\bigskip
\section*{\phantomsection Contents}\addcontentsline{toc}{section}{Contents}
\makeatletter\@starttoc{toc}\makeatother
\clearpage

\begin{table}[H]
\centering
\TABLESYM
\caption{Table~1 --- Symbols and abbreviations used throughout this paper. Most
entries are defined by the entry itself; where a symbol is fixed by a numbered equation
or in a particular section, that reference is given with it.}
\end{table}
\clearpage

\section*{\phantomsection Author's Note}
\addcontentsline{toc}{section}{Author's Note}
\label{sec:note}

This paper reaches the classical steady-precession relation of the heavy symmetric top. To the yes-or-no question of whether it adds new physics, the correct answer is no. This note records what that question does not measure: the order in which the work came about.

A physical phenomenon is what it is whether or not it has been explained. The same observable result can be reached by more than one mathematical description, and the ancillary comparison file follows three such routes to one relation. Two words are easily confused here: methodology and interpretation. A methodology answers the question of how a result is obtained: it describes a procedure, most often a mathematical one. An interpretation answers the question of why a phenomenon behaves as it does: it is a reading of the physics. A methodology can follow from an interpretation.

The work recorded in this paper was a physical inquiry carried out in steps. It began with an interpretation, not with a method. Its starting point was the question of why a spinning gyroscope, supported at one end of its shaft, turns slowly about the vertical at a constant tilt instead of falling. Put in terms of torques, the question was which torque holds the gravity torque in balance. What the author lacked at that stage was not a mathematical technique but a physical reading of that balance. The reading the author reached was a balance of torques about the pivot. Each mass element of the spinning, precessing body contributes, from the start, an infinitesimal inertia torque about the pivot (\S\ref{sec:5}). The sum of those contributions stands in dynamic equilibrium with the gravity torque (\S\ref{sec:7}).

The mathematics followed from that reading. Summing the inertia torques over the body called for integrals over the mass distribution. The definitions of equations~\eqref{e:1} and~\eqref{e:2}, the polar second moment of mass $J_0$ and the mass-vector operator $\vv{F}(\vv{v})$, which acts on any vector $\vv{v}$, came from that summation. Inertia about the center of mass enters the derivation through those two mass integrals alone. Keeping every contribution a torque vector about the pivot called for cross products carried through to the torque-vector balance. The vector derivation of \S\ref{sec:3} to \S\ref{sec:7} is that reading written out. The order was: the physical interpretation, then the definitions it required, then the mathematical method. The inquiry produced a mathematical model of the same observed motion, informationally equivalent to the classical description (\S\ref{sec:10}). The route to that model was set by the physical reading.

This work has one motivating cause and one tool: the interpretation is the cause, and the method is the tool of its explanation. The contribution of this paper is both, in that order. \S\ref{sec:1} states them in full, with what the interpretation shares with Brand~\cite{ref37} and with Klein and Sommerfeld~\cite{ref38} and what, among the features compared in the ancillary comparison file, is this paper's alone. This note is the author's account of how the work arose.

\clearpage
\hsection{1}{Introduction}\label{sec:1}

Gyroscope precession is counterintuitive to many students of classical mechanics. A spinning rotor, supported at one end of its shaft by a fixed pivot under gravity, turns slowly
about the vertical through the support at a constant tilt instead of falling as a non-spinning body
would --- the motion known as free precession. In this paper \emph{free precession} denotes the gravity-driven
steady precession of a pivoted gyroscope --- free in the sense that no drive or
constraint is imposed on the precession, not in the sense of the torque-free
Euler--Poinsot motion of an unsupported body, which lies outside the scope stated
above. The phenomenon is not a curiosity. It underlies inertial
navigation, gyrocompasses, spacecraft attitude control and vehicle stabilization, and its
mathematical description traces back to Euler's equations of rigid-body motion
(1765)~\cite{ref1}.

\begin{figure}[htbp]
\centering
\includegraphics[width=0.52\textwidth]{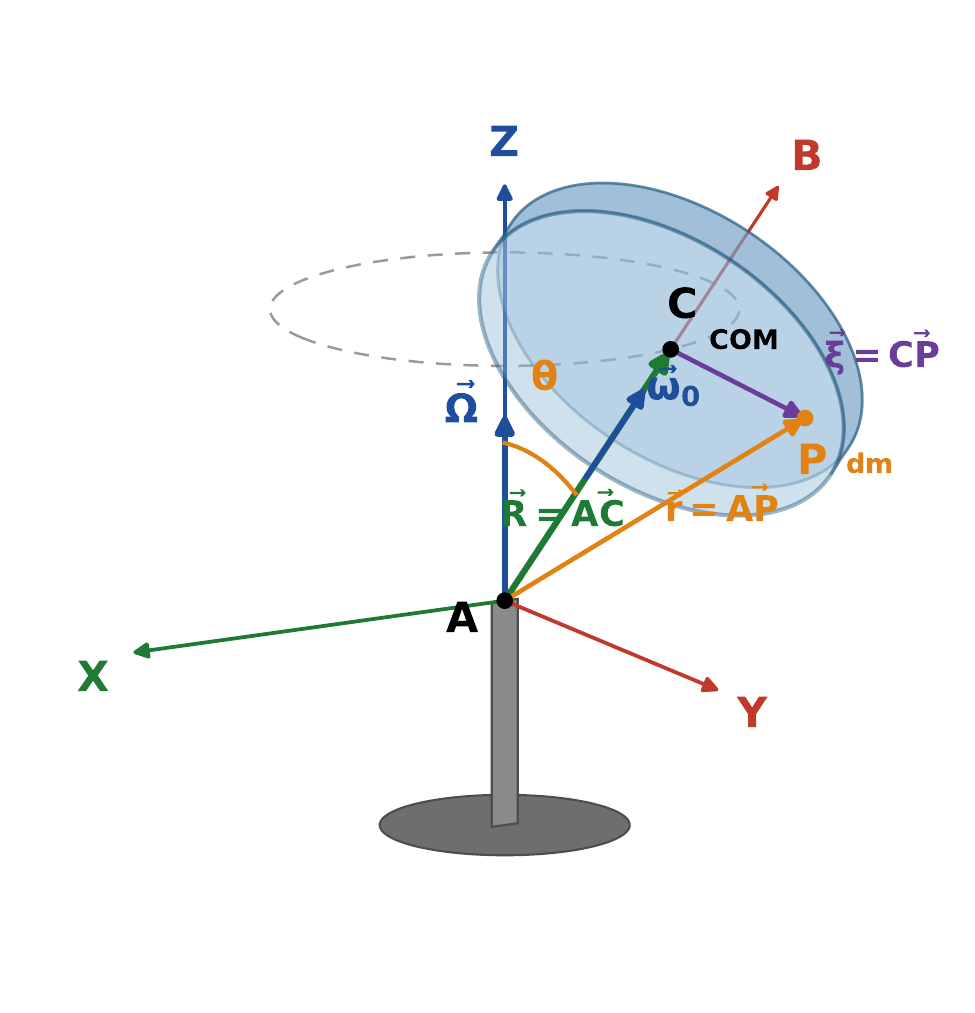}
\caption{Figure~1 --- A gyroscope precessing about a fixed pivot: the shaft, along the direction $B$, tilted from the upward vertical, the rotor spinning about the shaft, and the slow sweep of the
shaft about the upward vertical through the pivot.}\label{fig:1}
\end{figure}

Yet a persistent gap separates the maturity of that mathematical description from an
intuitive physical understanding of it. Complete treatments --- Euler angles, Lagrangian
methods, the inertia tensor and its principal axes --- resolve vectors into components
early, dissolving the geometric picture that makes the motion intelligible; introductory
treatments preserve the picture but, Bolina's short note~\cite{ref27} apart, deliver only the fast-top approximation for the
precession rate. Classical literature holds exact treatments as well: two referenced works ~\cite{ref37} and ~\cite{ref38} are compared feature by feature with this paper. Each of these two referenced works, as the ancillary comparison file shows, obtains the steady-precession relation at any non-vertical tilt, with both of its roots and without the fast-top approximation, and each sets the rate of change of the angular momentum, as a vector, equal to the moment of the external forces. Klein and Sommerfeld~\cite{ref38} lead that vector, which they call the impulse, around the vertical and read its change geometrically as the velocity of its end point; Brand~\cite{ref37} forms it as the mass integral $\vv{H} = \int \vv{r} \times \vv{v}\,dm$, his $\vv{v}$ being the velocity of the mass element, and writes the governing statement as the single vector equation $\vv{\Omega} \times \vv{H} = \vv{M}$. (In Brand's notation $\vv{H}$ is the moment of momentum and $\vv{M}$ the moment of the external forces; neither is the height $H$ or the mass $M$ of this paper.) Neither keeps the vectors to the balance: Brand writes $\vv{H}$ in components on principal axes chosen at the fixed point, Klein and Sommerfeld carry the impulse and its change in components on the body frame and the line of nodes, and each builds the scalar equation up from those components. Apart from Bolina's short note~\cite{ref27}, which keeps the angular-momentum picture for the steady-precession rate alone, the literature thus bifurcates: the treatments that keep the picture are incomplete, and the complete ones
leave it for components on the way to their result. The present work is addressed to that gap. It develops the full
dynamics of free precession in a single algebra --- vector integrals over the mass distribution and cross products of intrinsic vectors, with intrinsic dot products in a secondary role --- carried unbroken
from the first kinematic statement to the final equation of motion, with the direction
of every torque contribution readable at every step by the corkscrew (right-hand) rule.

The formulation rests on two objects defined directly as mass integrals about the
center of mass: the polar second moment of mass
\begin{equation}\tag{1}\label{e:1}
J_0 = \int_0^M (\vv{\xi}\cdot\vv{\xi})\,dm
\end{equation}
and a mass-vector operator
\begin{equation}\tag{2}\label{e:2}
\vv{F}(\vv{v}) = \int_0^M (\vv{\xi}\cdot\vv{v})\,\vv{\xi}\,dm
\end{equation}
which accepts any vector and returns a vector, with no reference direction, axis
choice, or component decomposition entering its definition. Directional inertia
information is carried by the operator itself and is extracted, per body, by
evaluating it --- directions emerge as outputs of the analysis, never as inputs to
the definitions.

Relation to the classical inertia tensor. A reader who knows the tensor description will recognize definitions~(1) and (2), so
the correspondence is worth stating once. The operator $\vv{F}$ of (2) is the action, on
a vector, of the body's second moment of mass: that quantity is usually written as a
standalone tensor that multiplies a vector, whereas (2) already carries its input vector
$\vv{v}$ --- so $\vv{v}$ is not missing: it is preloaded, encapsulating from the
beginning the vector the operator acts upon during the vectors-only derivations.
$\vv{F}$($\vv{v}$) is an integral entity that emerges in these vector calculations from
the calculations themselves, and it is identical to a tensor-calculus by-product. The
scalar $J_0$ of (1) can be interpreted equivalently as that same operator's trace ---
the single number formed by summing its diagonal, which equals
$\int_0^M (\vv{\xi}\cdot\vv{\xi})\,dm$ --- or, as is the case in this paper's
derivation, $J_0$ emerges from performing an integral and does not result from a tensor
trace. Taken together the two would reconstruct the classical inertia tensor, as
$\mathbf{I} = J_0\,1\!\!1 - \vv{F}$; the second moment of mass itself is what Binney and
Tremaine~\cite{ref28} call the moment of inertia tensor (their eq.~(8-11)), which they use for systems of freely moving particles in space, not for a rigid body in the dynamic equilibrium of steady gyroscopic precession; but this paper neither
forms nor uses I, and proceeds throughout on $J_0$ and $\vv{F}$ directly. The correspondence is a
recognition, not a route: $J_0$ and $\vv{F}$ coincide with these tensor by-products, yet
here they arise from the vector algebra alone and are never imported from any tensor
product, projection, or contraction. Retaining $\vv{F}$ in vector-valued form is what preserves the
three-dimensional representation: its output feeds directly into intrinsic cross products, whose geometric reading --- the right-hand rule, the corkscrew of Appendix~A
--- remains available until the final scalar reduction.

Working with material time derivatives referenced to the inertial observer, the
torque about the fixed pivot $A$ is assembled once from the complete acceleration of
each mass element, giving the exact statement
\begin{equation}\tag{3}\label{e:3}
\Mscr_A = -\,M\vv{R}\times\frac{d^{2}\vv{R}}{dt^{2}}
- \vv{R}\times\int_0^M \frac{d^{2}\vv{\xi}}{dt^{2}}\,dm
- \int_0^M \vv{\xi}\times\frac{d^{2}\vv{\xi}}{dt^{2}}\,dm
\end{equation}
the balance of the gravity torque $M\vv{R}\times\vv{g}$ against it, their resultant being
the zero vector, yields a unified vector equation of motion for steady precession, whose scalar reduction reproduces
the classical relation of the heavy symmetric top exactly --- both precession
branches from one quadratic, with the horizontal-shaft case holding exactly rather
than as a fast-spin approximation.

The scope of that method claim is worth stating plainly, and once. No tensor, matrix or dyadic formalism, and no Euler angle, enters the symbolic derivation presented here, or the
final exact symbolic solution it produces for dynamic-equilibrium free precession, at any stage. The
complete symbolic vector derivation culminating in the governing torque-vector equation~\eqref{e:35}
uses, in addition, no unit vectors, coordinate projections, principal-axis decomposition, or component
resolution; the physical vectors remain intrinsic vectors throughout. Unit vectors enter only
afterward --- the three of the trace identity~\eqref{e:12} in the auxiliary closure of \S\ref{sec:7.4},
and any others in the scalar reduction, the verification, or the numerical evaluation, where the
per-solid reduction of Appendix~C and the numerical verification of Appendix~F say so --- and those
operations are not part of the claimed unit-vector-free derivation of the governing torque-vector
equation. That identity enters the symbolic derivation once, in \S\ref{sec:7.4}, where it fixes the
coefficient across the shaft in the closure~\eqref{e:37}; it is also what fixes $J_R = J_0/3$ for the
isotropic bodies of Appendix~C, and elsewhere it serves only as an internal consistency check on
$J_0$. No vector of the motion is resolved along its three directions, there or anywhere else. Dot
products are used as intrinsic geometric operations, not as coordinate projections. The governing equations remain vector equations through (38); the scalar
relation~(39) is introduced only after \S\ref{sec:7.5} establishes geometrically that
every term in the torque-vector balance lies along the same physical direction. Numerical
evaluation is subsequent to that symbolic reduction. The claim is about the derivation
and its solution, and about nothing wider: the numerical verification of Appendix~F,
which exists to compare this work with the classical description on that description's
own terms, does work in components, and is presented there as verification rather than
as derivation.

The claim of this paper is deliberately calibrated. The physics recovered is classical,
and on the common ground of the axisymmetric rigid rotor the formulation is
informationally equivalent to the standard description. The steady-precession relation with both of its branches, holding at any non-vertical tilt without the fast-top approximation, and its horizontal-shaft case, holding exactly rather than as the fast-top limit, are prior results: Klein and Sommerfeld~\cite{ref38} stated them in 1897 and Brand~\cite{ref37} in 1930; they are reproduced here, not claimed. This work has one motivating cause and one tool. The cause is a physical interpretation: free precession read as a dynamic equilibrium of torque vectors about the pivot, held from the mass element to the governing equation. Each mass element enters from the start as an infinitesimal inertia torque. The governing statement is a torque-vector equality, complete before any scalar is taken. The constant tilt is the tilt at which the gravity torque and the inertia torque equalize. The tool is the vector-only method that carries out the explanation, with its economy of definition, its single algebra throughout and its geometric visualization preserved. The contribution of this paper is both, in that order: the interpretation, then the method. In detail, the economy of definition means that no unit vectors, axes, or perpendicularity constructions precede the definitions. The single algebra solves no eigenvalue problem and has no change of formalism anywhere; the principal directions of $\vv{F}$ enter only through the body's symmetry, first at the step to the governing equation~\eqref{e:35}, where $\vv{F}$ returns the symmetry-axis vector parallel to itself, and then in the closure of \S\ref{sec:7.4}. The method also reaches the exact results, shared with the classical exact treatments, with the familiar approximations recovered as special cases; and it gives a uniform two-integral computation recipe applied identically to a disk of any thickness, a sphere, and toruses of circular and rectangular cross-section.

The balance of the gravity torque against the inertia torque is shared with Brand and with Klein and Sommerfeld (rows~8 and 9 of the ancillary comparison file). Brand also starts the gyroscope equation from Newton's law applied to each mass element, taken as a moment about the fixed point (\S188~\cite{ref37} and \S214~\cite{ref37}; row~6). Klein and Sommerfeld's second derivation also builds the inertia turning from the accelerations of the mass elements (row~7). Among the features compared in the ancillary comparison file, this paper's alone are the formulation in intrinsic vectors, in which neither $J_0$ nor $\vv{F}$ presupposes an axis, a preferred direction, or a component resolution (row~4); the torque vectors carried through the whole balance with the direction of every term readable (row~5); the operator $\vv{F}(\vv{v})$, which accepts any vector and returns a vector (row~12); and, around the steady tilt at which all three works equalize the two torques, the restoring imbalance computed on either side of it (row~13). The literature reviewed in \S\ref{sec:2} contains no prior publication presenting this complete methodological combination. The rates this recipe produces have been checked against seven independently published exact formulations of the heavy symmetric top --- Klein and Sommerfeld~\cite{ref38}, Brand~\cite{ref37}, Massachusetts Institute of Technology (MIT) OpenCourseWare~\cite{ref24}, Tanr\i{}verdi~\cite{ref25}, Agashe~\cite{ref26}, Bolina~\cite{ref27} and Slade~\cite{ref33} --- which reduce to the same relation once notation is mapped; the comparisons are summarized in \S\ref{sec:8.4} and Appendix~D.5.

A first version of these derivations was completed by the author in 1975. That paper was privately
shared in 1975 with Professor Eric Laithwaite of Imperial College London, whose reply
acknowledged the notes, reported related work at Imperial toward almost the same
equations, and congratulated the author on the result; Appendix~B records the
correspondence. The author also gave a talk at a conference at the Romanian Academy
of Sciences in Bucharest in 1975; no proceedings from that conference are known to
have been published. The present paper is the complete, modern rewrite of that work,
with every symbolic solution re-verified. The historical fragment is reproduced in
Appendix~B (Figure~8).

This paper is organized as follows. \S\ref{sec:2} positions the work within the
literature. \S\ref{sec:3} fixes notation, the derivative convention, and the kinematic
foundations. \S\ref{sec:4} defines the operator framework and its closure identities.
\S\ref{sec:5} defines the infinitesimal torque of each mass element about the pivot;
\S\ref{sec:6} assembles the total torque about the pivot exactly once from the material
acceleration and establishes the two-route consistency theorem. \S\ref{sec:7} states the
dynamic-equilibrium balance and solves steady precession. \S\ref{sec:8} resolves the
balance in closed form for the three general bodies and their classical limits and
reports the independent
numerical confirmation. \S\ref{sec:9} discusses merits, scope, and limits, and
\S\ref{sec:10} concludes. Six appendices follow: the corkscrew-rule visualization
method (A), the 1975 historical fragment (B), extended per-solid algebra (C), the exact
continuity of the disk-to-rod transition with its external validation (D), a
claim-by-claim validation of the assertions this paper makes about itself (E), and the
complete numerical demonstration of torque equalization at dynamic equilibrium (F). The feature-level comparison with Brand's Vectorial Mechanics (1930) and Klein and Sommerfeld's The Theory of the Top (1897--1910) is provided as the ancillary comparison file, \texttt{Feature\_\allowbreak{}Comparison\_\allowbreak{}Three\_\allowbreak{}Works.pdf}.

\hsection{2}{Related Work and Literature Positioning}\label{sec:2}

The literature on gyroscope precession contains several overlapping traditions. For
present purposes the decisive distinction is not whether a source uses vectors --- Brand~\cite{ref37} does, and so do Klein and Sommerfeld~\cite{ref38} --- but whether vectors remain the working
three-dimensional representation of the derivation. The question is whether the physical
vectors and their cross-product directions are retained from the mass-element
description all the way to the exact torque balance, or whether the method changes
representation by projecting onto selected axes, replacing vectors by components, and
carrying the dynamics thereafter through projected inertia moments or another formalism.
That distinction separates the present derivation from Brand, and from Klein and Sommerfeld, while leaving their common mathematics, physics, and end results undisputed.

The pedagogical tradition emphasizes the immediate vector picture. Kleppner and
Kolenkow~\cite{ref6}, the Feynman Lectures~\cite{ref7}, and Kostov and Hammer's
continuation of Feynman's discussion~\cite{ref31} develop gyroscopic response through
torque and angular-momentum vectors, usually in the fast-top regime. Butikov's
inertial-frame treatment~\cite{ref11} is particularly close in spirit to the present
work because it argues against routing torque-induced precession through body-frame
Euler equations; it writes its precession law as a vector cross product, but with the angular momentum taken along the symmetry axis, and beyond that approximation it proceeds through scalar magnitudes, angles given in words, and a
decomposition of angular momentum and does not produce the exact operator balance
derived here.

Louis Brand's Vectorial Mechanics (1930)~\cite{ref37} is a close historical predecessor. Brand develops
rigid-body dynamics with vector products and mass integrals, treats the center of mass
and moments of inertia, and in \S214 (\emph{Kinetics of a Rigid Body with One Point
Fixed}, p.~498, \cite{ref37}) gives an exact vectorial
treatment of a symmetric body with one point fixed, carrying it into \S217
(\emph{Gyroscope}, p.~502, \cite{ref37}) and \S218 (\emph{Steady Precession}, p.~505,
\cite{ref37}). He begins the fixed-point angular momentum from the mass integral $\vv{H}$ =
$\int \vv{r}\times(\vv{\omega}\times\vv{r})\,dm$, uses the right-hand screw to discuss
gyroscopic direction, derives the exact steady-precession quadratic, obtains two, one,
or no real precession rates according to the discriminant, and recovers the
horizontal-shaft rate exactly. Those physical results and their vector treatment are
therefore prior work and are not claimed as new here. The agreement of the final
equations is not disputed; it is one of the principal checks that the two derivations
describe the same classical physics.

Brand's route is nevertheless structurally different from the present one. He begins
vectorially but then changes representation: a moving orthogonal triad of principal axes
is selected, angular velocity and angular momentum are resolved into components, and
$\vv{H}$ is written as $A\vv{\omega}_x + B\vv{\omega}_y + C\vv{\omega}_z$, with $\vv{\omega}_x$, $\vv{\omega}_y$, $\vv{\omega}_z$ the projections of $\vv{\omega}$ on the three axes and $A$, $B$, $C$ the conventional principal moments. The exact steady-precession equation is then obtained in that
principal-axis/component representation. The present derivation does not make that
conversion. $J_0$ = $\int$($\vv{\xi}\cdot\vv{\xi}$)dm and $\vv{F}$($\vv{v}$) =
$\int$($\vv{\xi}\cdot\vv{v}$)$\vv{\xi}$ dm are defined about the center of mass before
any direction is selected, $\vv{F}$ remains an arbitrary-input vector operator, $J_R$
emerges from $\vv{F}$($\vv{R}$)=$J_R\vv{R}$, and the separate torque contributions
remain physically legible cross-product terms through the operator balance~(35) and
cross products of the intrinsic vectors through the resolved torque-vector balance~(38). Only after those vector terms are shown to have one
common physical direction is the scalar relation~(39) taken. Brand and the present work
therefore use the same mathematics and recover the same physics, but the derivation
methods are distinctly different: Brand changes to a principal-axis/component
representation; the present method preserves the intrinsic vector representation
throughout.

Felix Klein and Arnold Sommerfeld's \emph{The Theory of the Top}~\cite{ref38}, published in German between 1897 and 1910 and used here in the English translation of R. J. Nagem and G. Sandri, reaches the same exact result earlier still; it is the earliest of the treatments compared here, and its authors trace the integral formulas of the heavy symmetric top to Lagrange, who first considered its general problem (Vol.~II, p.~224,~\cite{ref38}). In \S III.6 of Volume~I (\emph{The deviation resistance for regular precession of the symmetric top}, p.~174,~\cite{ref38}) they state the exact steady-precession condition as their equation~(3) (p.~177,~\cite{ref38}), give its two roots in closed form as their equation~(4), name them the slow and the fast precession, and state that a given spin and tilt admit two steady precessions or none (p.~178,~\cite{ref38}). They treat the horizontal shaft (p.~176,~\cite{ref38}), and they record that when $A = C$ their equation becomes linear and its second root becomes infinitely large (pp.~178--179,~\cite{ref38}), the degenerate case this paper meets as $K = 0$ in Appendix~D.7. In \S III.7 (\emph{A new derivation of the deviation resistance [\ldots] The Coriolis force}, p.~180,~\cite{ref38}) they obtain the deviation resistance a second time, for the horizontal case $\vartheta = 90^\circ$,
from the accelerations of the individual mass particles. These results, too, are prior work and are not claimed as new here. Their route, like Brand's, is carried in components: the impulse, their name for the angular momentum, is led around the vertical, and the deviation resistance is built up from the components of the impulse and of its change on the body frame and the line of nodes. Their work also carries an apparatus that this paper does not use: the rotation parameters and quaternions, the Euler angles, the Lagrange equations and elliptic functions. The present derivation differs from theirs as it differs from Brand's: it reaches the scalar relation~\eqref{e:39} by reduction from a complete torque-vector balance, not by building it up from components.

The comparison with both works has been carried out feature by feature rather than left as an impression. The ancillary comparison file compares this paper with Brand and with Klein and Sommerfeld on twenty-nine features: for each of the three works, the features that carry its own route to the steady precession and its result, thirteen of this paper's, twelve of Brand's and thirteen of Klein and Sommerfeld's; seven of them are on two or three of the lists and stand once. Each feature asks the same question of the three works. Each answer is Yes or No, all but four of them with what that work does, and each Yes gives the place in that work.

The broader literature supplies the remaining comparison points. Goldstein~\cite{ref2},
Landau and Lifshitz~\cite{ref3}, Thornton and Marion~\cite{ref4}, and Symon~\cite{ref5}
deliver the full heavy-top dynamics through Euler angles, Lagrangian methods, inertia
tensors and principal axes. Milne's Vectorial Mechanics (1948)~\cite{ref8}, Synge and
Griffith~\cite{ref9}, Rutherford~\cite{ref35}, Chorlton~\cite{ref36}, and
Kane~\cite{ref10} preserve a strong vector emphasis while retaining dyadics, inertia
tensors, reference triads, or component resolution. Hestenes~\cite{ref12} writes the rigid-body equations of motion without components through geometric algebra, keeping a body frame on the principal axes, while the Euler--Poincar\'e and
Lie--Poisson literature of Holm, Marsden and Ratiu~\cite{ref13} uses Lie-algebraic
structure; contemporary work --- Deriglazov on the heavy and dancing tops and the Grioli
gyroscope [14--17], Tanr\i{}verdi on an Euler-angle trajectory classification, its effective potential and, with Erbasan, the frictional casing study [18--20], Parker on the Ricci decomposition of the inertia
tensor~\cite{ref21}, Dragovi\'c, Gaji\'c and Jovanovi\'c on the heavy rigid body with a
gyroscope in n dimensions~\cite{ref22}, and Mityushov on the curvature
formulation~\cite{ref23} --- likewise proceeds through Euler--Poisson equations,
matrices, tensors, geometric algebra, integrable-systems machinery, Euler-angle Lagrangians with an effective potential, or one-dimensional friction models. Against that
record, the distinction claimed here is not vector notation, exact steady precession, or
visualization by itself. It is persistent intrinsic three-dimensional vector
representation: the COM-defined pair $J_0$ and $\vv{F}$ is retained throughout, the inertial-observer formulation is maintained, the scalar equation is
deferred until the exact torque-vector balance has a common direction, the same
geometry-to-solution method is actually used across finite thickness, and the torque
terms preserve continuous cross-product traceability. The conclusion is bounded to the searched record of Table~E.1, claim~24: a literature-search finding, not a
universal proof of priority.

\hsection{3}{Notation, Conventions, and Kinematic Foundations}\label{sec:3}

This section sets up the system and its vectors (\S\ref{sec:3.1}), the two time derivatives and the transport rule (\S\ref{sec:3.2}), rigid kinematics in inertial form (\S\ref{sec:3.3}) and the frame-bridge formula with its computation protocol (\S\ref{sec:3.4}).

\hsubsection{1}{System, vectors, and standing assumptions}\label{sec:3.1}

A single rigid body of constant total mass $M$ is pivoted at a point $A$ fixed in the
inertial (laboratory) frame. $C$ denotes the center of mass; $\vv{R} = \vv{AC}$ is the
pivot-to-COM vector, directed along the shaft; $\vv{g}$ is the uniform gravitational
field; and $P$ denotes a generic material point of the body, carrying mass element
$dm$ (see Figure~1). The position of $P$ relative to $A$ decomposes through the center
of mass as

\begin{equation}\tag{4}\label{e:4} \vv{r} = \vv{AP} = \vv{AC} + \vv{CP} = \vv{R} + \vv{\xi} \end{equation}

where $\vv{\xi} = \vv{CP}$ is the COM-relative material vector of $P$. The definition
of the center of mass supplies the identity used repeatedly below:

\begin{equation}\tag{5}\label{e:5} \int_0^M \vv{\xi}\,dm = \vv{0} \end{equation}

Standing assumptions for the core derivations: the pivot $A$ is fixed and
frictionless; the body is closed (constant mass, no flux); and gravity is uniform over the body.

\textbf{What this derivation does not use, and what it never assembles.}
Two restrictions hold throughout, and they are of different kinds.

\textbf{The first is representational:} it is the restriction stated in \S\ref{sec:1}. Inertia is
carried by two scalar mass integrals and one operator on vectors, not by an array of
components. The axial and transverse moments $I_\parallel$ and $I_\perp$ enter only in \S\ref{sec:8.3}, Appendix~D and Appendix~F (where the textbooks' $I_3$ is $I_\parallel$ and their $I_1$, taken about the fixed point, is $I_\perp + MR^2$), as the vocabulary in which the result is checked against the classical description, and never as a step in obtaining it.

\textbf{The second is dynamical, and it is the stronger of the two. At the level of the
body there is no force equation.} Each mass element enters the derivation only through its infinitesimal torques about $A$, each its
position crossed with its partner: gravity's, $\vv{r}_P\times(\vv{g}\,dm)$; its inertia torque,
$\vv{r}_P\times(-\,\vv{a}_P\,dm)$; and the torques of internal cohesion, which cancel pairwise, each
equal-and-opposite pair of internal actions lying along the line joining its two elements (Newton's
third law). The pivot contributes no torque about itself and is never solved for. No force resultant
is assembled anywhere in this work. What is assembled is a single torque balance, and it is exact
rather than an equilibrium assumption: macroscopically there are only torques, the infinitesimal
torques of the mass elements about $A$, integrated into one compact macroscopic torque balance.

Brand's treatment does otherwise, and the difference is part of the substance of the
ancillary comparison file: a force-sum equation written together with the moment equation
and an explicit solution for the pivot reaction (\S214~\cite{ref37} and \S217~\cite{ref37}) are apparatus this formulation is
constructed to do without; Klein and Sommerfeld carry neither. Nor does it use any force of a mass element by itself: \S\ref{sec:5} pairs each element's reversed
mass-acceleration with its position from the start, as the infinitesimal inertia torque~\eqref{e:20}.

The principal symbols and abbreviations are collected in Table~1, in the front matter; the rest are defined at or near their first use.

\hsubsection{2}{The two time derivatives and the transport rule}\label{sec:3.2}

Two time derivatives appear in this work, and the distinction between them governs
everything that follows. The symbol $d/dt$ denotes the inertial (absolute) time
derivative: the rate of change measured by the laboratory observer, following the same
material element $dm$ through time. This is the derivative that carries physical
meaning in Newton's law --- velocity is $\vv{v}_P = d\vv{r}/dt$ and acceleration is
$\vv{a}_P = d^{2}\vv{r}/dt^{2}$. The symbol $\partial/\partial t$ denotes the companion
(relative) derivative: the bookkeeping rate of change defined by the transport rule
below, computed from inertial data and never independently measured. The two are
connected, for any vector $\vv{A}$, by the transport rule

\begin{equation}\tag{6}\label{e:6} \frac{d\vv{A}}{dt} = \frac{\partial\vv{A}}{\partial t} + \vv{\omega}\times\vv{A} \end{equation}

Time derivatives of scalars are the same in both conventions; time derivatives of
vectors differ between them, and equation~\eqref{e:6} is the complete dictionary
between the two conventions. All derivatives in this paper are inertial ($d/dt$)
unless the symbol $\partial/\partial t$ appears explicitly; where it does appear ---
in the frame-bridge formula~\eqref{e:11} below --- it is computed from inertial data
via~\eqref{e:6}, never measured and never replaced by its inertial counterpart.

Applying~\eqref{e:6} twice to the same arbitrary vector $\vv{A}$ --- differentiating
\eqref{e:6} inertially, then using~\eqref{e:6} again on each $\partial$-quantity that
appears --- gives the second-derivative transport rule:

\begin{equation}\tag{7}\label{e:7} \frac{d^{2}\vv{A}}{dt^{2}} = \vv{\varepsilon}\times\vv{A} + \vv{\omega}\times(\vv{\omega}\times\vv{A}) + 2\vv{\omega}\times\frac{\partial\vv{A}}{\partial t} + \frac{\partial^{2}\vv{A}}{\partial t^{2}} \end{equation}

The derivation is two lines. Differentiating~\eqref{e:6}:
$d^{2}\vv{A}/dt^{2} = d/dt(\partial\vv{A}/\partial t) + (d\vv{\omega}/dt)\times\vv{A}
+ \vv{\omega}\times(d\vv{A}/dt)$. The first term, by~\eqref{e:6} applied to the vector
$\partial\vv{A}/\partial t$, is $\partial^{2}\vv{A}/\partial t^{2} +
\vv{\omega}\times(\partial\vv{A}/\partial t)$; the last term, by~\eqref{e:6} applied to
$\vv{A}$ itself, is $\vv{\omega}\times(\partial\vv{A}/\partial t) +
\vv{\omega}\times(\vv{\omega}\times\vv{A})$. Collecting, with
$\vv{\varepsilon} = d\vv{\omega}/dt$, produces~\eqref{e:7}. The factor 2 of the
Coriolis term is born here: one $\vv{\omega}\times(\partial\vv{A}/\partial t)$ arises
from~\eqref{e:6} acting on the vector $\partial\vv{A}/\partial t$, the other
from~\eqref{e:6} acting on $\vv{A}$ itself --- two distinct algebraic sources
delivering identical terms.

Like~\eqref{e:6}, equation~\eqref{e:7} is a pure kinematic identity connecting two
derivative conventions of one and the same vector: it holds for any $\vv{A}$, with no
physical assumption, no rigidity, and no restriction on how $\vv{A}$ moves. Its left
side, $d^{2}\vv{A}/dt^{2}$, is the inertial-material second rate --- the physical,
measurable quantity. Its right side re-expresses that same measurable quantity in four
named terms, each built algebraically from the inertial motion: the angular (Euler)
term $\vv{\varepsilon}\times\vv{A}$, from the rate of change of $\vv{\omega}$; the
centripetal term $\vv{\omega}\times(\vv{\omega}\times\vv{A})$; the Coriolis term
$2\vv{\omega}\times(\partial\vv{A}/\partial t)$; and the relative term
$\partial^{2}\vv{A}/\partial t^{2}$. The $\partial$-slots obey the same protocol as
in~\eqref{e:6}: they are computational bookkeeping terms defined by the transport
relation itself, never independently measured, and never to be filled with inertial
values --- that substitution double-counts the $\vv{\omega}$-physics. As
with~\eqref{e:6}, scalars are untouched, and $\vv{\varepsilon}$ is unambiguous since
$\vv{\omega}\times\vv{\omega} = \vv{0}$ makes
$d\vv{\omega}/dt = \partial\vv{\omega}/\partial t$.

Equation~\eqref{e:7} is the general-vector parent of two equations below: substituting
$\vv{A}\to\vv{r}$ yields the frame-bridge formula~\eqref{e:11} verbatim, and
substituting $\vv{A}\to\vv{\xi}$ yields the general internal kinematics.

\hsubsection{3}{Rigid kinematics in inertial form}\label{sec:3.3}

Let $P$ be a general material point with position vector $\vv{r}$ measured from the
fixed pivot $A$, and let $\vv{R}$ be the position vector of the center of mass. By the
decomposition already established,

\[ \vv{r} = \vv{R} + \vv{\xi} \]

with $\vv{\xi}$ the position of $dm$ relative to the center of mass. Nothing below
assumes the body is rigid: $\vv{\xi}(t)$ is an arbitrary --- possibly time-varying,
possibly deforming --- material vector, and $\vv{\omega}$ is the angular velocity of
the chosen frame of the transport rule~\eqref{e:6}, not necessarily the body's own
rotation rate; for the rigid body, from~\eqref{e:14} on, it is the body's total
angular velocity. All
derivatives that follow are material (they follow the same $dm$ through time, per the
standing convention), and the role of each is fixed permanently by its symbol: $d/dt$
is the inertial-material derivative --- the physical, measurable rate, the only one
ever assigned a value by measurement; $\partial/\partial t$ is the companion
bookkeeping derivative, defined by the transport relation~\eqref{e:6}, used only for
internal computation and never independently measured. Every equation displayed below
has its left-hand side in the inertial-material derivative; each is labeled
accordingly.

\textbf{First time differential, $d\vv{r}/dt = \vv{v}_P$.} It is the inertial
velocity. Apply the transport theorem directly to $\vv{r}$ --- no assumption about
$\vv{r}$'s motion is made or needed:

\[ \vv{v}_P = \frac{d\vv{r}}{dt} = \frac{\partial\vv{r}}{\partial t} + \vv{\omega}\times\vv{r} \qquad \text{(inertial left-hand side; general --- no rigidity assumed)} \]

The left side, $d\vv{r}/dt$, is the inertial-material velocity of $P$ --- the
physical, measurable quantity. The right side re-expresses that same measurable
velocity algebraically: $\partial\vv{r}/\partial t$ is the companion bookkeeping term
defined by the transport relation (left completely general --- not set to zero, not
assumed small, not assumed absent), and $\vv{\omega}\times\vv{r}$ is the term built
directly from $\vv{\omega}$ and $\vv{r}$. Because $\partial/\partial t$ is linear (it
commutes with vector addition, exactly as $d/dt$ does), the decomposition~\eqref{e:4}
carries through the derivative unchanged:

\[ \frac{\partial\vv{r}}{\partial t} = \frac{\partial\vv{R}}{\partial t} + \frac{\partial\vv{\xi}}{\partial t} \]

Nothing has been canceled and nothing has been assumed rigid: this is the complete,
general, inertial statement of the velocity of $P$.

\textbf{Second time differential, $d^{2}\vv{r}/dt^{2} = \vv{a}_P$.} It is the inertial
acceleration. Differentiate the first result inertially --- apply $d/dt$ to both sides
of the velocity equation above:

\[ \vv{a}_P = \frac{d^{2}\vv{r}}{dt^{2}} = \frac{d}{dt}\!\left(\frac{\partial\vv{r}}{\partial t}\right) + \left(\frac{d\vv{\omega}}{dt}\right)\times\vv{r} + \vv{\omega}\times\frac{d\vv{r}}{dt} \qquad \text{(product rule, inertial)} \]

Three substitutions, each already established and none requiring rigidity, resolve the
right side term by term.

\textbf{First term.} Apply the transport theorem once more, this time to the vector
$\partial\vv{r}/\partial t$ itself (the theorem holds for any vector):

\[ \frac{d}{dt}\!\left(\frac{\partial\vv{r}}{\partial t}\right) = \frac{\partial^{2}\vv{r}}{\partial t^{2}} + \vv{\omega}\times\frac{\partial\vv{r}}{\partial t} \]

\textbf{Second term.} By definition, the angular acceleration of the chosen frame is

\[ \vv{\varepsilon} = \frac{d\vv{\omega}}{dt} \]

\textbf{Third term.} Substitute the first-differential result
$d\vv{r}/dt = \partial\vv{r}/\partial t + \vv{\omega}\times\vv{r}$ obtained above:

\[ \vv{\omega}\times\frac{d\vv{r}}{dt} = \vv{\omega}\times\frac{\partial\vv{r}}{\partial t} + \vv{\omega}\times(\vv{\omega}\times\vv{r}) \]

Collecting all three substitutions:

\[ \frac{d^{2}\vv{r}}{dt^{2}} = \vv{\varepsilon}\times\vv{r} + \vv{\omega}\times\frac{\partial\vv{r}}{\partial t} + \vv{\omega}\times(\vv{\omega}\times\vv{r}) + \frac{\partial^{2}\vv{r}}{\partial t^{2}} + \vv{\omega}\times\frac{\partial\vv{r}}{\partial t} \]

The two identical terms $\vv{\omega}\times(\partial\vv{r}/\partial t)$ combine, giving
the general, fully inertial result:

\begin{equation}\tag{8}\label{e:8} \frac{d^{2}\vv{r}}{dt^{2}} = \vv{\varepsilon}\times\vv{r} + \vv{\omega}\times(\vv{\omega}\times\vv{r}) + 2\vv{\omega}\times\frac{\partial\vv{r}}{\partial t} + \frac{\partial^{2}\vv{r}}{\partial t^{2}} \qquad \text{(inertial LHS; general --- no rigidity assumed)} \end{equation}

This is the complete inertial-material acceleration of the general material point $P$,
obtained by starting from $\vv{r} = \vv{R} + \vv{\xi}$ and applying the transport
theorem twice, with no rigid-body assumption introduced at any step. The left side,
$d^{2}\vv{r}/dt^{2}$, is the physical, measurable acceleration. The right side's four
terms are the same measurable acceleration expressed algebraically --- the angular
(Euler), centripetal, Coriolis, and relative contributions, in that order --- with
$\partial\vv{r}/\partial t$ and $\partial^{2}\vv{r}/\partial t^{2}$ left completely
general throughout: at no point in this derivation has either been set to zero,
assumed small, or otherwise specialized. The rigid body is not addressed in this
section; its evaluation, when undertaken, is a distinct and later step, as this
general result places no restriction on how $\vv{\xi}$ --- and hence $\vv{r}$ ---
moves.

\textbf{Resolving $\vv{a}_P$ in terms of $\vv{R}$ and $\vv{\xi}$.} Insert the
decomposition $\vv{r} = \vv{R} + \vv{\xi}$~\eqref{e:4} into the result above,
replacing every occurrence of $\vv{r}$:

\[ \vv{a}_P = \frac{d^{2}\vv{r}}{dt^{2}} = \vv{\varepsilon}\times(\vv{R}+\vv{\xi}) + \vv{\omega}\times[\vv{\omega}\times(\vv{R}+\vv{\xi})] + 2\vv{\omega}\times\frac{\partial(\vv{R}+\vv{\xi})}{\partial t} + \frac{\partial^{2}(\vv{R}+\vv{\xi})}{\partial t^{2}} \]

and, using the linearity of $\partial/\partial t$ already established
($\partial\vv{r}/\partial t = \partial\vv{R}/\partial t + \partial\vv{\xi}/\partial t$,
$\partial^{2}\vv{r}/\partial t^{2} = \partial^{2}\vv{R}/\partial t^{2} +
\partial^{2}\vv{\xi}/\partial t^{2}$), together with the distributivity of the cross
product over vector addition, expand every term:

\begin{multline}\tag{9}\label{e:9} \vv{a}_P = \vv{\varepsilon}\times\vv{R} + \vv{\varepsilon}\times\vv{\xi} + \vv{\omega}\times(\vv{\omega}\times\vv{R}) + \vv{\omega}\times(\vv{\omega}\times\vv{\xi}) \\ + 2\vv{\omega}\times\frac{\partial\vv{R}}{\partial t} + 2\vv{\omega}\times\frac{\partial\vv{\xi}}{\partial t} + \frac{\partial^{2}\vv{R}}{\partial t^{2}} + \frac{\partial^{2}\vv{\xi}}{\partial t^{2}} \\ \text{(inertial LHS; general --- no rigidity assumed)} \end{multline}

This is $\vv{a}_P$ fully resolved: eight terms, each built only from $\vv{R}$,
$\vv{\xi}$, $\vv{\varepsilon}$, $\vv{\omega}$, and the $\partial$-derivatives of
$\vv{R}$ and $\vv{\xi}$ --- no occurrence of $\vv{r}$ remains anywhere. Regrouping by
which position vector each term carries (the $\vv{R}$-terms first, the
$\vv{\xi}$-terms second) makes the structure transparent:

\begin{multline}\tag{9A}\label{e:9A} \vv{a}_P = \left[\, \vv{\varepsilon}\times\vv{R} + \vv{\omega}\times(\vv{\omega}\times\vv{R}) + 2\vv{\omega}\times\frac{\partial\vv{R}}{\partial t} + \frac{\partial^{2}\vv{R}}{\partial t^{2}} \,\right] \\ + \left[\, \vv{\varepsilon}\times\vv{\xi} + \vv{\omega}\times(\vv{\omega}\times\vv{\xi}) + 2\vv{\omega}\times\frac{\partial\vv{\xi}}{\partial t} + \frac{\partial^{2}\vv{\xi}}{\partial t^{2}} \,\right] \end{multline}

Each bracket is recognizable: it is the same four-term transport-theorem pattern
derived above for $\vv{r}$, now applied instead to $\vv{R}$ and to $\vv{\xi}$
individually. That is, the $\vv{R}$-bracket is exactly $d^{2}\vv{R}/dt^{2}$ and the
$\vv{\xi}$-bracket is exactly $d^{2}\vv{\xi}/dt^{2}$ --- which is also the direct,
one-line consequence of differentiating $\vv{r} = \vv{R} + \vv{\xi}$ twice,
inertially, term by term:

\begin{equation}\tag{10}\label{e:10} \vv{a}_P = \frac{d^{2}\vv{R}}{dt^{2}} + \frac{d^{2}\vv{\xi}}{dt^{2}} \end{equation}

confirming that the substitution above is exact and that the two routes ---
substituting first and expanding, versus differentiating the sum directly --- agree
termwise.

A note on the form ``function of $\{\vv{R}, \vv{\xi}, \vv{\varepsilon},
\vv{\omega}\}$''. The expansion above is $\vv{a}_P$ written as an explicit function of
$\vv{R}$, $\vv{\xi}$, $\vv{\varepsilon}$, and $\vv{\omega}$ together with the first
and second companion ($\partial/\partial t$) derivatives of $\vv{R}$ and $\vv{\xi}$
--- $\partial\vv{R}/\partial t$, $\partial\vv{\xi}/\partial t$,
$\partial^{2}\vv{R}/\partial t^{2}$, $\partial^{2}\vv{\xi}/\partial t^{2}$. Those four
derivative quantities cannot be removed from the general result: nothing in this
section, or in the standing project rules, permits setting them to zero. Doing so
would be exactly the rigid-body assumption --- every material vector's companion
derivative vanishing --- that this section explicitly excludes. A literal function of
only the four bare symbols $\{\vv{R}, \vv{\xi}, \vv{\varepsilon}, \vv{\omega}\}$ ---
with no derivative terms --- exists only once that further assumption is
introduced, as a distinct later step.

\hsubsection{4}{The frame-bridge formula and the computation protocol}\label{sec:3.4}

The classical transport theorem, obtained by applying~\eqref{e:6} twice to the
position vector, expresses the same inertial, measurable acceleration algebraically in
terms of the companion ($\partial/\partial t$) bookkeeping derivatives:

\begin{equation}\tag{11}\label{e:11} \vv{a}_P = \vv{\varepsilon}\times\vv{r} + \vv{\omega}\times(\vv{\omega}\times\vv{r}) + 2\vv{\omega}\times\frac{\partial\vv{r}}{\partial t} + \frac{\partial^{2}\vv{r}}{\partial t^{2}} \end{equation}

Equation~\eqref{e:11} is a bridge between frames, and its four right-hand terms carry
the historical names Euler (angular), centripetal, Coriolis, and relative
acceleration. Its correct reading is fixed by the following protocol, observed
throughout this paper. The left side is the inertial, measurable acceleration; the
right side as a whole equals it identically and is therefore equally inertial; the
right side's individual terms, however, are frame-referred computational quantities,
evaluated from inertial data via the transport rule~\eqref{e:6}, and are used only in
computation --- never as separately measurable accelerations. In particular, the
$\partial/\partial t$ slots of~\eqref{e:11} are never to be filled with inertial
derivatives: substituting $d/dt$ values into those slots re-adds rotation kinematics
already contained in the other terms, and a one-line check exposes the error --- for
steady rotation at constant $\vv{\omega}$, such substitution returns four times the
true centripetal acceleration.

\hsection{4}{The Operator Framework}\label{sec:4}

This section defines the primary pair with its properties (\S\ref{sec:4.1}), the internal spin integral (\S\ref{sec:4.2}), the closure identities (\S\ref{sec:4.3}) and the rigid evaluation of the internal torque integral (\S\ref{sec:4.4}).

\hsubsection{1}{The primary pair and its properties}\label{sec:4.1}

The polar second moment $J_0$ and the mass-vector operator $\vv{F}$, defined
in~\eqref{e:1} and~\eqref{e:2} about the center of mass, are the primary carriers of
inertia information in this work. Their properties, each immediate from the defining
integrals: $J_0$ is a non-negative scalar (units
$\mathrm{kg}\cdot\mathrm{m}^{2}$), frame-invariant, and additive over sub-bodies.
$\vv{F}$ is linear in its argument,
$\vv{F}(\alpha\vv{u} + \beta\vv{v}) = \alpha\vv{F}(\vv{u}) + \beta\vv{F}(\vv{v})$,
likewise additive over sub-bodies, and satisfies the trace identity: for any three
mutually perpendicular unit vectors $\hat{e}_1$, $\hat{e}_2$, $\hat{e}_3$,

\begin{equation}\tag{12}\label{e:12} \hat{e}_1\cdot\vv{F}(\hat{e}_1) + \hat{e}_2\cdot\vv{F}(\hat{e}_2) + \hat{e}_3\cdot\vv{F}(\hat{e}_3) = J_0 \end{equation}

since the three projections of $\vv{\xi}\cdot\vv{\xi}$ sum pointwise under the
integral. The triad serves only to prove the identity; \eqref{e:12} itself holds for any triad, resolves no vector of the motion, and enters the derivation once, in \S\ref{sec:7.4}. There it fixes the coefficient across the shaft in the closure~\eqref{e:37}; everywhere else it serves as an internal consistency check. Neither object presupposes
an axis, a preferred direction, or a component resolution: the body's directional
response is extracted from $\vv{F}$ by evaluation, so that directions emerge as
outputs of the analysis.

\hsubsection{2}{The internal spin integral}\label{sec:4.2}

It is convenient to introduce also the internal spin integral

\begin{equation}\tag{13}\label{e:13} \vv{S} = \int_0^M \vv{\xi}\times\frac{d\vv{\xi}}{dt}\,dm \end{equation}

None of these definitions assumes rigidity; they apply verbatim to a deforming body.

\emph{Two things are worth noting about how $\vv{S}$ sits inside the framework
developed here, offered as context rather than as a further claim:}

\textbf{First,} for the rigid body, $\vv{S}$ resolves entirely into the operators
already defined in this paper. Substituting the rigid law
$d\vv{\xi}/dt = \vv{\omega}\times\vv{\xi}$~\eqref{e:14} and expanding the triple
product ---
$\vv{\xi}\times(\vv{\omega}\times\vv{\xi}) = \vv{\omega}(\vv{\xi}\cdot\vv{\xi}) -
\vv{\xi}(\vv{\xi}\cdot\vv{\omega})$, the standard BAC--CAB identity --- gives

\[ \vv{S} = J_0\vv{\omega} - \vv{F}(\vv{\omega}) \]

the coordinate-free version of the textbook relation ``angular momentum equals the
inertia tensor times angular velocity'' ($\vv{S} = \mathbf{I}\cdot\vv{\omega}$),
produced here without ever introducing a tensor. It is the same
$J_0(\cdot) - \vv{F}(\cdot)$ pattern that already appears elsewhere in this paper ---
for instance the $J_0\vv{\varepsilon} - \vv{F}(\vv{\varepsilon})$ piece of the engine
identity~\eqref{e:19} --- here with $\vv{\omega}$ in the slot that~\eqref{e:19} fills
with $\vv{\varepsilon}$.

\textbf{Second,} Proposition~3's claim ---
$\int_0^M \vv{\xi}\times(d^{2}\vv{\xi}/dt^{2})\,dm = d\vv{S}/dt$ --- is exactly
Euler's second law in kinematic form: the rate of change of spin angular momentum
equals the net moment of the mass elements' accelerations about the center of mass.
This is a standard, textbook identity, not a novel claim of this work; its value here
is that it lets \S\ref{sec:7.6} and C13 state the deforming-body form of the governing
balance compactly, in one recognized symbol, rather than as a bare integral.

\S\ref{sec:7.6} (``Scope and validity tiers'') is the paper's explicit statement of
exactly how far the governing balance extends: it asserts that the balance~\eqref{e:32}
is exact for any closed body pivoted at a fixed, frictionless point in uniform gravity
--- rigid or deforming --- and it uses $d\vv{S}/dt$ precisely so that the
deforming-body form can be written in one line,
``$M\vv{R}\times\vv{g} = M\vv{R}\times(d^{2}\vv{R}/dt^{2}) + d\vv{S}/dt$,'' rather
than as an unexpanded integral. The same section then identifies exactly where
rigidity enters when the body is rigid --- solely through the kinematic substitutions~\eqref{e:14}--\eqref{e:15} and~\eqref{e:26}--\eqref{e:27} and the internal-torque evaluation~\eqref{e:19} --- so a reader can see precisely which steps would need to
change for a deforming body, rather than being told rigidity is assumed throughout.
C13, in \S\ref{sec:9.1} (``Catalog of contributions''), restates this same fact among the results, credited there to Brand's theorem of moment of momentum: that the governing balance ``holds exactly beyond rigidity\dots{} with rigidity confined to the closed-form solution'' --- that
is, the physics of the balance itself requires no rigid-body assumption; only the
final closed-form scalar solution, and the worked examples built on it, specialize to
rigid bodies. In both places, $\vv{S}$ is what makes this generality statement
expressible in a single recognizable symbol.

\textbf{Rigid specialization of the internal kinematics.} The general kinematics
of \S\ref{sec:3.3} place no restriction on $\vv{\xi}$. The propositions and results of
the remainder of this section specialize to the rigid body, for which every
COM-relative material vector is carried bodily by the rotation, and its
inertial-material rate is the pure rotation law:

\begin{equation}\tag{14}\label{e:14} \frac{d\vv{\xi}}{dt} = \vv{\omega}\times\vv{\xi} \end{equation}

Differentiating once more, inertially, by the product rule:
$d^{2}\vv{\xi}/dt^{2} = (d\vv{\omega}/dt)\times\vv{\xi} +
\vv{\omega}\times(d\vv{\xi}/dt) = \vv{\varepsilon}\times\vv{\xi} +
\vv{\omega}\times(\vv{\omega}\times\vv{\xi})$, using
$\vv{\varepsilon} = d\vv{\omega}/dt$ and substituting the velocity law just obtained:

\begin{equation}\tag{15}\label{e:15} \frac{d^{2}\vv{\xi}}{dt^{2}} = \vv{\varepsilon}\times\vv{\xi} + \vv{\omega}\times(\vv{\omega}\times\vv{\xi}) \end{equation}

These are the $\vv{\xi}$-specific rigid instances used by the propositions below.

\hsubsection{3}{Closure identities: Propositions}\label{sec:4.3}

The framework verifies its own bookkeeping through three identities, each with a
one-line proof.

\textbf{Proposition~1 (constancy of $J_0$, rigid body).} For rigid motion,
$d\vv{\xi}/dt = \vv{\omega}\times\vv{\xi}$ by~\eqref{e:14}, and
$\vv{\xi}\cdot(\vv{\omega}\times\vv{\xi}) = 0$ pointwise; hence

\begin{equation}\tag{16}\label{e:16} \frac{dJ_0}{dt} = 2\int_0^M \vv{\xi}\cdot\frac{d\vv{\xi}}{dt}\,dm = 0 \end{equation}

\textbf{Proposition~2 (vanishing of the net internal acceleration, closed body).}
Differentiating the COM identity~\eqref{e:5} twice --- in either frame, since the
transport rule maps the zero vector to itself --- gives, with no rigidity assumption,

\begin{equation}\tag{17}\label{e:17} \int_0^M \frac{d^{2}\vv{\xi}}{dt^{2}}\,dm = \frac{d^{2}}{dt^{2}}\int_0^M \vv{\xi}\,dm = \vv{0} = \int_0^M \frac{\partial^{2}\vv{\xi}}{\partial t^{2}}\,dm \end{equation}

\textbf{Proposition~3 (the internal torque integral as a rate; general).}
Differentiating~\eqref{e:13} inertially, the $d\vv{\xi}/dt \times d\vv{\xi}/dt$ term
vanishes identically, leaving --- again with no rigidity assumption ---

\begin{equation}\tag{18}\label{e:18} \int_0^M \vv{\xi}\times\frac{d^{2}\vv{\xi}}{dt^{2}}\,dm = \frac{d\vv{S}}{dt} \end{equation}

Proposition~1 is the rigid specialization; Propositions~2 and 3 hold for any closed
body, rigid or deforming.

\hsubsection{4}{Rigid evaluation of the internal torque integral}\label{sec:4.4}

The central rigid-body identity of this paper follows by inserting the
kinematics~\eqref{e:15} into the internal torque integral of Proposition~3 and
expanding the double cross product
$\vv{\xi}\times(\vv{\varepsilon}\times\vv{\xi}) =
\vv{\varepsilon}(\vv{\xi}\cdot\vv{\xi}) - \vv{\xi}(\vv{\xi}\cdot\vv{\varepsilon})$
together with
$\vv{\xi}\times[\vv{\omega}\times(\vv{\omega}\times\vv{\xi})] =
(\vv{\omega}\cdot\vv{\xi})(\vv{\xi}\times\vv{\omega})$:

\begin{equation}\tag{19}\label{e:19} \int_0^M \vv{\xi}\times\frac{d^{2}\vv{\xi}}{dt^{2}}\,dm = J_0\vv{\varepsilon} - \vv{F}(\vv{\varepsilon}) + \vv{F}(\vv{\omega})\times\vv{\omega} \end{equation}

Equation~\eqref{e:19} is the engine of the dynamics developed below: its first pair
carries the angular-acceleration inertia --- note that $J_0$ never acts alone, but
always in the pair $J_0\vv{\varepsilon} - \vv{F}(\vv{\varepsilon})$, which retains full
directional information for a body of any shape --- and its last term carries the
gyroscopic coupling. Both pairs are pure cross-product mass integrals, exact for any
rigid body with no thin-body or symmetry idealization.

\hsection{5}{The Infinitesimal Torque About the Pivot A}\label{sec:5}

Each infinitesimal mass element $dm$ located at point $P$ is being accelerated, and by
its own inertia resists that acceleration. The element never enters alone: it is paired from the start with its position $\vv{r}_P$ from the
fixed reference point $A$ --- the pivot-end of the gyroscope shaft --- and the pair is the
infinitesimal inertia torque

\begin{equation}\tag{20}\label{e:20} d\Mscr_P = \vv{r}_P\times(-\,\vv{a}_P\,dm) \end{equation}

where $\vv{a}_P$ is defined in~\eqref{e:9},~\eqref{e:10},~\eqref{e:11}. The reversal of the mass-acceleration, carried here and nowhere else in this work, is what makes
$d\Mscr_P$ the torque the element's inertia produces about $A$. The paper never uses the element's
mass-acceleration alone; it uses only this pair.

\medskip\noindent\textbf{The sign convention.} Both torques in this work act about the
horizontal axis at right angles to the vertical plane containing the shaft --- the only
axis about which a torque can change the tilt at all --- and each is reported by a single
signed scalar along that axis. \emph{The sign records the effect on the tilt.} A torque
whose scalar is negative works to increase the tilt angle and lay the shaft down; a torque
whose scalar is positive works to reduce the tilt angle and bring the shaft up toward the
vertical. Gravity lays the shaft down: the scalar of $M\vv{R}\times\vv{g}$ is $-MgR\sin\theta$, negative at every tilt in the physical range. At dynamic equilibrium the inertia torque $\Mscr_A$, assembled below from the infinitesimal inertia torques of the mass elements, works to bring the shaft up, and its scalar is positive; these two torque vectors are equal in magnitude and are diametrically opposed in direction, so their torque vector sum is the zero vector.
Neither torque disappears there and neither is canceled by the other: both are present at
all times, each with its own magnitude and its own effect on the tilt, and it is only the
arithmetic of adding two opposed vectors of equal length that reaches zero. Every torque in this work is a position vector from $A$ crossed with its partner, in that order ---
$M\vv{R}\times\vv{g}$ for gravity and $\vv{r}_P\times(-\,\vv{a}_P\,dm)$ for each mass element --- and
no cross product anywhere in this work is written in reversed order. The convention serves only to report the two torques: the derivation carries them as vectors through the balance~\eqref{e:38} and uses none of these scalars, which are first taken in the scalar reduction of \S\ref{sec:7.5}.

The infinitesimal inertia torques of all the mass elements add to the total inertia torque about $A$,

\begin{equation}\tag{21}\label{e:21} \Mscr_A = \int_0^M d\Mscr_P \end{equation}

where $\vv{r}_P$ is the position vector of point $P$ measured from $A$ (written simply
$\vv{r}$ below where no confusion arises), decomposed per~\eqref{e:4} as

\begin{equation}\tag{22}\label{e:22} \vv{r}_P = \vv{R} + \vv{\xi} \end{equation}

with $\vv{R}$ the position vector of the center of mass and $\vv{\xi}$ the local position of the mass
element $dm$ relative to the COM. Written out, the total inertia torque is

\begin{equation}\tag{23}\label{e:23} \Mscr_A = \int_0^M \vv{r}_P\times(-\,\vv{a}_P\,dm) = -\int_0^M \vv{r}_P\times\vv{a}_P\,dm \end{equation}

No force is used in this derivation, only torques: every mass element enters solely through its
infinitesimal inertia torque~\eqref{e:20} about $A$, and no force resultant is assembled anywhere in
this work. \S\ref{sec:6} carries out the
integration~\eqref{e:23} exactly once.

\hsection{6}{Total Torque About the Pivot A}\label{sec:6}

This section assembles the total torque about the pivot once, from the material acceleration (\S\ref{sec:6.1}), and checks the assembly by a second route (\S\ref{sec:6.2}).

\hsubsection{1}{Single assembly from the material acceleration}\label{sec:6.1}

The inertia torque about the fixed pivot is the mass integral~\eqref{e:23} of the infinitesimal
inertia torques~\eqref{e:20}, each element's position crossed with its reversed mass-acceleration,
assembled exactly once. Expanding
$\vv{r} = \vv{R} + \vv{\xi}$ by~\eqref{e:22}, one of the two mixed terms vanishes at once
by the COM identity~\eqref{e:5}, leaving

\begin{equation}\tag{24}\label{e:24} \Mscr_A = -\int_0^M \vv{r}\times\frac{d^{2}\vv{r}}{dt^{2}}\,dm = -\,M\vv{R}\times\frac{d^{2}\vv{R}}{dt^{2}} - \vv{R}\times\int_0^M \frac{d^{2}\vv{\xi}}{dt^{2}}\,dm - \int_0^M \vv{\xi}\times\frac{d^{2}\vv{\xi}}{dt^{2}}\,dm \end{equation}

With the remaining mixed term erased by Proposition~2 --- $\vv{R}$ factors out of
$\int_0^M (d^{2}\vv{\xi}/dt^{2})\,dm = \vv{0}$ --- this is, for any closed body, rigid or
deforming,

\begin{equation}\tag{25}\label{e:25} \Mscr_A = -\,M\vv{R}\times\frac{d^{2}\vv{R}}{dt^{2}} - \int_0^M \vv{\xi}\times\frac{d^{2}\vv{\xi}}{dt^{2}}\,dm \end{equation}

--- transport torque plus internal torque, and nothing else, for any closed body,
rigid or deforming.

\textbf{For the rigid body,} since the pivot $A$ and the center of mass $C$ are both fixed relative to the body, whether or not $C$ lies in its material, $\vv{R}$ is body-fixed exactly as $\vv{\xi}$ is (\S\ref{sec:4.2}), and obeys the identical pure rotation law:

\begin{equation}\tag{26}\label{e:26} \frac{d\vv{R}}{dt} = \vv{\omega}\times\vv{R} \end{equation}

Differentiating once more, inertially, by the product rule --- exactly the step already
carried out for $\vv{\xi}$ in~\eqref{e:15}:

\begin{equation}\tag{27}\label{e:27} \frac{d^{2}\vv{R}}{dt^{2}} = \vv{\varepsilon}\times\vv{R} + \vv{\omega}\times(\vv{\omega}\times\vv{R}) \end{equation}

Inserting this rigid evaluation of $\vv{R}$ together with equation~\eqref{e:19} (for
the internal torque integral) into~\eqref{e:25} yields the exact rigid form:

\begin{equation}\tag{28}\label{e:28} \Mscr_A = -\,M\vv{R}\times(\vv{\varepsilon}\times\vv{R}) - J_0\vv{\varepsilon} + \vv{F}(\vv{\varepsilon}) - M\vv{R}\times[\vv{\omega}\times(\vv{\omega}\times\vv{R})] - \vv{F}(\vv{\omega})\times\vv{\omega} \end{equation}

Every symbol in~\eqref{e:28} is a scalar mass integral or a cross product of intrinsic vectors; the expression is exact for any rigid body of any shape, with no
symmetry or thinness assumption.

\hsubsection{2}{The decomposition table and the two-route consistency theorem}\label{sec:6.2}

As in \S\ref{sec:6.1}, integrating $\vv{r}\times(\cdot)$ against each term of the
frame-bridge formula~\eqref{e:11} produces a decomposition table of the single
torque~\eqref{e:25}. The two frame-correction rows, evaluated with the operator
algebra, are

\begin{equation}\tag{29}\label{e:29} \Mscr_{A1} = -\,M\vv{R}\times(\vv{\varepsilon}\times\vv{R}) - J_0\vv{\varepsilon} + \vv{F}(\vv{\varepsilon}) \qquad \text{(Angular row)} \end{equation}

\begin{equation}\tag{30}\label{e:30} \Mscr_{A2} = -\,M\vv{R}\times[\vv{\omega}\times(\vv{\omega}\times\vv{R})] - \vv{F}(\vv{\omega})\times\vv{\omega} \qquad \text{(Centripetal row)} \end{equation}

while the Coriolis and Relative rows carry only $\partial/\partial t$ slots; comparing
the rigid laws~\eqref{e:14} and~\eqref{e:26} with the transport relation~\eqref{e:6}
applied to $\vv{\xi}$ and $\vv{R}$ shows $\partial\vv{\xi}/\partial t = \vv{0}$ and
$\partial\vv{R}/\partial t = \vv{0}$ for the rigid body, so both rows vanish. The
table's total is therefore $\Mscr_{A1} + \Mscr_{A2}$, and comparison
with~\eqref{e:28} establishes the two-route consistency theorem:

\begin{equation}\tag{31}\label{e:31} \Mscr_{A1} + \Mscr_{A2} = -\,M\vv{R}\times\frac{d^{2}\vv{R}}{dt^{2}} - \int_0^M \vv{\xi}\times\frac{d^{2}\vv{\xi}}{dt^{2}}\,dm = \Mscr_A \end{equation}

The direct material route~\eqref{e:25} and the frame-decomposition route agree term for
term --- an internal cross-validation of the assembly of the inertia torque, obtained without
appeal to any external theorem. The protocol caution of \S\ref{sec:3.4} applies in full here: the four rows are labels of one torque. Filling the
$\partial$-slots of the Coriolis and Relative rows with inertial derivatives and
summing all four rows counts the $\vv{\varepsilon}$-physics twice and the
$\vv{\omega}$-physics four times --- for steady rotation, four times the true torque
--- and any precession rate extracted from such a sum is wrong by order-one factors.
Assembled once, per~\eqref{e:25}, the torque is exact; and the row names (Angular,
Centripetal, Coriolis, Relative) remain available as the physically legible
decomposition of that one exact object.

\hsection{7}{Dynamic Equilibrium and the Steady-Precession Solution}\label{sec:7}

This section writes the balance as the exact equation of motion (\S\ref{sec:7.1}), carries the steady-precession kinematics (\S\ref{sec:7.2}) into the unified vector equation (\S\ref{sec:7.3}), closes it with the operator's own scalar (\S\ref{sec:7.4}), reduces it to both branches (\S\ref{sec:7.5}) and states its scope (\S\ref{sec:7.6}).

\hsubsection{1}{The balance as the exact equation of motion}\label{sec:7.1}

Newton's law for each mass element is taken from the start crossed with the element's position
$\vv{r}$, so that every element enters only through its torques about $A$: gravity's,
$\vv{r}\times(\vv{g}\,dm)$, its inertia torque~\eqref{e:20}, and the torques of internal cohesion.
Integrating, the internal torques cancel pairwise, each equal-and-opposite pair of internal actions
lying along the line joining its two elements (Newton's third law); the support sits at
$\vv{r} = \vv{0}$ and contributes no torque; and gravity's torques integrate, by~\eqref{e:5}, to
$(\int_0^M \vv{r}\,dm)\times\vv{g} = M\vv{R}\times\vv{g}$. What remains is exact:

\begin{equation}\tag{32}\label{e:32} M\vv{R}\times\vv{g} + \Mscr_A = \vv{0} \end{equation}

Equation~\eqref{e:32} is not an equilibrium assumption to be tested; it is the equation
of motion about $A$, derived from Newton's law per mass element with no
angular-momentum primitive invoked. Dynamic equilibrium then has a precise meaning: the
steady-precession kinematics satisfy~\eqref{e:32} identically at every instant.

\hsubsection{2}{Steady-precession kinematics}\label{sec:7.2}

The steady precession combines a spin $\vv{\omega}_0$ along the shaft
($\vv{\omega}_0 \parallel \vv{R}$, the body axisymmetric about it) with a slower
precession $\vv{\Omega}$ about the vertical through $A$, at constant magnitudes and
constant tilt:

\begin{equation}\tag{33}\label{e:33} \vv{\omega} = \vv{\Omega} + \vv{\omega}_0, \qquad \vv{\varepsilon} = \frac{d\vv{\omega}}{dt} = \vv{\Omega}\times\vv{\omega}_0 \end{equation}

since $\vv{\Omega}$ is constant and $\vv{\omega}_0$, of constant magnitude, is carried
around the vertical by the precession. Note
$\vv{\varepsilon} \perp \vv{R}$. The COM acceleration now follows
from~\eqref{e:27}, and the collinearity $\vv{\omega}_0\times\vv{R} = \vv{0}$ enters
twice. It empties the inner bracket, so that
$(\vv{\Omega} + \vv{\omega}_0)\times\vv{R} = \vv{\Omega}\times\vv{R}$ exactly --- the
spin about the shaft does not move the center of mass. It does not, however, reach the
outer factor, which leaves the surplus
$\vv{\omega}_0\times(\vv{\Omega}\times\vv{R})$; and the same collinearity turns
$\vv{\varepsilon}\times\vv{R} = (\vv{\Omega}\times\vv{\omega}_0)\times\vv{R}$ into
precisely $-\,\vv{\omega}_0\times(\vv{\Omega}\times\vv{R})$. The two cancel
identically, and the exact rigid acceleration reduces without approximation to:

\begin{equation}\tag{34}\label{e:34} \frac{d^{2}\vv{R}}{dt^{2}} = \vv{\varepsilon}\times\vv{R} + \vv{\omega}\times(\vv{\omega}\times\vv{R}) = \vv{\Omega}\times(\vv{\Omega}\times\vv{R}) \end{equation}

Both rates are magnitudes throughout this paper, and the direction of each is named
rather than carried as a sign. Each name is that of an ordinary right-hand screw, read
by the corkscrew rule of Appendix~A (Figures~6 and 7). For the precession the screw is
taken about the upward vertical through $A$: the precession is thread-in when its
vector advances downward, and thread-out when it advances upward. Up to the torque-vector balance~\eqref{e:38},
$\vv{\Omega}$ is an intrinsic vector and needs no sign: it points up for a thread-out precession and
down for a thread-in one. A sign appears only in the final scalar reduction of \S\ref{sec:7.5}, where
$\vv{\Omega}$ is taken on the upward vertical $Z$ axis: its single component is positive for a
thread-out precession and negative for a thread-in one, and the precession rate reported is always
that component's magnitude. Neither sense changes the tilt: $\theta$ is measured from the upward
vertical in every case, and the direction of $\vv{\Omega}$ never enters its definition. For the spin the
screw is taken along the shaft with its point at the pivot: the spin is thread-in when
its vector points into $A$, and thread-out when it points away from $A$, along $\vv{R}$
toward the center of mass. Each name is fixed to an axis and not to a viewpoint, so both hold unchanged at every tilt. Every configuration worked in this
paper has a thread-out spin and a thread-out precession; Appendix~D.7 exhibits the one
circumstance in which a precession is thread-in instead.

\hsubsection{3}{The unified vector equation of motion}\label{sec:7.3}

Substituting~\eqref{e:33} and~\eqref{e:34} into the balance~\eqref{e:32} with the
torque~\eqref{e:25} and the internal-torque identity~\eqref{e:19}, expanding
$\vv{F}(\vv{\omega})\times\vv{\omega}$ by the linearity of $\vv{F}$, and using
$M\vv{R}\times[\vv{\Omega}\times(\vv{\Omega}\times\vv{R})] =
M(\vv{\Omega}\cdot\vv{R})(\vv{R}\times\vv{\Omega})$ together with
$\vv{F}(\vv{\omega}_0)\times\vv{\omega}_0 = \vv{0}$ ($\vv{F}$ of the symmetry-axis
vector is parallel to it), the steady-precession balance reads, in pure operator form:

\begin{multline}\tag{35}\label{e:35} M\vv{R}\times\vv{g} - M(\vv{\Omega}\cdot\vv{R})(\vv{R}\times\vv{\Omega}) - J_0(\vv{\Omega}\times\vv{\omega}_0) + \vv{F}(\vv{\Omega}\times\vv{\omega}_0) \\ - \vv{F}(\vv{\Omega})\times\vv{\Omega} - \vv{F}(\vv{\Omega})\times\vv{\omega}_0 - \vv{F}(\vv{\omega}_0)\times\vv{\Omega} = \vv{0} \end{multline}

This is the paper's central result: one vector equation, valid exactly for every
axisymmetric rigid rotor, in which gravity's lever arm, the COM transport, the
angular-acceleration pair, and the gyroscopic couplings appear as physically legible
cross-product terms, each readable by the corkscrew rule. From~\eqref{e:20} to~\eqref{e:35} every term is an intrinsic vector, a mass
integral, or a cross product of intrinsic vectors --- its one dot product, $\vv{\Omega}\cdot\vv{R}$ in~\eqref{e:35}, comes from reducing the transport term $-M\vv{R}\times[\vv{\Omega}\times(\vv{\Omega}\times\vv{R})]$ --- and $\vv{F}$ is evaluated in one respect only, the direction it gives the symmetry-axis vector, which \S\ref{sec:7.4} takes up as~\eqref{e:36}.

\hsubsection{4}{Closure by the operator's own scalar}\label{sec:7.4}

For each rotor, one additional scalar closes the equation, and the formulation produces
it itself: feeding the shaft vector $\vv{R}$ into $\vv{F}$ returns a vector parallel to
$\vv{R}$, defining

\begin{equation}\tag{36}\label{e:36} \vv{F}(\vv{R}) = J_R\vv{R}, \qquad J_R = \frac{\vv{R}\cdot\vv{F}(\vv{R})}{\vv{R}\cdot\vv{R}} \end{equation}

$J_R$ is a standalone mass integral in the same family as $J_0$ --- not an imported axial moment. For
the axisymmetric body $\vv{F}$ returns $J_R\vv{R}$ along the shaft, by~\eqref{e:36}. The defining
integral~\eqref{e:2} shows that $\vv{F}$ is symmetric with respect to the dot product:
\[ \vv{u}\cdot\vv{F}(\vv{v}) = \int(\vv{\xi}\cdot\vv{v})(\vv{u}\cdot\vv{\xi})\,dm = \int(\vv{\xi}\cdot\vv{u})(\vv{v}\cdot\vv{\xi})\,dm = \vv{v}\cdot\vv{F}(\vv{u}). \]
Let $\vv{v}_\perp$ be any vector perpendicular to the shaft, so that $\vv{v}_\perp\cdot\vv{R} = 0$. Then
\[ \vv{R}\cdot\vv{F}(\vv{v}_\perp) = \vv{v}_\perp\cdot\vv{F}(\vv{R}) = J_R\,(\vv{v}_\perp\cdot\vv{R}) = 0. \]
Thus $\vv{F}$ maps the plane perpendicular to the shaft into itself. Because the body is unchanged by
any turn about the shaft, no direction in that transverse plane is distinguished; therefore one scalar
$J_\perp$ describes the response to every transverse vector: $\vv{F}(\vv{v}_\perp) = J_\perp\vv{v}_\perp$.

The trace identity~\eqref{e:12}, taken with one of its directions along the shaft and two
perpendicular to it, fixes that scalar. The axial contribution is $J_R$ and the two equal transverse
contributions are $J_\perp$, so $J_0 = J_R + 2J_\perp$, giving $J_\perp = \tfrac{1}{2}(J_0 - J_R)$.

For an arbitrary vector $\vv{v}$, write $\vv{v} = \vv{v}_\parallel + \vv{v}_\perp$. Its transverse part
can be written intrinsically, without introducing any coordinate components, as
$\vv{v}_\perp = [\vv{R}\times(\vv{v}\times\vv{R})]/(\vv{R}\cdot\vv{R})$. Therefore
\[ \vv{F}(\vv{v}) = J_R\vv{v}_\parallel + J_\perp\vv{v}_\perp = J_R\vv{v} + (J_\perp - J_R)\,\vv{v}_\perp = J_R\vv{v} + \tfrac{1}{2}(J_0 - 3J_R)\,\frac{\vv{R}\times(\vv{v}\times\vv{R})}{\vv{R}\cdot\vv{R}}. \]
The pair $(J_0, J_R)$ therefore determines $\vv{F}$ completely, in closed cross-product form:

\begin{equation}\tag{37}\label{e:37} \vv{F}(\vv{v}) = J_R\vv{v} + \tfrac{1}{2}(J_0 - 3J_R)\,\frac{\vv{R}\times(\vv{v}\times\vv{R})}{\vv{R}\cdot\vv{R}} \end{equation}

Substituting~\eqref{e:37} into~\eqref{e:35} evaluates every operator term and yields
the fully resolved balance:

\begin{equation}\tag{38}\label{e:38} M\vv{R}\times\vv{g} - M(\vv{\Omega}\cdot\vv{R})(\vv{R}\times\vv{\Omega}) - \tfrac{1}{2}(J_0 + J_R)(\vv{\Omega}\times\vv{\omega}_0) - \tfrac{1}{2}(3J_R - J_0)\,\frac{\vv{R}\cdot\vv{\omega}}{\vv{R}\cdot\vv{R}}\,(\vv{R}\times\vv{\Omega}) = \vv{0} \end{equation}

Three scalars --- M, $J_0$, and $J_R$ --- and cross products of the intrinsic vectors $\vv{R}$, $\vv{\Omega}$, $\vv{\omega}_0$, $\vv{g}$, with dot products only as scalar coefficients: nothing else appears.

The two inertia terms of~\eqref{e:38} are grouped as the substitution produces them,
with the total angular velocity $\vv{\omega} = \vv{\Omega} + \vv{\omega}_0$ in the last.
That grouping is not the only one. Because $\vv{\omega}_0$ is collinear with $\vv{R}$,
the $\vv{\omega}_0$ part of the last term equals
$\tfrac{1}{2}(3J_R - J_0)(\vv{\Omega}\times\vv{\omega}_0)$, and moving it into the
preceding term gives the equivalent form

\begin{equation}\tag{38A}\label{e:38A} M\vv{R}\times\vv{g} - M(\vv{\Omega}\cdot\vv{R})(\vv{R}\times\vv{\Omega}) - (J_0 - J_R)(\vv{\Omega}\times\vv{\omega}_0) - \tfrac{1}{2}(3J_R - J_0)\,\frac{\vv{R}\cdot\vv{\Omega}}{\vv{R}\cdot\vv{R}}\,(\vv{R}\times\vv{\Omega}) = \vv{0} \end{equation}

The forms~\eqref{e:38} and~\eqref{e:38A} are algebraically identical.
In~\eqref{e:38A}, the coefficient of $(\vv{\Omega}\times\vv{\omega}_0)$ is
$-(J_0 - J_R)$ because the companion factor $\vv{R}\cdot\vv{\omega}$ of~\eqref{e:38}
has been replaced by $\vv{R}\cdot\vv{\Omega}$ at the same time. The two changes belong
together: the coefficient $-(J_0 - J_R)$ cannot be combined with the factor
$\vv{R}\cdot\vv{\omega}$ without counting the spin contribution twice.

\hsubsection{5}{Scalar reduction: both branches, and an exact special case}\label{sec:7.5}

Every vector term of~\eqref{e:38} lies along the same horizontal direction (each is a
cross product involving the vertical $\vv{\Omega}$ and the shaft), so the balance of the two torque vectors reduces to one
scalar relation. In this final scalar reduction $\vv{\Omega}$ is taken on the upward vertical $Z$ axis, as Appendix~F
writes it, $\vv{\Omega} = \Omega\,(0, 0, 1)$: its single component $\Omega$ is positive for a thread-out
precession, $\vv{\Omega}$ pointing up, and negative for a thread-in precession, $\vv{\Omega}$ pointing
down, and its magnitude $|\Omega| = |\vv{\Omega}|$ is the precession rate. With that component $\Omega$,
$\theta$ the tilt of the shaft $\vv{R}$ from the upward vertical, the magnitudes $R = |\vv{R}|$,
$\omega_0 = |\vv{\omega}_0|$ of the thread-out spin, and $g = |\vv{g}|$, and the common factor
$\sin\theta$ divided out:

\begin{equation}\tag{39}\label{e:39} MgR = (J_0 - J_R)(\omega_0 + \Omega\cos\theta)\,\Omega - \left[\tfrac{1}{2}(J_0 + J_R) + MR^{2}\right]\Omega^{2}\cos\theta \end{equation}

Relation~\eqref{e:39} was obtained by dividing the balance through by $\sin\theta$, so
it holds wherever $\sin\theta \neq 0$ --- that is, at every non-vertical tilt. At the two
vertical configurations, $\theta = 0^\circ$ and $\theta = 180^\circ$, the balance itself
settles the matter. Every term of the torque-vector balance~\eqref{e:38} carries the factor
$\sin\theta$ --- which is why it could be divided out at all --- so at either vertical
configuration both torques of~\eqref{e:32} are the zero vector, \eqref{e:38} reads
$\vv{0} = \vv{0}$, and it is satisfied by any $\Omega$ whatever: no equation of the
balance constrains $\Omega$ there. Relation~\eqref{e:39}, which is what remains once
that factor has been divided away, does not apply at the vertical configurations and
must not be evaluated there.

The geometry says the same thing, and says why. The two vertical configurations lie
outside relation~\eqref{e:39} for a reason more basic than the division. Precession is the
turning of the shaft about the vertical, and that turning is measured by the azimuth
the shaft sweeps as it goes round. When the shaft is itself vertical there is no such
azimuth: the shaft and the vertical coincide, the angle has no value, and the phrase
``the rate at which the shaft precesses about the vertical'' has nothing left to refer
to. A precession rate at exact vertical alignment is therefore not a quantity this
treatment leaves undetermined; it is a quantity that has no meaning, since precession
is defined only in relation to an existing tilt. Consistently with that, no torque is
produced at either vertical configuration --- and nothing is canceled to bring that
about. At a tilt both torques of~\eqref{e:32} are present, equal in magnitude and
diametrically opposed, so their resultant is the zero vector and the tilt is held. At the
vertical there is nothing to equalize: the weight acts along the shaft, so its line of
action passes through the pivot and it has no arm to turn about, and the spin axis lies
along the vertical, so there is no azimuth for the shaft to sweep. Each torque
of~\eqref{e:32} is separately the zero vector because no turning is produced there, not
because two turnings were set against each other.

The relation is quadratic in $\Omega$ and therefore contains both classical precession
branches --- slow and fast --- in one equation, with the parallel-axis contribution
$MR^{2}$ arising automatically from the transport term rather than by a separate
theorem. At $\theta = 90^\circ$ (horizontal shaft) both $\cos\theta$ terms vanish and

\begin{equation}\tag{40}\label{e:40} MgR = (J_0 - J_R)\,\omega_0\,\Omega \qquad \text{(exactly, at }\theta = 90^\circ\text{)} \end{equation}

holds exactly --- not as a fast-spin approximation --- while the familiar textbook rate
$\Omega = MgR/[(J_0 - J_R)\omega_0]$ is recovered from~\eqref{e:39} as the slow root in
the fast-top limit. \S\ref{sec:8} verifies that~\eqref{e:39} coincides term for term
with the classical steady-precession relation of the heavy symmetric top. Appendix~F
exhibits the same balance entirely in numbers for one worked configuration, showing the
two torques of~\eqref{e:32} equal in magnitude and opposed digit for digit at the steady
tilt, so that their resultant is the zero vector, and separating in a restoring direction
on either side of it.

Collected as a quadratic in $\Omega$, relation~\eqref{e:39} carries for its
$\Omega^{2}$ coefficient the product of $\cos\theta$ with
$\tfrac{1}{2}(J_0 - 3J_R) - MR^{2}$, and equation~\eqref{e:40} is the case in which the
first of those two factors vanishes. The second can vanish as well. That bracket
depends on the shape of the body and on the length of the shaft together, so for any
body disk-like enough that $\tfrac{1}{2}(J_0 - 3J_R)$ is positive there is one pivot
distance at which the two terms cancel exactly. The relation is linear again there and
returns a single precession rate; approaching that geometry from either side the fast
branch grows without bound, and its sense is opposite on the two sides, while
the slow branch passes through it unaffected. Appendix~D.7 sets this out and locates
that geometry.

\hsubsection{6}{Scope and validity tiers}\label{sec:7.6}

The balance~\eqref{e:32}, with $\Mscr_A$ given by~\eqref{e:25} and the internal torque
equal to $d\vv{S}/dt$ (Proposition~3), is exact for any closed body pivoted at a fixed
frictionless point in uniform gravity --- regular or irregular, rigid or deforming: in
the deforming case it reads
$M\vv{R}\times\vv{g} = M\vv{R}\times(d^{2}\vv{R}/dt^{2}) + d\vv{S}/dt$, a governing law
that transcends rigidity. Rigidity enters only through the kinematic substitutions~\eqref{e:14}--\eqref{e:15} and~\eqref{e:26}--\eqref{e:27} and the internal-torque evaluation~\eqref{e:19}; axisymmetry with spin about the symmetry axis enters in the steady-precession kinematics~\eqref{e:33}--\eqref{e:34}, in the step to~\eqref{e:35}, where $\vv{F}$ returns the symmetry-axis vector parallel to itself, and in the closure~\eqref{e:36}--\eqref{e:37} --- an asymmetric body
obeys the same balance but generally tumbles instead of precessing steadily. The
physical limits are those stated in \S\ref{sec:3.1}: a fixed frictionless pivot,
uniform gravity, closed constant mass, and no other external actions; within them, no
approximation is made anywhere in the chain from~\eqref{e:32} to~\eqref{e:40}. What
remains outside the closed-form scope of this paper is general non-steady motion: the
same balance~\eqref{e:32} governs it, but no closed-form solution of it is offered
here.

\hsection{8}{Worked Examples: Per-Solid Resolutions}\label{sec:8}

This section applies one two-integral recipe to every solid (\S\ref{sec:8.1}), reads the special cases off the scalars (\S\ref{sec:8.2}), and checks the results against the classical relation (\S\ref{sec:8.3}) and by independent numbers (\S\ref{sec:8.4}).

\hsubsection{1}{The uniform two-integral recipe}\label{sec:8.1}

Every solid is processed by the identical procedure: compute the two scalar mass
integrals $J_0$ = $\int$($\vv{\xi}\cdot\vv{\xi}$)dm and $J_R$ =
$\int$($\vv{\xi}\cdot\vv{R}$)$^2$ dm / ($\vv{R}\cdot\vv{R}$) from the body's geometry
--- elementary volume integrals in which odd terms cancel by symmetry --- and insert the
pair into the one balance~(38)--(39). No shape requires a different formalism and no
thickness regime requires a different precession equation. Table~2 collects the results
for three general homogeneous bodies, each symmetric about the shaft through its center
of mass: a hollow sphere of outer diameter D and inner diameter E; a torus of circular
cross-section with outer diameter D and inner diameter E; and a torus of rectangular
cross-section with outer diameter D, inner diameter E, and height H; these bodies and
their dimensions, with the thin disk, are drawn in Figures~2 to 5. The classical solid sphere, disk, and
cylinder or rod are recovered from these three as their $E = 0$ and $H \to 0$ or $D \to 0$ limits.
The practical point is not that classical mechanics lacks continuous mass properties ---
it does not --- but that the same two integrals and the same solver interface remain valid while the geometry is swept continuously.

\afterpage{\begin{table}[p]
\centering
\TABLEONE
\caption{Table~2 --- The two closure scalars per solid. Each $J_R$ except the hollow sphere's is computed directly from the $\vv{F}(\vv{R})$ integral, and the hollow sphere's follows from isotropy and the trace identity; derivation highlights appear in Appendix~C.}
\end{table}}

\hsubsection{2}{Special cases read off the scalars}\label{sec:8.2}

Structural facts that classically require principal-moment comparisons drop out of the
pair $(J_0, J_R)$ by inspection. For any isotropic body --- the hollow sphere of
Table~2 and its solid-sphere limit alike --- full isotropy requires $J_R = J_0/3$, so
by~\eqref{e:37} the operator collapses to a single-scalar law
$\vv{F}(\vv{v}) = (J_0/3)\,\vv{v}$, the coefficient $\tfrac{1}{2}(3J_R - J_0)$ vanishes
identically, and the balance loses its third term:

\begin{equation}\tag{41}\label{e:41} M\vv{R}\times\vv{g} - M(\vv{\Omega}\cdot\vv{R})(\vv{R}\times\vv{\Omega}) - \tfrac{2J_0}{3}(\vv{\Omega}\times\vv{\omega}_0) = \vv{0} \qquad \text{(isotropic body)} \end{equation}

For the thin disk --- the rectangular torus of Table~2 at $E = 0$, $H \to 0$ --- the
plane-lamina limit gives $J_R \to 0$, and the balance closes through the polar second
moment of mass alone:

\begin{equation}\tag{42}\label{e:42} M\vv{R}\times\vv{g} - M(\vv{\Omega}\cdot\vv{R})(\vv{R}\times\vv{\Omega}) - \tfrac{1}{2}J_0(\vv{\Omega}\times\vv{\omega}_0) + \tfrac{1}{2}J_0\,\frac{\vv{R}\cdot\vv{\omega}}{\vv{R}\cdot\vv{R}}\,(\vv{R}\times\vv{\Omega}) = \vv{0} \qquad \text{(lamina)} \end{equation}

And for the cylinder --- the rectangular torus of Figure~4 at $E = 0$ --- the coefficient $\tfrac{1}{2}(3J_R - J_0)$ changes sign at $H = D\sqrt{3}/2$, the exact transition
between disk-like and rod-like gyroscopic response, produced automatically by the two
standalone scalars with no eigen-analysis; the crossing is
continuous, every physical quantity remaining smooth through the point (Appendix~D, Table~D.1). Brand's
conventional principal-moment framework~\cite{ref37} can also represent this same
finite-cylinder continuity when its general inertia formulas are retained. The
distinction here is computational organization: one pair ($J_0$,$J_R$) is evaluated from
geometry and inserted unchanged into one precession relation over the whole sweep.
Appendix~D compares the two descriptions explicitly.

\begin{figure}[htbp]
\centering
\begin{minipage}[t]{1.818in}\vspace{0pt}%
\includegraphics[trim=28.08bp 24.48bp 113.76bp 7.2bp,clip,width=1.818in]{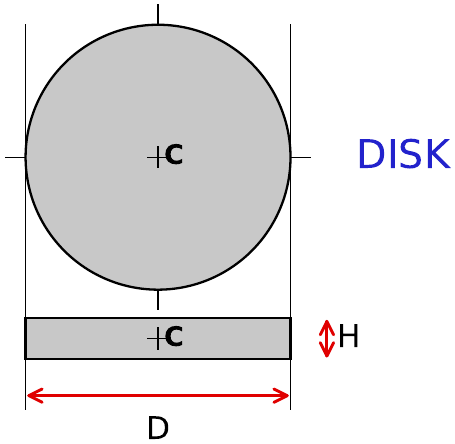}%
\end{minipage}\hspace{1.2in}%
\begin{minipage}[t]{2.094in}\vspace{0pt}%
\includegraphics[trim=11.88bp 9.72bp 96.84bp 11.88bp,clip,width=2.094in]{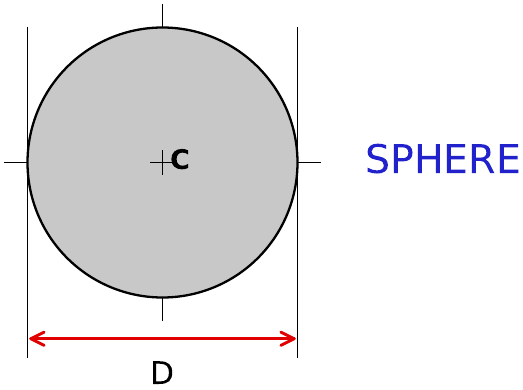}%
\end{minipage}
\caption{Figures~2 and 3 --- The geometry of the disk and of the sphere, and the
position of their center of mass, C.}
\end{figure}

\begin{figure}[htbp]
\noindent\hspace*{0.7696in}%
\begin{minipage}[t]{2.5008in}\vspace{0pt}%
\includegraphics[trim=46.44bp 18.72bp 91.44bp 12.6bp,clip,width=2.5008in]{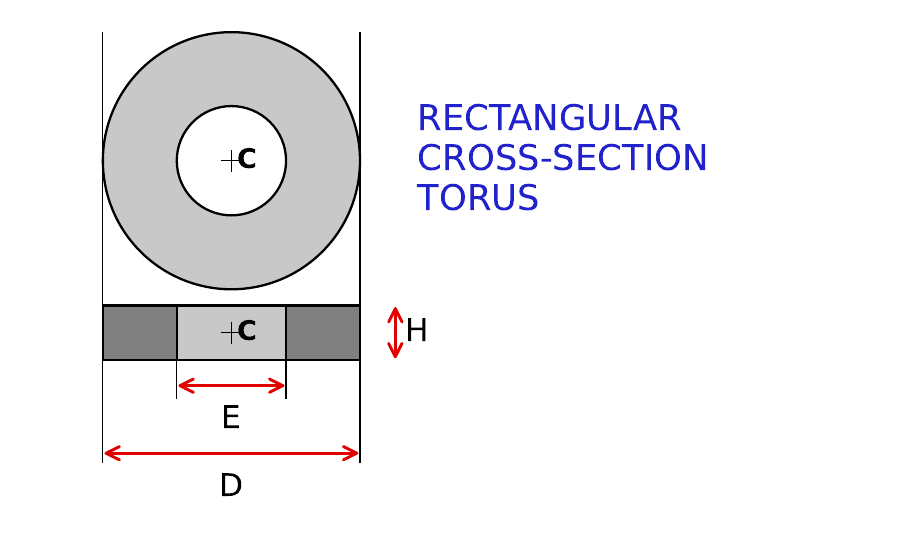}%
\end{minipage}\hspace{0.4921in}%
\begin{minipage}[t]{2.217in}\vspace{0pt}%
\includegraphics[trim=18.72bp 8.64bp 93.96bp 18.72bp,clip,width=2.217in]{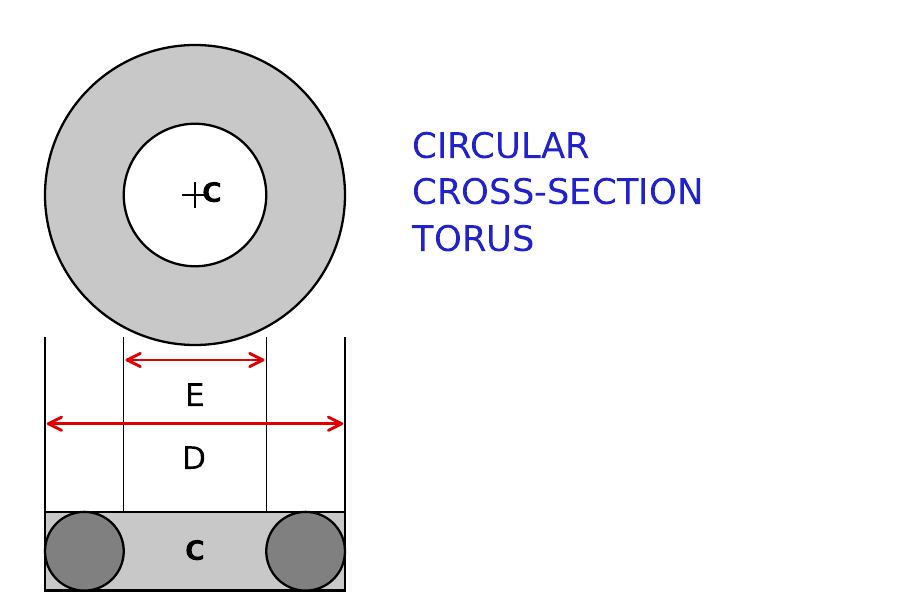}%
\end{minipage}
\caption{Figures~4 and 5 --- The torus of rectangular cross-section and the torus of
circular cross-section, with their center of mass, C.}
\end{figure}

\hsubsection{3}{Verification against the classical relation}\label{sec:8.3}

Before the correspondence, a note on symbols: the axial and transverse moments
$I_\parallel$ and $I_\perp$ appearing in this section and in Appendix~D, like the textbooks' $I_3$ and $I_1$ of Appendix~F, are projected quantities of the classical description. They enter only as the
vocabulary of verification --- a result can be checked against the classical relation
and its published values only by writing it, at that moment, in the classical language
--- and they are formed nowhere in the derivation, which runs throughout on the
direction-free $J_0$ and $J_R$ alone. For readers versed in the standard description,
the correspondence is a one-line dictionary --- stated here as a consequence of the
formulation, never as a definition within it: $J_0 - J_R$ coincides with the spin
moment of inertia about the symmetry axis (commonly $I_\parallel$), and
$\tfrac{1}{2}(J_0 + J_R)$ with the transverse moment through the center of mass
(commonly $I_\perp$); the transport term supplies the parallel-axis $MR^{2}$.
Substituting into~\eqref{e:39}, with the combination $\omega_0 + \Omega\cos\theta$
written out:

\begin{equation}\tag{43}\label{e:43} MgR = I_\parallel\,(\omega_0 + \Omega\cos\theta)\,\Omega - (I_\perp + MR^{2})\,\Omega^{2}\cos\theta \end{equation}

--- the classical steady-precession relation of the heavy symmetric top, matched term
for term against the three independent statements of it consulted here --- MIT
OpenCourseWare~\cite{ref24}, Tanr\i verdi~\cite{ref25} and Agashe~\cite{ref26} --- and
against the exact two-branch form of Bolina~\cite{ref27} in Appendix~D.5. The combination $\omega_0 + \Omega\cos\theta$ that multiplies
$I_\parallel$ in~\eqref{e:43} is what the classical texts write as a single axial rate:
it is the projection of the total angular velocity onto the symmetry axis. This paper
writes it out in its own two rates and gives it no symbol of its own; it is an object of
the classical description, brought in here solely to cast the result in the standard form
for comparison, and it enters no part of the derivation of~\eqref{e:39}, which
constructs no such projection. The correspondence therefore runs one way:~\eqref{e:43}
follows from equation~\eqref{e:39} through the dictionary and this single substitution,
whereas nothing in the paper is derived from~\eqref{e:43} --- the classical relation is
reproduced by the formulation, never used by it. The formulation is thus
informationally equivalent to the standard description on their common ground, while
having been derived without the classical axial and transverse moments $I_\parallel$ and $I_\perp$.

Brand~\cite{ref37} supplies an earlier exact vectorial route to this same end relation.
In his notation $C$ is the principal moment about the symmetry axis and $A$ the transverse
principal moment about the fixed point; neither is the center of mass $C$ or the pivot $A$ of this paper. For the body and geometry used here the
dictionary is $C = J_0 - J_R$ and $A = \tfrac{1}{2}(J_0 + J_R) + MR^2$. Brand's
exact gravity-precession equation, $(A - C)\cos\theta\cdot\Omega^2 - C\omega'\Omega
+ Wb = 0$, becomes the quadratic form of (39) identically under $W = Mg$, $b = R$ and
$\omega' = \omega_0$, with no adjustment of sign or convention required. Brand
therefore confirms the exact physics, both branches and the horizontal special case. The
methodological difference lies before this common scalar endpoint: Brand reaches it
through a moving principal-axis triad and component angular momentum, whereas (35)
retains $J_0$ and $\vv{F}$ as intrinsic objects and keeps the separate torque
contributions cross-product-readable until the final reduction.

\hsubsection{4}{Independent numerical confirmation}\label{sec:8.4}

The dictionary of \S\ref{sec:8.3} makes a direct numerical comparison possible, and
three general bodies were computed to carry it out: a hollow sphere and toruses of
circular and rectangular cross-section, each of mass 1 kg spinning at 1000 rpm,
evaluated at seven pivot distances, with the solid sphere recovered exactly at $E = 0$ and two finite cylinders as the disk and the rod; the tilts examined span 60$^\circ$,
42$^\circ$, 90$^\circ$ and 31.416$^\circ$. The numerical inputs are these:
$g = 9.8$~m/s$^2$ throughout; the seven pivot distances are $R = 0.02$, 0.03, 0.04,
0.05, 0.06, 0.07 and 0.08~m; the hollow sphere has $D = 0.100$~m and $E = D/3$ at
$60^\circ$, the torus of circular cross-section $D = 0.300$~m and $E = 0.100$~m at
$31.416^\circ$, and the torus of rectangular cross-section the same two diameters with
$H = 0.100$~m at the same tilt; the classical bodies are taken at $E = 0$ and $D = 0.100$~m --- the solid sphere at $60^\circ$, the thin disk with $H = 0.010$~m at
$42^\circ$, and the rod with $H = 0.200$~m at $90^\circ$. Each was resolved from its Table~2 pair and
the scalar reduction~(39) alone, then compared against three independently published
exact formulations of the heavy symmetric top --- MIT OpenCourseWare~\cite{ref24},
Tanr\i{}verdi~\cite{ref25} and Agashe~\cite{ref26}. After notation is mapped, those
three reduce to the relation~(39) of this paper, the same quadratic --- or, at $\theta$ = 90$^\circ$, to the same exact linear relation~(40)
--- and reproduce the same precession rates to the displayed precision, the residual
differences being decimal rounding. The comparison also exhibits the structural features
the formulation predicts: both branches retained, the coefficient zero at which the
second branch reverses sign --- set out in Appendix~D.7 --- and the boundary
of the domain of definition at which the discriminant vanishes. This three-body sweep is not the whole
of the external confirmation offered here. Appendix~D adds further comparisons ---
the exact two-branch forms of Bolina~\cite{ref27} and of Slade~\cite{ref33}, each valid at every non-vertical tilt except the horizontal shaft, beside the exact relations of Brand~\cite{ref37} and of Klein and
Sommerfeld~\cite{ref38}, and the minimum-spin condition of Goldstein~\cite{ref2} and of
Thornton and Marion~\cite{ref4}. The three-body studies
and their source-by-source equation mapping are collected in the supporting numerical results (\texttt{Part2\_\allowbreak{}Supporting\_\allowbreak{}Numerical\_\allowbreak{}Results.pdf}), an ancillary file provided with this paper, as are a forward-and-inverse precession solver (described in \texttt{Part3\_\allowbreak{}Forward\_\allowbreak{}and\_\allowbreak{}Inverse\_\allowbreak{}Precession\_\allowbreak{}Solver.pdf}), a verification package (gyro\_\allowbreak{}verification\_\allowbreak{}suite) implementing the checks~\#1--\#31 that Appendix~E cites, and a sweep program (Sweeper\_\allowbreak{}Optimizer) that evaluates the relation across a range of one
variable and reports where steady precessions exist, how many, and on which branch, and
the ancillary comparison file.

\hsection{9}{Discussion: Contributions, Merits, and Limits}\label{sec:9}

This section lists the contributions (\S\ref{sec:9.1}), weighs the merits against the classical description (\S\ref{sec:9.2}) and states the limits (\S\ref{sec:9.3}).

\hsubsection{1}{Catalog of contributions}\label{sec:9.1}

The contributions of this work, enumerated for precision and grouped by kind. Read in the order \S\ref{sec:1} gives them, the interpretation, which is the motivating cause of the work, is carried by C4, C5, C7 and C17. The method, which is the tool of that interpretation, is carried by C1--C3, C6, C8, C9 and C14--C16. C10--C13 are the results the two yield. \S\ref{sec:1} states what the interpretation shares with Brand and with Klein and Sommerfeld, and what, among the features compared in the ancillary comparison file, is this paper's alone.

\textbf{Definitional (C1--C4).} C1: the polar scalar $J_0$ serves as a primary inertia
descriptor of the dynamics, defined directly by~\eqref{e:1} and never assembled from
axial or planar moments. C2: a single mass-vector operator $\vv{F}$~\eqref{e:2} carries
all directional inertia information, replacing the apparatus of moments and products of
inertia; where the classical description needs a symmetric array and an eigenvalue
problem, this one needs one integral rule applied on demand. C3: the framework's
internal bookkeeping closes on itself through Propositions~1--3 --- the constancy of
$J_0$ for the rigid body, the vanishing of the net internal acceleration for any closed
body, and the identity of the internal torque integral with $d\vv{S}/dt$ --- with no
external tensor identities invoked anywhere. C4: the entire dynamics is formulated with
material derivatives referenced to the inertial, measurable frame --- the
$\partial/\partial t$ bookkeeping derivative never appears as an independent physical input, the one other frame named, the chosen frame of the transport rule~\eqref{e:6}, serves only that bookkeeping and carries no dynamics, and no force is used
anywhere in the derivation: every dynamic quantity is a torque, each mass element entering only through
its infinitesimal inertia torque~\eqref{e:20}.

Methodological (C5--C9). C5: the governing balance~(32) is derived from Newton's law per mass element, taken from the start as a torque about the pivot --- the
internal torques canceling pairwise, the pivot contributing no torque --- with no angular-momentum primitive invoked. C6: inertia about the center of mass enters the torque exclusively through the inseparable pairs $J_0\vv{\varepsilon} - \vv{F}(\vv{\varepsilon})$
and $\vv{F}$($\vv{\omega}$) $\times\vv{\omega}$, each a pure mass-integral vector
expression retaining full directional information for a body of any shape, which is why
no thin-body idealization is needed anywhere. C7: the dynamics culminates in one unified
vector equation of motion~(35), each of whose terms is physically legible and can be visualized by the corkscrew mnemonic; the distinguishing property is not vector notation alone but
persistence of the intrinsic three-dimensional vector representation --- the governing vectors are not replaced by coordinate or principal-axis components before the final geometrically justified scalar reduction, and their cross-product directions remain traceable. C8: the two-route consistency theorem~(31)
validates the assembly of the inertia torque internally --- the direct material route and the
frame-decomposition route agree term for term --- and simultaneously exposes and
forecloses the multiple-counting error that arises if frame-decomposition rows are
summed as independent torques. C9: for each solid, the closing scalar $J_R$ is produced
by the formulation itself, as the operator's own response to the shaft vector~(36) --- the direction of $\vv{F}(\vv{R})$ and that scalar are outputs, never inputs.

Results (C10--C13). C10: the results are exact where stated --- in particular the
horizontal-shaft relation~(40) holds exactly, with the familiar textbook rate recovered
only afterward as a limiting root; these exact physical results are not claimed as new
and are independently present in Klein and Sommerfeld~\cite{ref38} (1897), Brand~\cite{ref37} (1930) and other classical treatments. C11:
the single quadratic~(39) contains both classical precession branches, with the
parallel-axis contribution M R$^2$ arising automatically from the transport term. C12:
structural facts --- any isotropic body's single-scalar collapse~(41), the lamina
closure~(42), the disk-to-rod crossover at $H = D\sqrt{3}/2$ --- are read off the scalar
pair by inspection, with no eigen-analysis. Appendix~D establishes the crossover's exact continuity and includes Brand's exact relation and finite-cylinder inertia formulas
as an external classical comparison. C13: the balance itself --- the theorem of moment of momentum that Brand states for any system of particles (\S183~\cite{ref37} and \S184~\cite{ref37}), a textbook identity, as \S\ref{sec:4.2} notes --- holds exactly beyond rigidity, $M\vv{R}\times\vv{g} = M\vv{R}\times(d^{2}\vv{R}/dt^{2}) + d\vv{S}/dt$, with rigidity confined to the closed-form solution.

Computational and pedagogical (C14--C16). C14: one uniform two-integral recipe resolves
every solid geometry identically (Table~2) and is actually used as one unchanged
geometry-to-solver interface across finite thickness, including the continuous
disk-to-rod sweep of Appendix~D; no classification as disk-like, cylinder-like, or
rod-like changes the equations. C15: a single algebra --- mass integrals and intrinsic cross products, with dot products secondary, appearing inside the two mass integrals and as reductions of cross products --- runs unbroken from the first kinematic statement to the
exact torque-vector balance, and each torque contribution retains a traceable cross-product sense until the final geometrically justified scalar reduction.
C16: the formulation's own boundary conditions are stated as results rather than left
tacit: the standing assumptions are enumerated (\S\ref{sec:3.1}), the scope of the steady solution is stated (\S\ref{sec:7.6}), and its existence condition at a given spin is set out in Appendix~D.6 and D.7. Dynamic equilibrium constant tilt angle (C17). C17:
the dynamic equilibrium constant tilt angle occurs as the equalization of the two
torques about the pivot --- the
gravity torque and the inertia torque assembled from (35) --- which Klein and Sommerfeld~\cite{ref38} and Brand~\cite{ref37} already state at the steady tilt, and which is shown here to be broken in a restoring direction on either side of it. Appendix~F
works one configuration completely in numbers and reads that equalization, and its
restoring signature, directly off a table in N$\cdot$m. For the claim-by-claim
validation of this catalog, see Appendix~E.

\hsubsection{2}{Merits relative to the classical description}\label{sec:9.2}

The structural difference is present at the level of the definitions, before any
dynamics: $J_0$ and $\vv{F}$ reference nothing outside the body itself --- no unit
vectors, no chosen axes, no collinearity or perpendicularity constructions, no prior
decomposition of space into `along' and `across'. The classical axial and transverse
moments cannot be written down until an axis has been selected; here, directions are
never inputs to definitions and appear only as outputs of evaluation. That intrinsic representation is not temporary. The torque-vector balance is retained through (38), and
\S\ref{sec:7.5} takes the scalar relation only after all terms are shown to lie along
one common physical direction. From this follow the practical merits: one operator rule instead of parallel/perpendicular case analysis up to~\eqref{e:35}; no matrix or diagonalization for any body treated; identical definitions across regular,
irregular, and deforming bodies; and continuous geometric traceability of every torque
contribution through explicit cross products.

The difference is one of representation and method rather than of physics.
Brand~\cite{ref37} is the clearest historical example: he starts from vector mass
integrals and uses the right-hand screw, then deliberately changes representation by
selecting a moving principal-axis triad, resolving angular momentum into components, and
carrying the inertia as A, B, C. That route is correct and gives the same exact
steady-precession quadratic, the same branches, and the same horizontal special case. Klein and Sommerfeld~\cite{ref38}, earlier still, reach the same quadratic, its two roots, the same horizontal case and the degenerate case $A = C$ by a route carried in the components of the impulse and of its change, within a work that also uses the Euler angles, the Lagrange equations and elliptic functions.
The present formulation does not make either representational change. Inertia remains in
the COM-defined scalar $J_0$, the mass-vector operator $\vv{F}$, and the closure scalar
$J_R$ the operator produces on its own; the torque remains an intrinsic vectorial-torque balance
whose cross-product directions can be traced term by term. The three routes use the same mathematics and physics and end at equivalent scalar results, but their derivational
methodologies are distinctly different.

The correspondence of \S\ref{sec:8.3} should be read in that light. The identities
$I_\parallel$ = $J_0-J_R$ and $I_\perp$ = $\tfrac{1}{2}$($J_0$ + $J_R$) are outputs of
the formulation, offered so that a reader fluent in the classical description can follow
the comparison; they are never inputs to it. Brand's A and C map to the same physical
inertia information once the pivot shift is included, so his exact equation and (39) are
algebraically identical. That agreement establishes equivalence of the physics, not
equivalence of the derivation. Brand's principal-axis components are his working
representation; $J_0$, $\vv{F}$, $J_R$ and the intrinsic dot/cross-product balance are
the working representation here. This persistence of representation --- including the
fact that the scalar relation is deferred until the vector geometry itself supplies one
common direction --- is the methodological distinction.

\textbf{Capability versus methodology.} Brand's formulas contain enough classical information for
a reader to construct a continuous finite-cylinder calculation: \S193~\cite{ref37} supplies
the axial cylinder inertia, \S196~\cite{ref37} supplies a transverse result, and the
transfer theorem supplies the required shift. Combined with Brand's exact
\S218~\cite{ref37} precession equation, those ingredients can be assembled into a
continuous disk-to-rod sweep. His actual method requires the user to
select the relevant axes, assemble the corresponding principal moments and shifts, and
then evaluate the precession relation. This paper does something different in fact, not
merely in possibility: the same definitions $J_0(H)$ and $J_R(H)$, the same operator
closure, and the same exact relation~(39) are used without alteration from the thin-disk
limit through every finite cylinder to the rod limit. The distinction is therefore
precisely the difference between what the classical formulas can be made to do and what
the present methodology does by construction.

The same distinction applies to present-day numerical software: it establishes capability; it does not establish that a gyroscope-specific workflow, teaching
simulator, or user implementation actually presents one symbolic disk-to-rod
methodology, avoids regime-dependent approximations, or preserves the intrinsic
cross-product representation of the derivation. This paper therefore makes no blanket
claim that all modern simulators are discontinuous. Its narrower and directly
demonstrated usability claim is about what this method does: for the axisymmetric
precession problem, one geometry routine returns $J_0$ and $J_R$, and one unchanged
dynamics relation consumes them across the entire finite-thickness family without
classifying the body as thin disk, thick disk, cylinder, or rod.

Brand's gyroscope derivation terminates
in the conventional principal-moment variables A, B, C and does not propose a separate
geometry-sweep interface. The present work proposes and demonstrates a different computational
organization: preserve the $J_0$--$\vv{F}$ vector-operator language through the
derivation, close the axisymmetric geometry by $J_R$, and pass the same pair to the same
steady-precession equation for every member of the family. Whether this compact
interface is advantageous in a particular software system is an engineering question;
the ancillary forward, inverse, and sweep programs of Sweeper\_\allowbreak{}Optimizer (its point solver, solver2, and its sweep program, sweeper) make the interface directly testable.

\hsubsection{3}{Assessment and limits}\label{sec:9.3}

On their common ground --- the axisymmetric rigid rotor --- the operator description,
Brand's principal-moment description~\cite{ref37}, and the standard tensor description
are informationally equivalent: they encode the same inertia information and predict the
same steady-precession physics. The contributions cataloged above are therefore contributions of interpretation and of formulation, not new physical content: the torque-vector reading of the classical balance, and the method that carries it out --- persistent intrinsic-vector
representation, economy of definition, uniformity of method, explicit geometry continuity, internal verifiability, and cross-product traceability, with an exactness it shares with the classical treatments, the familiar approximations included as special cases. Continuity as a mathematical capability
is not exclusive to this formulation: Brand's finite-cylinder formulas can be assembled into a continuous calculation. The methodological claim is different: that capability does not by itself demonstrate the same unchanged geometry-to-dynamics workflow. This paper actually carries the same primary
pair and the same solver relation through the disk-to-rod family without a change of
representation. That methodological claim is not externally validated, and the gap is
stated rather than passed over: the searched literature supplies the exact end equation
and the finite-height inertia ingredients, but no published worked precession sweep
across the disk-to-rod transition was found, so the continuous sweep itself is confirmed
only internally; the two-scalar machinery is checked against outside sources at
individual geometries, including ones in which $J_R$ is not negligible. Appendix~E records that claim as
open and Appendix~D.8 sets out what Brand does and does not supply. The limits are
equally explicit: the balance presumes a fixed
frictionless pivot, uniform gravity, and closed constant mass; pivot friction, gravity
gradients over large bodies, and aerodynamic torques lie outside the core derivations;
general non-steady motion, while governed by the same balance~(32), is not solved in
closed form here; and the steady-precession solution exists as such only for symmetric
rotors spun about their symmetry axis --- an asymmetric body obeys the same equation and generally tumbles. The restoring signature of C17 is constrained in the same explicit way: the spin
and the precession are held at their magnitude equalization values while the tilt alone is
displaced, so it is a result about the two torques and not a stability analysis, as
Appendix~F.12 sets out in full. Where the two descriptions part company is not in what they predict but in how
they are carried, and the ancillary comparison file reports that difference feature by feature.

\textbf{Validity of the general (non-rigid) formulation.} The limits above concern the
rigid closed-form solution. The general formulation --- the kinematics of
\S\ref{sec:3.3} and the balance whose scope \S\ref{sec:7.6} states --- carries three
standing conditions, stated at their points of use and gathered here in one place. The
body is closed: $M$ is constant and no mass crosses its boundary, which excludes
rockets, ablating bodies, and any system exchanging mass with its surroundings. The
vector $\vv{\xi}$ is measured from the instantaneous center of mass, so that
identity~\eqref{e:5} and its time differentials hold at every instant however the body
deforms. And every derivative that carries physical meaning is material, taken
following the same mass element $dm$; the companion derivative $\partial/\partial t$
enters only as the bookkeeping object defined by the transport rule~\eqref{e:6}, never
as an independently measured rate.

\textbf{What the general formulation does not supply.} It is a torque accounting, not a
theory of deformation. Nothing in it determines what causes $\vv{\xi}$ to change: no
constitutive law --- elastic, plastic, viscous, or fluid --- is included, so applying
the balance to a real deforming body requires that physics to be supplied separately.
The pairwise cancellation of internal moments in \S\ref{sec:7.1} likewise assumes that each equal-and-opposite pair of internal actions lies along the line joining its two elements, equivalently that the internal stress is symmetric with no
couple stresses. A medium carrying internal couples lies outside the accounting, not
because the balance fails but because its internal moment no longer integrates to zero.
Energy is not tracked either, so the framework alone cannot certify that a prescribed
deformation is energetically consistent or physically realizable; given the
deformation, it returns the torque that deformation entails. Within these conditions
the balance~\eqref{e:32} holds without approximation for any closed deforming body, as
\S\ref{sec:7.6} states.

\hsection{10}{Conclusions}\label{sec:10}

A complete, self-contained, vector-only formulation of gyroscope precession about a
fixed pivot has been presented. Its foundations are two direction-free mass integrals
--- the polar second moment of mass $J_0$ and the mass-vector operator $\vv{F}$ ---
together with a small companion family whose closure identities are proven inside the
calculus itself. Assembling the torque exactly once from the material acceleration of
each mass element yields the exact statement
$\vv{\mathscr{M}}_A = -\,M\,\vv{R}\times(d^2\vv{R}/dt^2)
- \int_0^M \vv{\xi}\times(d^2\vv{\xi}/dt^2)\,dm$, validated
internally by the two-route consistency theorem; balancing the gravity torque against it
produces a unified vector equation of motion for steady precession~(35), resolved in
closed form by the emergent scalar $J_R$. The scalar reduction~(39) contains both
classical precession branches, holds exactly at the horizontal shaft, and coincides term
for term with the classical relation of the heavy symmetric top. Klein and Sommerfeld~\cite{ref38} stated that relation in 1897 and Brand~\cite{ref37} in 1930, each at any non-vertical tilt, with both branches and without the fast-top approximation, and each with the horizontal-shaft case holding exactly. Both take the angular momentum, as a vector, as the primary carrier of the rotation and build their scalar equation up from its components on chosen axes, and both equations map directly to \eqref{e:39} under the same identities, $C = J_0 - J_R$ and $A = \tfrac{1}{2}(J_0 + J_R) + M R^2$. No novelty is
therefore claimed for those physical results, for the mathematics they share, or for
vector treatment itself. The contribution claimed here is, in order, an interpretation and a method. The interpretation, the motivating cause of the work, reads free precession as a dynamic equilibrium of torque vectors about the pivot, held from the mass element to the governing equation. The method, the tool of that interpretation, is the derivational methodology and its persistence of representation: physical vectors are not replaced by coordinate or principal-axis components before the final scalar reduction; dot products are intrinsic geometric operations rather
than coordinate projections; no tensor, matrix or dyadic formalism and no Euler angle enters the derivation, and no unit
vector, principal-axis construction, or component resolution enters the derivation of the governing
torque-vector equation~\eqref{e:35}; and the direction and
sense of the separate torque contributions remain explicit intrinsic cross products
through the exact torque-vector balance. The balance of the gravity torque against the inertia torque is itself shared with Brand and with Klein and Sommerfeld. \S\ref{sec:1} names what, among the features compared in the ancillary comparison file, is this paper's alone. The scalar equation is taken only after that balance
has established one common physical direction. The same distinction applies
computationally. Brand's formulas can be assembled into a continuous finite-cylinder
calculation. Here the same $J_0(H)$, $J_R(H)$, operator closure,
and exact precession relation are actually used from the thin-disk limit through finite
cylinders to the rod limit without a regime switch in equations or solver interface.
General-purpose simulators may be capable of the same continuous physical modeling when
supplied full inertia, but capability is not the methodological claim made here. The
demonstrated distinction is that the continuity is built into and exercised by the same
intrinsic-vector/two-integral representation that produces the precession calculation.
The claims extend no further than what is demonstrated: on the common ground of the
axisymmetric rigid rotor the formulation is informationally equivalent to Klein and Sommerfeld, to Brand and to the classical description, in that $J_0$ and $J_R$ carry the same inertia information as the principal moments that Brand and Klein and Sommerfeld write $A$ and $C$, and \eqref{e:39} is the same steady-precession relation. The formulation's contribution is, in order, the torque-vector interpretation and the method that is its tool --- persistent
three-dimensional vector integrity, economy, uniformity, explicit continuity of method, internal verifiability, and continuous cross-product traceability, with an exactness it shares with the classical treatments. Beyond the
steady state itself, the same balance and the same two scalars follow the torque equalization away from the steady tilt: it is exact at that tilt, where Klein and Sommerfeld and Brand already state it, and broken in a restoring direction on either side. Appendix~F demonstrates that equalization completely in
numbers for one configuration. A first version of these derivations was completed by the author in 1975; they are published here in full for the first time.

\section*{\phantomsection Acknowledgments}
\addcontentsline{toc}{section}{Acknowledgments}
\label{sec:ack}

I am deeply grateful to the mathematics and science teachers of my school and
university years --- the faculty of Colegiul Na\textcommabelow{t}ional ,,Sf\^antul
Sava'' in Bucharest, and my professors at the Bucharest Polytechnic Institute, Faculty
of Electronics. In high school physics we were taught to carry a problem symbolically
from beginning to end, tracking both the algebra and the dimensional consistency of
every variable, and only at the very end converting the symbolic result into numbers.
That principle --- work as generally as the problem permits, and turn to particular
numerical cases only at the close --- has guided my working life since, as an
electronic design engineer in Silicon Valley and as an independent researcher. My
special thanks to my high school physics teacher, Ms.~Georgeta Nicolov, and my undergraduate
college mechanics professor, for their efforts in educating me and for providing the
intellectual ability and discipline to ask questions and investigate systematically and
thoroughly. My gratitude to my wife Bonnie, and to my family for encouraging and
supporting me during this research effort.

% arXiv version 2: the AI-use disclosure follows the Acknowledgments directly.
\medskip
\noindent\textbf{Use of artificial-intelligence (AI) tools.} Multiple AI systems assisted
in preparing this work, in the following capacities: symbolic and algebraic verification
of the derivations; numerical evaluation of the results; assistance in drafting and
revising the text; writing the ancillary software (gyro\_\allowbreak{}verification\_\allowbreak{}suite and Sweeper\_\allowbreak{}Optimizer) and creating its tests and stress
tests; organizing related items into lists, identifying the features compared and
arranging the comparison tables; systematizing the numbering of equations, figures and
tables; searching, gathering and formatting the cited references and checking the paper's statements against their contents; reviewing the manuscript for errors; and proposing the software's terminology and drafting its specification. Errors found in the course of the work,
whether by the author or by an AI, were listed and corrected in sequence. Every example,
model and comparison in the paper is either the paper's own result or drawn from a cited
published source; none is an invention of an AI. No figure was produced by an
image-generating model; figures drawn with AI assistance follow the author's own sketches
and specifications, or the paper's own computed values. No AI tool is an author, and the
author takes full responsibility for the paper.

% The reference list sits between the Acknowledgments and Appendix A, as in the
% Word master. thebibliography supplies the heading; the contents line is added here.
\phantomsection

\section*{\phantomsection Appendix~A --- The Corkscrew Rule: Visualizing Cross Products}
\addcontentsline{toc}{section}{Appendix~A --- The Corkscrew Rule: Visualizing Cross Products}
\label{sec:appA}

The corkscrew mnemonic is a mechanical analog of the right-hand rule, used throughout
this paper to read torque directions off a figure without component calculations. To
evaluate $\vv{V}_3 = \vv{V}_1 \times \vv{V}_2$: place an imagined corkscrew
perpendicular to the plane of the ordered pair $\{\vv{V}_1, \vv{V}_2\}$; rotate
$\vv{V}_1$ toward $\vv{V}_2$ along the shortest path (strictly less than
$180^\circ$); stand on the side from which this rotation appears clockwise; a clockwise
twist advances the corkscrew along $\vv{V}_3$. The mnemonic encodes perpendicularity
and sense at once, agrees exactly with the right-hand rule, and --- being tied to the
shortest rotation --- avoids ambiguous long-way-round readings. Boundary cases:
parallel and antiparallel vectors span no unique plane, the shortest rotation is
undefined, and the cross product is the zero vector, consistent with
$|\vv{V}_1 \times \vv{V}_2| = |\vv{V}_1||\vv{V}_2|\sin\theta$. Practical procedure for
a torque diagram: draw both arrows from one origin; mark the smaller angle; choose the
side from which it looks clockwise; twist; draw the result in the advance direction;
label the order, since $\vv{V}_2 \times \vv{V}_1$ reverses the direction. In the torque
constructions of this paper the ordered pair is always the moment arm first --- as in $\vv{R} \times (M\vv{g})$ and $\vv{r}_P\times(-\,\vv{a}_P\,dm)$ --- so the local plane, the sense, and the
resulting arrow capture the turning influence about $A$ in one mental move. Figures~6
and 7 show the construction: the ordered pair of vectors, and the sense in which the
screw advances.

\begin{figure}[htbp]
\centering
\begin{minipage}[t]{2.3068in}\vspace{0pt}%
\includegraphics[trim=46.08bp 18.0bp 63.72bp 7.2bp,clip,width=2.3068in]{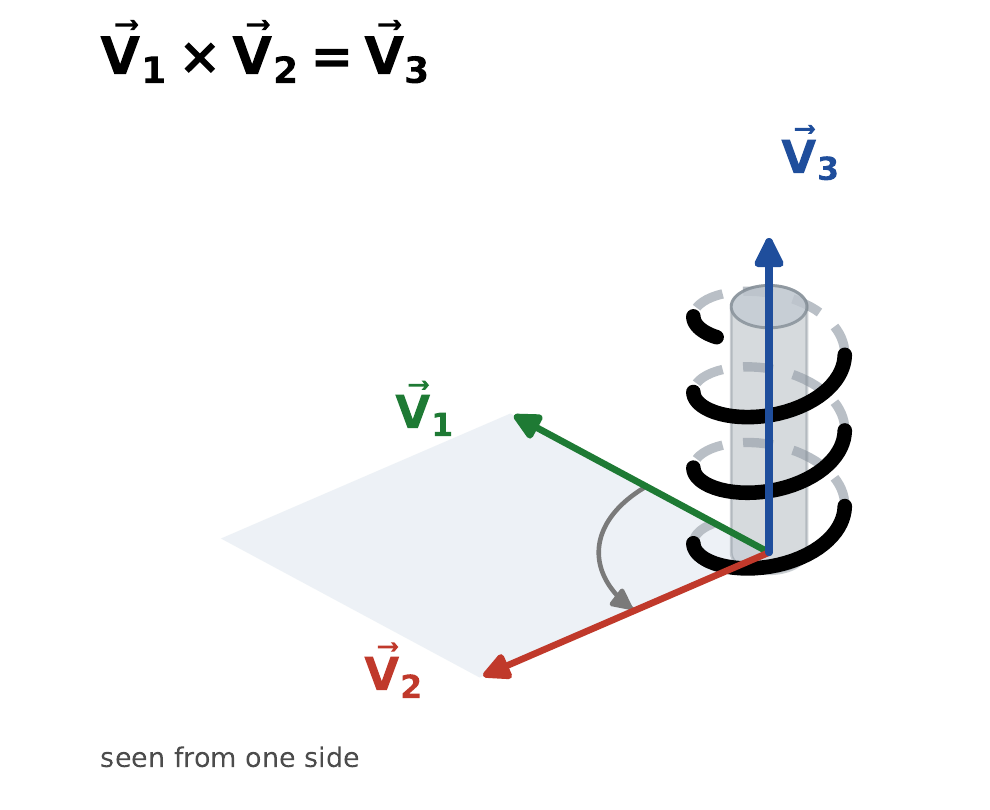}%
\end{minipage}\hspace{1.14in}%
\begin{minipage}[t]{2.1795in}\vspace{0pt}%
\includegraphics[trim=49.32bp 15.48bp 80.64bp 7.2bp,clip,width=2.1795in]{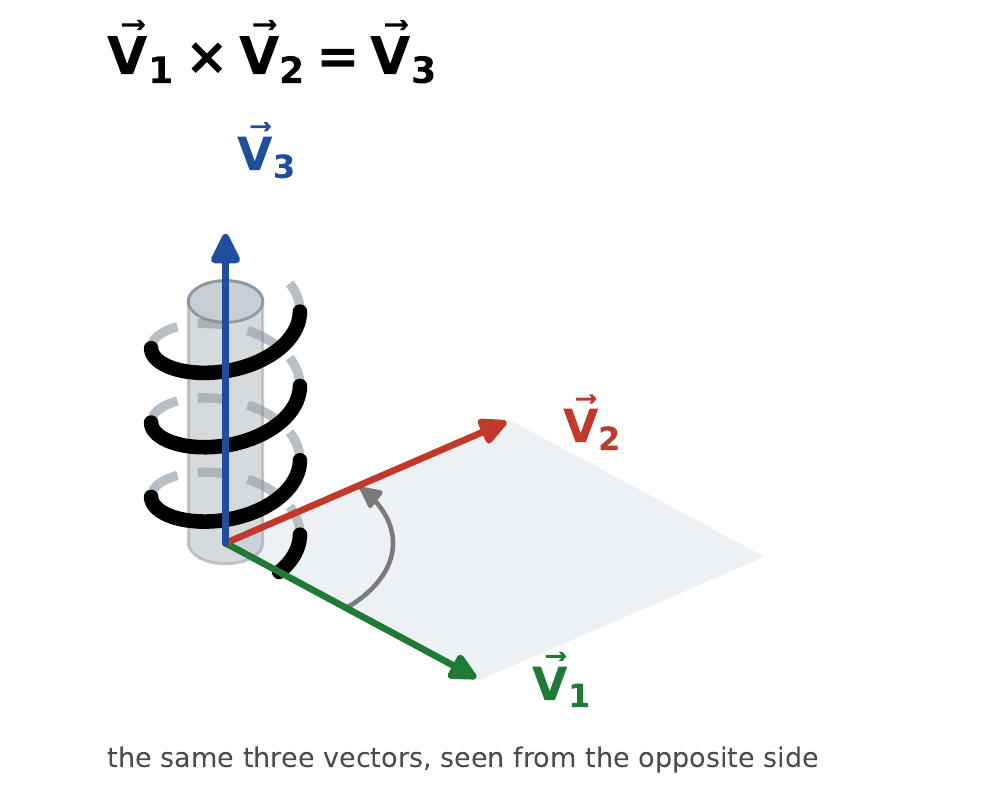}%
\end{minipage}
\caption{Figures~6 and 7 --- The corkscrew rule: the ordered pair of vectors and the
sense in which the screw advances, read off the figure without component calculations.}
\end{figure}

\section*{\phantomsection Appendix~B --- Historical Fragment (1975 Typewritten Manuscript)}
\addcontentsline{toc}{section}{Appendix~B --- Historical Fragment (1975 Typewritten Manuscript)}
\label{sec:appB}

The page reproduced below is from the author's original 1975 typewritten manuscript.

\begin{figure}[htbp]
\centering
\includegraphics[width=3.0in]{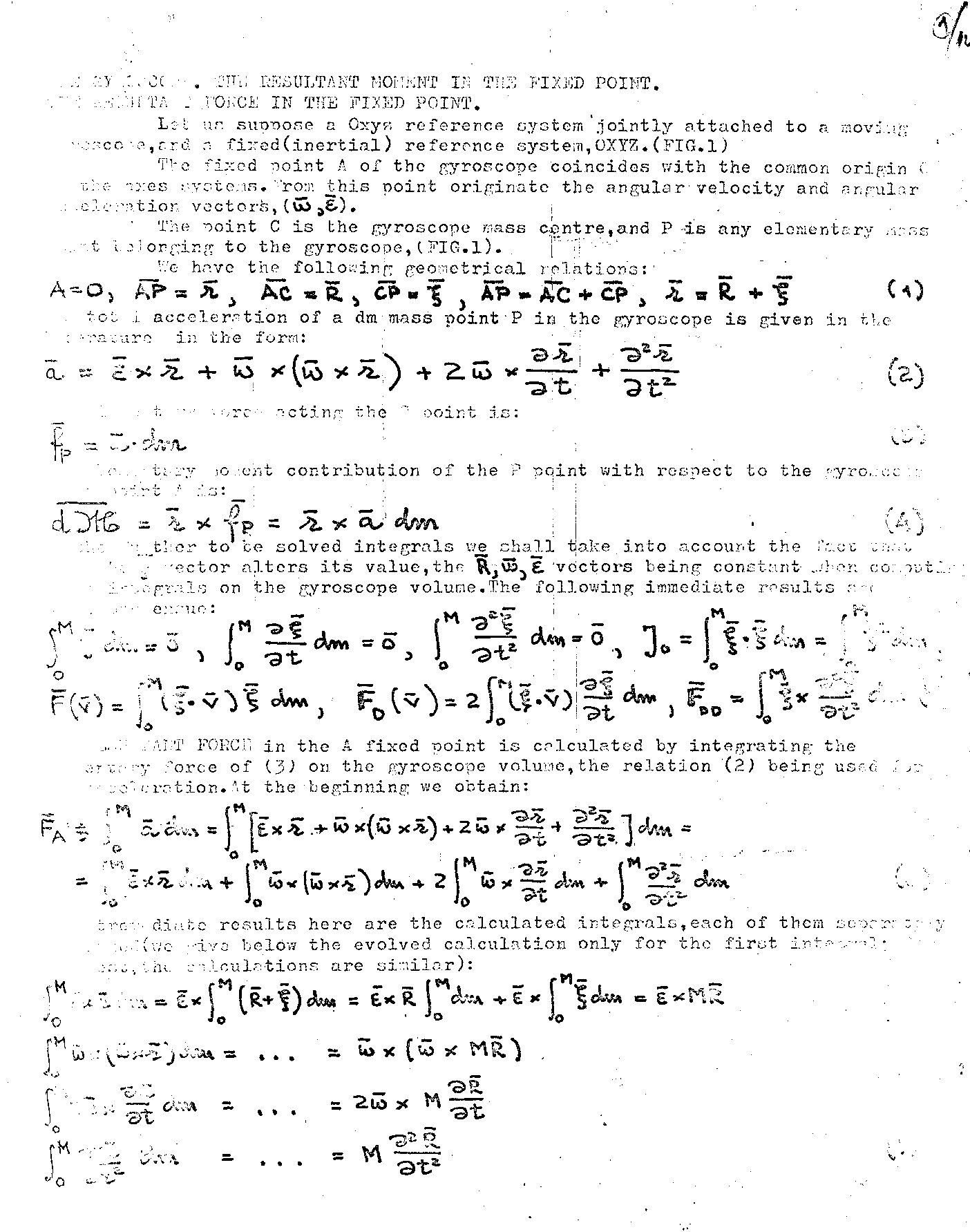}
\caption{Figure~8 --- A page of the author's 1975 typewritten manuscript.}
\end{figure}

That manuscript was sent, with a covering letter dated 10 August 1975, to Professor
Eric Laithwaite (1921--1997), then Professor of Heavy Electrical Engineering at
Imperial College of Science and Technology, London. He replied on 1 September 1975, on
Department of Electrical Engineering letterhead under the reference ERL/EMB. His reply
acknowledged the letter and notes, recorded that a colleague of his at Imperial had
been working toward almost the same equations and had not so far completed them, and
congratulated the author on the result. The original letter is held by the author. It is described here rather than reproduced.

\FloatBarrier
\section*{\phantomsection Appendix~C --- Derivation Highlights for Table~2}
\addcontentsline{toc}{section}{Appendix~C --- Derivation Highlights for Table~2}
\label{sec:appC}

Each $J_R$ in Table~2 is the direct evaluation of
$\int(\vv{\xi}\cdot\vv{R})^{2}dm/(\vv{R}\cdot\vv{R})$; the projection
$(\vv{\xi}\cdot\vv{R})/|\vv{R}|$ is the shaft-parallel coordinate of the mass element,
so the integrand reduces per solid as follows.

\textbf{Disk $(D, H)$ and rectangular torus $(D, E, H)$.} The shaft-parallel coordinate
runs over the height; $\int z^{2}dm$ over $z \in [-H/2, +H/2]$ with the uniform density
gives $J_R = MH^{2}/12$ in both cases, the radial extent contributing nothing to the
axial projection.

\textbf{Hollow sphere $(D, E)$.} By full isotropy --- which a uniform spherical shell
shares with the solid sphere --- $\int(\vv{\xi}\cdot\hat{e})^{2}dm$ is the same for
every unit $\hat{e}$, and the trace identity~\eqref{e:12} requires each to equal
$J_0/3$; hence $J_R = J_0/3$. The polar integral is
$J_0 = \int(\vv{\xi}\cdot\vv{\xi})dm = (3/20)M(D^{5}-E^{5})/(D^{3}-E^{3})$, which
reduces to the solid sphere's $(3/20)MD^{2}$ as $E \to 0$.

\textbf{Circular torus $(D, E)$.} With major radius $R_{\mathrm{M}} = (D+E)/4$ and tube radius
$r_{\mathrm{T}} = (D-E)/4$, the tube parametrization, with $\chi$ the angle around the tube's cross-section and $\eta$ the azimuth about the symmetry axis, gives shaft-parallel coordinate $\rho\sin\chi$ and mass element proportional to $(R_{\mathrm{M}} + \rho\cos\chi)\,\rho\,d\rho\,d\chi\,d\eta$; the odd $\cos\chi$ term cancels over $\chi$, $\int\sin^{2}\chi\,d\chi = \pi$, and the $\rho$-integral
yields $J_R = Mr_{\mathrm{T}}^{2}/4 = M(D-E)^{2}/64$.

\textbf{Verification.} For each solid, $J_0$ as listed in Table~2 was computed
independently from $\int(\vv{\xi}\cdot\vv{\xi})dm$ and cross-checked against the trace
identity; and the pair $(J_0, J_R)$ reproduces, through the dictionary of
\S\ref{sec:8.3}, the standard tabulated moments of inertia for all three bodies and
their classical limits. Each pair $(J_0, J_R)$ was confirmed three independent ways: by
the analytic closed form, by deterministic numeric quadrature (agreement below
$10^{-10}$), and by stochastic Monte-Carlo integration (agreement to about four
significant digits).

\section*{\phantomsection Appendix~D --- The Disk-to-Rod Transition: Exact Continuity of the Two-Scalar
Closure, with External Validation}
\addcontentsline{toc}{section}{Appendix~D --- The Disk-to-Rod Transition: Exact
Continuity of the Two-Scalar Closure, with External Validation}
\label{sec:appD}

This appendix follows the two closure scalars from the disk to the rod: its purpose (D.1), the scalars across the range (D.2 to D.4), the external validation (D.5 to D.7), and the limitations, results and sources (D.8 to D.10).

\subsection*{\phantomsection D.1 Purpose of the transition study}
\addcontentsline{toc}{subsection}{\protect\numberline{D.1}Purpose of the transition study}

The body of this paper resolves every solid through the same two mass integrals, $J_0$
and $J_R$, and remarks in \S\ref{sec:8.2} that for the finite cylinder --- the rectangular torus of Table~2 at $E = 0$ --- the coefficient $\tfrac{1}{2}(3J_R - J_0)$ changes sign at $H = D\sqrt{3}/2$, the transition between disk-like and rod-like
gyroscopic response. This appendix substantiates that remark and records external checks
of the steady-precession relation. Three questions are answered in order: whether the
transition is genuinely continuous (algebraically, D.2--D.3), whether it is continuous
in the numbers a user would compute (D.4), and whether the relation those numbers come
from agrees with independently published results (D.5--D.6). Brand~\cite{ref37} is included both as an exact historical precession comparator and, through his finite-cylinder inertia formulas, as evidence that the classical principal-moment framework can also span the full height range continuously. Klein and Sommerfeld~\cite{ref38} are included as the earliest exact comparator, and as a published statement of the degenerate case that D.7 examines. D.7 records the distinct
geometry at which the fast branch ceases to exist. D.8 then states the limitations and the precise usability claim; D.9 summarizes the checks, and D.10 records
the external sources.

\subsection*{\phantomsection D.2 The two closure scalars across the whole range}
\addcontentsline{toc}{subsection}{\protect\numberline{D.2}The two closure scalars across the whole range}

For the homogeneous solid cylinder of diameter $D$ and height $H$ --- the rectangular
torus of Table~2 at $E = 0$ --- Table~2 gives

\begin{equation}\tag{D1}\label{e:D1} J_0 = M\!\left(\frac{D^{2}}{8} + \frac{H^{2}}{12}\right), \qquad J_R = \frac{MH^{2}}{12} \end{equation}

so the two combinations that enter the balance~\eqref{e:38}--\eqref{e:39} are, for
every $H$ and $D$,

\begin{equation}\tag{D2}\label{e:D2} I_\parallel = J_0 - J_R = \frac{MD^{2}}{8}, \qquad I_\perp = \tfrac{1}{2}(J_0 + J_R) = \frac{M(3D^{2} + 4H^{2})}{48} \end{equation}

Both are polynomials in D and H. They have no pole, no branch point, no domain
restriction, and no case distinction anywhere in $0 \le H < \infty$: the same two expressions
serve the flattest disk and the longest rod. Two structural facts follow immediately and
are useful because the operator pair makes them visible without first selecting
principal axes. First, $I_\parallel$ is independent of H altogether --- the spin inertia
of a uniform cylinder about its own symmetry axis does not change as the cylinder is
lengthened at fixed diameter and fixed mass. Second, the coefficient whose sign defines
the transition is

\begin{equation}\tag{D3}\label{e:D3} \tfrac{1}{2}(3J_R - J_0) = M\!\left(\frac{H^{2}}{12} - \frac{D^{2}}{16}\right), \qquad \text{vanishing exactly at } H = \tfrac{\sqrt{3}}{2}D \approx 0.8660\,D \end{equation}

This is a smooth quadratic in $H$ crossing zero transversally. The ``transition'' is
therefore a sign change of a continuous function, not a discontinuity, not a
switch between two models, and not a fitted junction: nothing in the formulation
changes as the crossover is passed, and no quantity is undefined on either side of it
or at it.

\subsection*{\phantomsection D.3 Exact recovery of both classical limits}
\addcontentsline{toc}{subsection}{\protect\numberline{D.3}Exact recovery of both classical limits}

The two limiting cases the literature tabulates separately are recovered from the
single pair above by taking limits, with no residual terms and no fitting constants.
Computer-algebra evaluation (SymPy) gives:

\begin{equation}\tag{D4}\label{e:D4} H \to 0 \ (\text{thin disk},\ R_{\mathrm{DISK}} = D/2): \quad I_\parallel \to \frac{MD^{2}}{8} = \tfrac{1}{2}MR_{\mathrm{DISK}}^{2}, \quad I_\perp \to \frac{MD^{2}}{16} = \tfrac{1}{4}MR_{\mathrm{DISK}}^{2} \end{equation}

\begin{equation}\tag{D5}\label{e:D5} D \to 0 \ (\text{thin rod},\ L_{\mathrm{ROD}} = H): \quad I_\parallel \to 0, \quad I_\perp \to \frac{MH^{2}}{12} = \frac{ML_{\mathrm{ROD}}^{2}}{12} \end{equation}

All four limiting values reproduce the standard tabulated moments exactly ---
zero discrepancy, no approximation invoked, and no separate derivation required for
either end. The classical results are thus not merely consistent with the two-scalar
closure; they are two evaluations of it.

\subsection*{\phantomsection D.4 Numeric transition through the crossover}
\addcontentsline{toc}{subsection}{\protect\numberline{D.4}Numeric transition through the crossover}

Algebraic smoothness guarantees numeric smoothness, but the point of the claim is
practical, so the transition is also exhibited in computed precession rates. A
representative cylinder is held fixed in mass, diameter, pivot distance, spin, and
tilt, and only the height is varied through and far beyond the crossover: $M = 1$ kg,
$D = 0.10$ m, $R = 0.06$ m, $g = 9.8$ m/s$^{2}$, $\omega_0 = 200$ rad/s,
$\theta = 70^\circ$. For each $H$ the two scalars are computed from Table~2 and the
slow precession root is obtained from the scalar reduction~\eqref{e:39}. The crossover
sits at $H/D = \sqrt{3}/2 \approx 0.866$ (shaded row). Moments are in
kg$\cdot$m$^{2}$; $\Omega$ in rad/s.

\begin{table}[H]
\centering
\TABLEDONE
\caption{Table~D.1 --- Disk-to-rod sweep at fixed $M = 1$~kg, $D = 0.10$~m, $R = 0.06$~m, $\omega_0 = 200$~rad/s and $\theta = 70^\circ$ ($g = 9.8$~m/s$^{2}$). Column~5 is this paper's exact result, the slow root of~\eqref{e:39}; the shaded row marks the crossover, $H/D = \sqrt{3}/2$. Every
column is quoted to a fixed number of decimal places rather than to a fixed number of
significant digits, so that entries of different magnitude carry the same absolute
precision.}
\end{table}

One thing is visible. The exact column is smooth and monotone across the crossover and far past it: $\Omega$ rises steadily from $2.374957$ rad/s at $H = 0$ to
$2.566557$ rad/s at $H/D = 5$, passing $H/D = 0.866$ with no kink or jump where $\tfrac{1}{2}(3J_R - J_0)$ changes sign. The sign change
is real --- the coefficient column passes through zero exactly there --- but it is a
change in the internal balance of terms, not a discontinuity in the observable.

\subsection*{\phantomsection D.5 External validation of the steady-precession relation}
\addcontentsline{toc}{subsection}{\protect\numberline{D.5}External validation of the
steady-precession relation}

The transition results above rest on equations~(38)--(39). Those equations were separately checked against six independently published treatments of the same physical system. Where a source prints a worked example of the top, its own stated numbers are reported below; where it states its relation for the top only, as Brand and Klein and Sommerfeld do (Sources~5 and 6; the one numerical example Brand works in \S218~\cite{ref37}, p.~507, is the pendulum mill, where a wall force acts besides gravity; the torus rotor Klein and Sommerfeld work in Vol.~II \S V.2~\cite{ref38}, p.~299, concerns another motion, its numbers giving the transit time and the height of a single cycloidal arc of pseudoregular precession, not a precession rate), its equation is evaluated at the configuration of Appendix~F. Each is followed by the same
physical inputs run through this paper's relations --- (39) in general, or its exact
horizontal-shaft special case~(40), $MgR = (J_0 - J_R)\,\omega_0\Omega$, where the
source's configuration is $\theta$ = 90$^\circ$.

\subsubsection*{Source~1 \cite{ref29} --- OpenStax University Physics, Vol.~1,
\S11.4~\cite{ref29}, ``Precession of a Gyroscope''}

\textbf{They report: }a disk of mass $0.30$ kg spinning at $20$ rev/s ($125.66$ rad/s),
center of mass $5.0$ cm from the pivot, disk radius $5.0$ cm;
$I = \tfrac{1}{2}mr^{2} = 3.75\times10^{-4}$ kg$\cdot$m$^{2}$; precession
$\omega_P = rMg/(I\omega) = 3.12$ rad/s, period $2.0$ s.

\textbf{This paper: }the same inputs, with the shaft taken horizontal, $\theta = 90^\circ$ (OpenStax states no tilt; at $\theta = 90^\circ$ its slow-precession formula is exact), thin-disk limit $J_R = 0$,
$J_0 = 3.75\times10^{-4}$ kg$\cdot$m$^{2}$; equation~\eqref{e:40} gives
$\Omega = (0.30)(9.8)(0.05)/[(3.75\times10^{-4})(125.66)] = $
$3.12$ rad/s --- exact agreement to the precision quoted.

\subsubsection*{Source~2 \cite{ref30} --- Jensen and Poling, ``Teaching Rotational
Physics with Bivectors'', arXiv:2207.03560}

\textbf{They report: }a gyroscope disk of mass $0.2$ kg, $I = 0.001$ kg$\cdot$m$^{2}$,
spin $117$ rad/s, on a horizontal massless rod with center $12$ cm from a frictionless
pivot; torque $0.2352$ N$\cdot$m, angular momentum $0.117$ kg$\cdot$m$^{2}$/s, precession
$\approx 2.0$ rad/s.

\textbf{This paper: }same inputs, $\theta = 90^\circ$, $J_R = 0$,
$J_0 = 0.001$ kg$\cdot$m$^{2}$; equation~\eqref{e:40} gives
$\Omega = (0.2)(9.8)(0.12)/[(0.001)(117)] = $ $2.01$ rad/s --- agreement to
the precision quoted, with one further digit resolved.

\textbf{Note on both of the above, stated plainly: }at $\theta = 90^\circ$ with
$J_R = 0$, this paper's exact~\eqref{e:40} and the textbook expression for the
rate --- the torque divided by the angular momentum --- are the same equation. These two checks therefore confirm that the formulation reduces
correctly to the known special case; they do not, by themselves, test anything beyond
it.

\subsubsection*{Source~3 \cite{ref27} --- Bolina, ``The Precessing Top'',
arXiv:physics/0005025}

\textbf{They report: }dropping the fast-top simplification, an exact two-branch precession rate valid at every non-vertical tilt except the horizontal shaft ($\cos\theta = 0$), where the relation becomes linear with a single root,

\begin{equation}\tag{D6}\label{e:D6} \omega_z = \frac{I\,\omega_c \pm \sqrt{I^{2}\omega_c^{2} + 4(I - I_n)\,m g l\cos\theta}}{2(I_n - I)\cos\theta} \end{equation}

with $I$ the moment about the symmetry axis, $I_n$ the moment about a normal axis at
the fixed point $O$, $l$ the pivot-to-COM distance, $\omega_c$ the spin.

\textbf{This paper: }under the natural correspondence
$I \leftrightarrow I_\parallel = J_0 - J_R$ and
$I_n \leftrightarrow I_\perp + MR^{2} = \tfrac{1}{2}(J_0 + J_R) + MR^{2}$
(Bolina~\cite{ref27}'s $I_n$ is already referred to the pivot, so it already carries the parallel-axis shift this paper produces from the transport term), $l \leftrightarrow R$ and $\omega_c \leftrightarrow \omega_0$, his spin being reckoned relative to the precessing axes (p.~2~\cite{ref27}), solving~\eqref{e:39} for $\Omega$ symbolically returns Bolina~\cite{ref27}'s expression
term for term. The check was run in SymPy on the difference of the two closed
forms, which simplifies to exactly zero for both roots --- an identity, not a
numerical coincidence.

\textbf{This is one of four load-bearing external checks. Klein and Sommerfeld~\cite{ref38}, Brand~\cite{ref37}, Bolina~\cite{ref27} and Slade~\cite{ref33} are the four of the six that are themselves exact and $\theta$-dependent, matching the generality of \eqref{e:39}; Sources~1 and 2 are $\theta = 90^\circ$ reductions.} Two exact treatments of the same system, reached by
different routes --- Bolina~\cite{ref27} through angular-momentum vector geometry, this
paper through the mass-integral operator formalism --- agree identically.

\subsubsection*{Source~4 \cite{ref33} --- Slade, ``Classical Symmetric Top in a
Gravitational Field''}

This source treats a heavy wheel on a light stem by the Lagrangian route, and reaches,
before any fast-spin simplification is introduced, an exact two-branch precession rate valid at every non-vertical tilt except the horizontal shaft ($\cos\theta = 0$), where the relation becomes linear with a single root,

\begin{equation}\tag{D7}\label{e:D7} \dot{\varphi} = \frac{p_\psi}{2I_0\cos\theta}\left[\,1 \pm \sqrt{1 - \frac{4mglI_0\cos\theta}{p_\psi^{2}}}\,\right] \end{equation}

with $I_\psi$ the moment about the symmetry axis, $I_0$ the moment about a normal axis
at the fixed pivot, $l$ the pivot-to-COM distance, and
$p_\psi = I_\psi(\dot{\psi} + \dot{\varphi}\cos\theta)$ the conserved spin momentum, in
kg$\cdot$m$^{2}$/s.

Rearranged, that expression is the quadratic
$I_0\cos\theta \cdot \dot{\varphi}^{2} - p_\psi\dot{\varphi} + mgl = 0$. Under the
correspondence $I_\psi \leftrightarrow I_\parallel = J_0 - J_R$ and
$I_0 \leftrightarrow I_\perp + MR^{2} = \tfrac{1}{2}(J_0 + J_R) + MR^{2}$, with $l \leftrightarrow R$ and $p_\psi \leftrightarrow I_\parallel(\omega_0 + \Omega\cos\theta)$, this
is~\eqref{e:43}, and hence~\eqref{e:39}, term for term. The check was run in SymPy on
the difference of the two quadratics, which is identically zero for all admissible
values of the symbols --- an identity, not a numerical coincidence. The same
correspondence carries the existence condition: this source requires
$|p_\psi| \ge \sqrt{4mglI_0\cos\theta}$ for a well-defined uniform precession, which is
exactly the minimum-spin threshold recorded in \S D.6 against Goldstein~\cite{ref2} and
Thornton and Marion~\cite{ref4}.

The worked example is a thin disk on a stem, and it exercises the two mass integrals
directly. The body is a disk of mass $M = 0.1$ kg and diameter $D = 0.1$ m carried at
$R = 0.15$ m, tilted at $\theta = 45^\circ$, with the combination $\omega_0 + \Omega\cos\theta$ held at $100\pi$ rad/s (the source fixes $p_\psi = I_\psi(\dot{\psi} + \dot{\varphi}\cos\theta)$), and with $g = 9.81$ m/s$^{2}$. The recipe of \S\ref{sec:8.1} gives
$J_0 = MD^{2}/8 = 1.250000\times10^{-4}$ kg$\cdot$m$^{2}$ and $J_R \to 0$ for the thin
disk, whence $I_\parallel = J_0 - J_R = 1.250000\times10^{-4}$ kg$\cdot$m$^{2}$ and
$I_\perp + MR^{2} = 2.3125000\times10^{-3}$ kg$\cdot$m$^{2}$. The source's own two
constants, formed independently as $I_\psi = ma^{2}/2$ and $I_0 = ml^{2} + ma^{2}/4$,
are $1.250000\times10^{-4}$ and $2.3125000\times10^{-3}$ kg$\cdot$m$^{2}$. The
agreement is exact in every digit, and it is worth being precise about why: this paper
never forms an axial or a normal moment, yet the two scalars it does form reproduce
both of that source's, from the same body, by a different route.

With $p_\psi = 3.926991\times10^{-2}$ kg$\cdot$m$^{2}$/s the exact relation gives a
slow root of $4.645916070$ rad/s ($44.36523$ rpm) and a fast root of $19.369667379$
rad/s ($184.96670$ rpm). Both expressions return those figures, differing by
$9\times10^{-16}$ rad/s --- machine precision.

Two further observations belong here, because both bear on claims made elsewhere in
this paper.

The first concerns reach. The source also quotes the familiar slow-precession
approximation $\dot{\varphi} \approx mgl/p_\psi$, which for this configuration returns
$3.747144$ rad/s. Against the exact slow root of $4.645916070$ rad/s that is $19.35$ percent low. The reason is visible in the configuration itself: the minimum spin for this body and tilt is $\omega_0 + \Omega\cos\theta = 248.18955$ rad/s, and the example runs at
$314.15927$ rad/s, only $1.27$ times that threshold, outside the fast-spin regime the approximation assumes. The sphere of Appendix~F runs at a similar margin, $1.29$ times its own threshold (Appendix~F.9). This example shows that the approximate forms have short reach, while~\eqref{e:39} holds across the whole range.

The second concerns three arithmetic inconsistencies in that source's parameter table,
recorded here so that a reader who consults it is not misled. Its Table~I prints
$I_\psi = 1.25\times10^{-1}$ kg$\cdot$m$^{2}$, whereas the source's own formula
$I_\psi = ma^{2}/2$ gives $1.25\times10^{-4}$ kg$\cdot$m$^{2}$, and its own printed
$p_\psi = 0.0392$ kg$\cdot$m$^{2}$/s requires $1.25\times10^{-4}$; the accompanying
computer code uses $1.25\times10^{-4}$, so the printed table entry is a typesetting
slip of a factor of one thousand. The same table prints
$I_0 = 2.33\times10^{-3}$ kg$\cdot$m$^{2}$ where the source's own equation for $I_0$
gives $2.3125\times10^{-3}$, a difference of about $0.76$ percent. It also prints $p_\varphi = 0.0273$ kg$\cdot$m$^{2}$/s, where the source's own (5), with $\dot{\varphi}_0 = 0$ and $\theta_0 = \pi/4$, gives $p_\psi\cos\theta_0 = 0.0278$ kg$\cdot$m$^{2}$/s. None of the three affects the
comparison made here, which rests on the source's formulas rather than on its printed
constants. Every other entry in its three parameter tables was recomputed: the other conserved momenta (Table~I's $p_\psi = 0.0392$ by truncation of $0.03927$, the rest by rounding), the total energies and Table~I's tilt angles reproduce to the printed digits --- for instance $0.785$ and $1.258$ radians against the printed $0.785$ and $1.26$ --- while the upper tilt angles of Tables~II and III, printed $0.969$ and $1.504$ radians, reproduce with the printed $I_0 = 2.33\times10^{-3}$ kg$\cdot$m$^{2}$ and $g = 9.8$ m/s$^{2}$ but not with the source's own $2.3125\times10^{-3}$, which gives $0.966$ and $1.498$, so neither of the two values of $I_0$ reproduces every entry.

What this source establishes is therefore twofold. The two mass integrals of this paper
reproduce, exactly, both of the inertia constants that an independent Lagrangian
treatment forms for the same body; and the exact precession relation that treatment
reaches is the relation~\eqref{e:39}, reached here without Lagrangians, Euler angles,
or generalized momenta.

\subsubsection*{Source~5 \cite{ref37} --- Brand, \emph{Vectorial Mechanics} (1930)}

Brand's \S214 (\emph{Kinetics of a Rigid Body with One Point Fixed}, p.~498,
\cite{ref37}), \S217 (\emph{Gyroscope}, p.~502, \cite{ref37}) and \S218
(\emph{Steady Precession}, p.~505, \cite{ref37}) provide an exact vectorial treatment
of a homogeneous solid of revolution with one point fixed. The derivation first takes principal axes at the fixed point and resolves
$\vv{\omega}$ into its projections on them; it then forms $\vv{H} = \int \vv{r}\times(\vv{\omega}\times\vv{r})\,dm$
as the sum of the parts due to those projections, writes $\vv{H}$ in terms of the conventional moments $A$, $B$, $C$, and lets the triad turn as the motion proceeds, its transverse axes kept principal. For steady precession under gravity Brand obtains, in \S218~\cite{ref37}, $(A - C)\cos\theta\cdot\Omega^2 - C\omega'\Omega + Wb = 0$ and the corresponding two-root
formula, stating explicitly that two, one, or no real precession rates occur according
to the discriminant; at $\theta$ = 90$^\circ$ he obtains $\Omega$ = Wb/(C$\omega'$)
exactly, in \S217~\cite{ref37}.

Under $C = J_0 - J_R$, $A = \tfrac{1}{2}(J_0 + J_R) + MR^2$, $W = Mg$, $b = R$ and
$\omega' = \omega_0$, Brand's equation is identically the quadratic form of (39);
the symbolic difference of the two reduces to exactly zero, with no adjustment of sign
or convention required. His spin variable is this paper's own: he writes $\vv{\omega}$ =
$\vv{\Omega} + \vv{\omega}'$ with $\omega'$ the spin about the shaft, not the axial
projection, so no intermediate rate has to be introduced to make the comparison. The
check is also numerical. Given the configuration of Appendix~F --- the solid sphere of $M = 1$ kg, $D = 0.2$ m, $R = 0.2$ m, $\omega_0 = 1500$ rpm, $\theta = 45^\circ$, $g = 9.8$ m/s$^2$ --- Brand's equation returns $\Omega$ = 3.753732625419 rad/s on the slow branch
and 18.460682065373 rad/s on the fast, agreeing with the rates of Appendix~F.4 and Appendix~F.9 to every digit printed there. Given the thin disk of Source~4, with his axial rate $\omega' + \Omega\cos\theta$ held at that source's $100\pi$ rad/s, so that his spin $\omega'$, like this paper's $\omega_0$, differs between the two branches, it returns 4.645916070 and 19.369667379 rad/s, the two roots printed above. Brand therefore supplies an
independent historical confirmation of the exact end relation, both branches and the
horizontal special case.

Brand also supplies the ingredients for a continuous finite-cylinder sweep in
conventional notation. His \S193~\cite{ref37} gives $k^2 = r^2/2$ about the cylinder axis; his $r$ is the radius of the cylinder, not the position vector $\vv{r}$ of this paper. His
\S196~\cite{ref37} gives $k_x^2 = r^2/4 + h^2/3$ about a diameter of the base; applying
his \S194~\cite{ref37} transfer theorem to move that axis by $h/2$ to the center of mass gives
$k_{\perp,\mathrm{COM}}^2 = r^2/4 + h^2/12$. These are continuous in $h$ and map
directly to the classical $I_\parallel$ and $I_\perp$ used in \S\ref{sec:8.3}.
His method is fully capable of the explicit disk-to-rod sweep performed here. The comparison therefore relocates the contribution from
``continuity exists'' to ``continuity is exposed and automated by one unchanged
two-integral interface.''

\subsubsection*{Source~6~\cite{ref38} --- Klein and Sommerfeld, \emph{The Theory of the Top} (1897--1910)}

Klein and Sommerfeld's \S III.6 (\emph{The deviation resistance for regular precession of the symmetric top}, Vol.~I, p.~174,~\cite{ref38}) treats the regular precession of the heavy symmetric top about its support point, setting the gravity turning-force, their name for a torque, against the deviation resistance, the resistance the top offers to being guided along the precession. Their equation~(3), p.~177~\cite{ref38}, is the exact steady-precession condition $P = C \mu \nu + (C - A) \nu ^2 \cos \vartheta$, with $A$ and $C$ the moments of inertia about the support point, $P$ the product of the weight and the distance of the center of gravity from the support point, $\mu$ the spin, $\nu$ the precession and $\vartheta$ the tilt; their $A$, $C$ and $P$ are not the pivot $A$, the center of mass $C$ or the material point $P$ of this paper. Their equation~(4), p.~178~\cite{ref38}, gives its two roots in closed form, $\nu = [C \mu \mp \sqrt{C^2\mu ^2 - 4P(A - C) \cos \vartheta}]/\allowbreak [2(A - C) \cos \vartheta ]$, names them the slow and the fast precession, and states that at given $\mu$ and $\vartheta$ both are real or neither is, as the quantity under the root of (4) is positive or negative. For $A = C$ their equation becomes linear in $\nu$ and gives one root, the second having become infinitely large (pp.~178--179~\cite{ref38}). Their \S III.7 (\emph{A new derivation of the deviation resistance [\ldots] The Coriolis force}, p.~180,~\cite{ref38}) obtains the deviation resistance a second time, for the horizontal case $\vartheta = 90^\circ$.

Under $C = J_0 - J_R, A = \tfrac{1}{2}(J_0 + J_R) + MR^2, P = MgR, \mu = \omega_0, \nu = \Omega$ and $\vartheta = \theta$, their equation~(3) is identically \eqref{e:39}; the symbolic difference of the two reduces to exactly zero, with no adjustment of sign or convention required. Their spin has the same value as this paper's: $\mu$ is the rotation about the figure axis reckoned relative to the precessing axes, which is $\omega_0$, obtained here as $\vv{\omega}_0 = \vv{\omega} - \vv{\Omega}$ with both vectors observed from the inertial frame; they print the axial projection separately as $r = \mu + \nu \cos \vartheta$ (p.~50~\cite{ref38}; their $r$ is not the position vector $\vv{r}$ of this paper), so no intermediate rate has to be introduced to make the comparison. For the same reason their reality condition, taken at a given $\mu$, is the quantity under the square root of \eqref{e:D12} --- the existence condition of the note in \S D.6, not the minimum-spin condition. Their $A - C$ is $K$, so their case $A = C$ is the condition $K = 0$ of \S D.7, where the fast branch ceases to exist: a published statement of that geometry independent of Brand's.

The check is also numerical. Given the configuration of Appendix~F --- the solid sphere of $M = 1$ kg, $D = 0.2$ m, $R = 0.2$ m, $\omega_0 = 1500$ rpm, $\theta = 45^\circ, g = 9.8$ m/s$^2$ --- their equation~(4) returns $\Omega = 3.753732625419$ rad/s on the slow branch and 18.460682065373 rad/s on the fast, the rates of Appendix~F.4 and Appendix~F.9 to every digit printed there. Klein and Sommerfeld therefore supply the earliest of the independent confirmations collected here of the exact end relation, both branches and the degenerate case.

\subsection*{\phantomsection D.6 The existence condition against the canonical texts}
\addcontentsline{toc}{subsection}{\protect\numberline{D.6}The existence condition
against the canonical texts}

Throughout this paper, and in every comparison with the literature, the minimum spin is
one and the same thing: the smallest value of $\omega_0 + \Omega\cos\theta$ at which a
steady precession exists for the given body and tilt. That combination is what the
classical texts write as a single axial rate of their own; this paper writes it out in its
two rates and gives it no symbol. In the paper's own symbols the condition reads
$(\omega_0 + \Omega\cos\theta)^{2} \ge 4(I_\perp + MR^{2})\,MgR\cos\theta / I_\parallel^{2}$,
which is~\eqref{e:D8} below under the dictionary of \S\ref{sec:8.3}. Relation~\eqref{e:39},
collected as a quadratic in $\Omega$ at a fixed $\omega_0$ --- which is~\eqref{e:D10} of
\S D.7 --- carries a second and different statement: its two rates~\eqref{e:D12} are real
only when $I_\parallel^{2}\omega_0^{2} - 4K\cdot MgR\cos\theta$ is not negative. That is the
existence condition at a fixed $\omega_0$, and it is nowhere called a minimum spin.

Goldstein, Poole and Safko~\cite{ref2} (Classical Mechanics, 3rd ed., Ch.~5, ``The Heavy
Symmetrical Top with One Point Fixed'', \S5.7~\cite{ref2}) and Thornton \&
Marion~\cite{ref4} (Classical Dynamics of Particles and Systems, 5th ed., Ch.~11, ``Motion of a Symmetric Top with One Point Fixed'', \S11.11~\cite{ref4}) state the same threshold, writing that combination as a single axial
rate of their own. Goldstein, Poole and Safko write it with the strict inequality ``$>$'' (their equations~(5.78) and (5.79), p.~219~\cite{ref2}), which leaves out the equality, where the two branches merge into a double root; Thornton and Marion write it with ``$\ge$'' (their equations~(11.169) and (11.170), p.~458~\cite{ref4}), as~\eqref{e:D8} below does:

\begin{equation}\tag{D8}\label{e:D8} (\omega_0 + \Omega\cos\theta)^{2} \ \ge\ \frac{4I_1\,M g l\cos\theta}{I_3^{2}} \end{equation}

with $I_3$ about the symmetry axis and $I_1$ transverse through the fixed point.
As a check of the correspondence, this paper's~\eqref{e:43}, which is~\eqref{e:39}
written in the textbook variables, was collected as a quadratic in $\Omega$ with the combination $\omega_0 + \Omega\cos\theta$ held fixed, as the textbooks hold their axial rate, and its discriminant taken symbolically:

\begin{equation}\tag{D9}\label{e:D9} \mathrm{disc} = I_\parallel^{2}\,(\omega_0 + \Omega\cos\theta)^{2} - 4(I_\perp + MR^{2})\,MgR\cos\theta \end{equation}

Under $I_3 \leftrightarrow I_\parallel$, $I_1 \leftrightarrow I_\perp + MR^{2}$,
$l \leftrightarrow R$ (Thornton and Marion write $h$ for $l$), this differs from the textbook condition by exactly zero (SymPy).
The minimum spin of this paper is therefore the canonical one, written in the paper's own
two rates. The existence condition at a fixed $\omega_0$, stated in the note above and
used in \S D.7, is a separate statement about a fixed spin, and it is not a minimum spin.

\subsection*{\phantomsection D.7 The fast branch, and the geometry at which it ceases to exist}
\addcontentsline{toc}{subsection}{\protect\numberline{D.7}The fast branch, and the
geometry at which it ceases to exist}

\S D.2 to \S D.4 answer the question this appendix was written for: the disk-to-rod
transition is continuous, algebraically and in the numbers a reader would compute. This
subsection records the one place where the steady-precession relation does become
unbounded, so that the two are not confused --- it is a different zero, of a different
quantity, and it has nothing to do with the crossover. \S\ref{sec:8.4} already names it
in passing, among the structural features reproduced against the three published
formulations: the coefficient zero at which the second branch reverses sign.

Relation~\eqref{e:39} is written with the gravity term standing alone. Collected
instead as a quadratic in $\Omega$, with the combination
$\omega_0 + \Omega\cos\theta$ resolved so that only this paper's own spin $\omega_0$
appears, and with the signs taken so that the leading term is positive
for an ordinary body, it reads

\begin{equation}\tag{D10}\label{e:D10} K\cos\theta\cdot\Omega^{2} \ -\ I_\parallel\,\omega_0\,\Omega \ +\ MgR \ =\ 0 \end{equation}

where $I_\parallel = J_0 - J_R$ and $I_\perp = \tfrac{1}{2}(J_0 + J_R)$ are the
verification vocabulary of \S\ref{sec:8.3}, and where the leading coefficient is named
once:

\begin{equation}\tag{D11}\label{e:D11} K \ \equiv\ I_\perp + MR^{2} - I_\parallel \ =\ MR^{2} + \tfrac{1}{2}(3J_R - J_0) \qquad [\mathrm{kg}\cdot\mathrm{m}^{2}] \end{equation}

Equation~\eqref{e:F3} of Appendix~F is this same form written out for the isotropic
sphere, where $J_R = J_0/3$ and $K$ reduces to $MR^{2}$. The bracket that appears in
\S\ref{sec:7.5} as the $\Omega^{2}$ coefficient of~\eqref{e:39} is $-K$, the sign
difference being the reversal just described.

Both roots of~\eqref{e:D10} satisfy the balance and both are steady motions of the same
body under the same gravity torque; the condition for them to be real is that the quantity under the square
root of~\eqref{e:D12} is not negative --- the existence condition at a fixed $\omega_0$
set out in the note of \S D.6. Written
out,

\begin{equation}\tag{D12}\label{e:D12} \Omega_\pm \ =\ \frac{I_\parallel\,\omega_0 \pm \sqrt{I_\parallel^{2}\omega_0^{2} - 4K\,MgR\cos\theta}}{2K\cos\theta} \end{equation}

The slow root is the familiar gyroscopic precession, the one the correspondence of
\S\ref{sec:8.3} reduces to. The fast root is not an approximation to it and not a
second reading of it; it is the other steady motion.

$K$ is not a quantity that may be assumed positive. By~\eqref{e:D11} it is the
disk-to-rod crossover coefficient of \S D.2 shifted by the transport term $MR^{2}$, so
it depends on the shape of the body and on the length of the shaft together. It
vanishes when $MR^{2} = \tfrac{1}{2}(J_0 - 3J_R)$, which requires $J_0 > 3J_R$ --- a
body on the disk-like side of the crossover. For a thin disk of diameter $D$,
$J_R = 0$ and $J_0 = MD^{2}/8$, so $K = M(R^{2} - D^{2}/16)$ and the zero falls
at $R = D/4$ exactly, whatever the mass. For the isotropic sphere $J_R = J_0/3$
identically, the right side is zero, and no shaft length produces the
degeneracy: $K = MR^{2}$ for every arm.

At $K = 0$ the $\Omega ^2$ term of \eqref{e:D10} is absent. The relation is linear in $\Omega$ and returns a single precession rate: the fast branch does not exist there. Regarded as a function of the shaft length $R$, at fixed body, spin and tilt, the fast root has a simple pole at the degenerate length: it grows without bound on either side, and it is not defined at that length itself. Klein and Sommerfeld state the same collapse for their case $A = C$, which under the correspondence of \S D.5, Source~6, is $K = 0$: their equation becomes linear in $\nu$ and gives one root, the second having become infinitely large (Vol.~I, pp.~178--179~\cite{ref38}). They also examine the neighborhood of that case, qualitatively and in their variable $A - C$: their schema on p.~179 carries the second root through infinity, with its sense reversed on the two sides of $A = C$~\cite{ref38}. What is set out here from~\eqref{e:D10} is that neighborhood in the shaft length, as a simple pole; the reversal of sense with the sign of the leading coefficient is also stated by Brand in \S218~\cite{ref37}.

The sense reverses through the pole. At shaft lengths below the degenerate one the fast
precession is thread-in, and at lengths above it thread-out, the magnitude running to
infinity on both approaches; at a tilt beyond the horizontal the two senses exchange,
since $\cos\theta$ changes sign. The slow precession is continuous through the
same length and stays thread-out, so nothing about the body or the motion is singular
there --- only this one branch of the solution is.

None of this qualifies the continuity established in \S D.2 and \S D.3, because the two
statements concern different quantities and different zeros. The crossover of \S D.2 is the zero of $\tfrac{1}{2}(3J_R - J_0)$ alone, at
$H = D\sqrt{3}/2$ for a cylinder, and is a property of the shape alone. The degeneracy
of this subsection is the zero of $MR^{2} + \tfrac{1}{2}(3J_R - J_0)$, which involves
the shaft as well. At the crossover itself $K = MR^{2}$, strictly positive for every
shaft length greater than zero, so no body has a pole at its own crossover, and the two
zeros never coincide. What acquires a pole is not the closure scalars, which are
polynomials in the geometry with no pole of their own, but the solution built from
them. The sweep of
Table~D.1 is clear of the degeneracy by a wide margin: it holds $R = 0.06$ m on a body
of $D = 0.10$ m, and its $K$ runs from $0.002975$ to $0.023808$ kg$\cdot$m$^{2}$ across
the whole range of heights, never approaching zero; its thin-disk end would require
$R = 0.025$ m.

Close to the degenerate length the fast root, although it remains a correct solution
of~\eqref{e:D10}, becomes larger than any real shaft and pivot could sustain: the result
is correct mathematics, but the motion it describes cannot be realized by a physical
body.

There is a second, different limit, and the two should not be conflated. Much closer still to
the degenerate length, $K$ is computed as the difference of
two comparable quantities and its own leading digits are lost to rounding; below that
level the fast root is set by the rounding error in $K$ rather than by the physics. The
solver supplied with this paper states that level explicitly and withholds the fast
root beneath it.

Finally, a negative $K$ does not prevent a steady precession from existing. Whether one
exists is decided by the quantity under the square root of~\eqref{e:D12} and by nothing else. When
$K\cos\theta$ is negative, the term $-4K\cdot MgR\cos\theta$ under that square root is
positive, so it is positive at every spin and both roots of~\eqref{e:D12} are real.

\subsection*{\phantomsection D.8 Limitations and differences}
\addcontentsline{toc}{subsection}{\protect\numberline{D.8}Limitations and differences}

\textbf{1. Two of the six sources test only a special case. }Jensen and Poling~\cite{ref30} state a horizontal shaft, $\theta$ = 90$^\circ$, and OpenStax~\cite{ref29} a thin disk with no tilt stated and a formula derived for slow precession; both are compared at $\theta$ = 90$^\circ$, the regime in which this paper's exact relation and the textbook formula coincide algebraically. Their
agreement is a correct reduction check, not evidence of added generality. The
observation is worth stating in the other direction as well. The textbook formula is
what the relation tested here becomes when cos $\theta$ is set to zero; the $\theta$ =
90$^\circ$ result is read off the same single expression with one term identically
vanishing. No separate formula is invoked and no case is switched to obtain it.

2. Brand~\cite{ref37} supplies an important classical comparison, but the distinction
between capability and published methodology must be kept explicit. His exact
precession equation of \S218~\cite{ref37} is the same quadratic as (39) after the
dictionary of \S\ref{sec:8.3}, and \S\S193, 194 and 196~\cite{ref37} ---
\emph{Moment of Inertia of Solids of Revolution} (p.~436), \emph{Transfer Theorem}
(p.~439) and \emph{Application of Transfer Theorem} (p.~444) --- supply the axial cylinder
inertia, transfer theorem, and finite-height transverse inertia from which a reader can
construct $A(H)$ and $C(H)$ for any cylinder height. Those ingredients are sufficient to
construct a continuous disk-to-rod calculation in the principal-moment representation.
The present appendix carries such a sweep, identifies the $H = D\sqrt{3}/2$ crossover as a continuous family property, and formulates an unchanged geometry-to-precession interface spanning disk to rod, using the same $J_0(H)$, $J_R(H)$ pair and the same relation~(39). Brand therefore
validates the classical end equation and the finite-height inertia ingredients; the
continuous methodology demonstrated here is not something this paper attributes to
Brand.

\textbf{3. One consulted source contributes no number. }MIT OpenCourseWare
8.09~\cite{ref32}, Assignment~3, is a reading assignment and problem set on rigid bodies that points to the heavy symmetric top in Goldstein \S5.7~\cite{ref2} and offers no worked value to test against; it is listed in
Table~D.3 for completeness and scope clarity only.

The first two points above isolate the usability claim as a distinction between
capability and formulation. Classical principal-moment mechanics is mathematically
capable of continuous finite-height modeling if the full geometry-dependent moments are
assembled at every height; Brand supplies the required ingredients. That fact is not the
same as presenting and using one continuous numerical methodology. The present
formulation does so explicitly: $J_0$ and $J_R$ are evaluated from the same
mass-integral definitions at every geometry, passed through the same operator closure,
and inserted into the same relation~(39) without a switch of body category, inertia
representation, or solver interface. Every numerical result in the disk-to-rod sweep is
generated by that unchanged path. This is the usability claim made here.

\subsection*{\phantomsection D.9 Summary of results}
\addcontentsline{toc}{subsection}{\protect\numberline{D.9}Summary of results}

Table~D.2 collects every check performed for this appendix, naming the reference each
was tested against and the outcome.

\TABLEDTWO

Conclusion. The disk-to-rod transition is continuous in both senses demonstrated here:
algebraically, because the closure pair ($J_0$, $J_R$) consists of polynomials valid
without restriction over the whole range and reproducing both classical limits exactly;
and numerically, because the precession rate computed from (39) passes through $H = D\sqrt{3}/2$ smoothly and monotonically, with no feature at the crossover.
Brand~\cite{ref37} provides an exact classical precession equation, which maps identically onto~\eqref{e:39},
and finite-cylinder inertia formulas. No disk-to-rod sweep is attributed to him. The
specific result demonstrated by this appendix is that the present two-integral
representation itself remains unchanged and is actually used continuously across the
family.

\subsection*{\phantomsection D.10 External sources accessed}
\addcontentsline{toc}{subsection}{\protect\numberline{D.10}External sources accessed}

Table~D.3 records the sources of this appendix and of Appendix~F's external checks, with what each was consulted for and whether it contributed a validation datum; the two further Tanr\i{}verdi works named below were encountered and not used as validation sources.

\begin{table}[H]
\centering
\TABLEDTHREE
\caption{Table~D.3 --- External sources accessed in preparing this appendix, keyed to
the paper's reference list, with what each supplied and whether it counts as a
validation datum.}
\end{table}

Two further works by Tanr\i{}verdi were encountered during the search: the Euler-angle motion classification~\cite{ref18}, not used as a validation datum, and the small-oscillation study~\cite{ref34}, which treats a phenomenon this paper does not address.

\section*{\phantomsection Appendix~E --- Claims Validation}
\addcontentsline{toc}{section}{Appendix~E --- Claims Validation}
\label{sec:appE}

Table~E.1 is a registry of the manuscript's principal interpretive, technical, methodological,
computational, and literature-positioning assertions. It is not intended to enumerate
every descriptive sentence in the abstract or prose; each registered claim cites its
supporting equations and the specific verification or source comparison that bears on
it.

Support basis codes --- (i): the support is shown in the paper itself (a derivation, a
textual property confirmed by direct symbol audit, or a stated axiom). (ii): the support
is formally or externally checked --- check numbers~\#1--\#31 refer to the independent re-derivation of the main symbolic results, by computer algebra except for the numerical checks~\#4 to \#9, \#19 and \#23 (on random particle data, by a central difference, on point clouds or by numerical quadrature), and Appendix~D references
identify external comparisons. (iii): literature-search bounded --- a negative
literature-positioning statement is supported only as a report of the searched
accessible record and is not a universal proof of priority. (iv): still open --- the
support required is not available, and the paper says where. A mathematical claim is
marked SUPPORTED only when its mathematical content is covered by (i) and/or (ii); a
literature-positioning claim is explicitly labeled search-bounded.

\TABLEE

\noindent\textbf{Summary}

Of the 26 registered principal claims, one is the interpretive claim of \S\ref{sec:1} (claim~\#26), supported by derivation and by the ancillary comparison file. Another 23 are technical or methodological claims
supported by derivation, formal symbolic checks, external equation comparisons, or
direct textual audit. Brand~\cite{ref37} closes the external comparison at the
equation-and-finite-cylinder-inertia level: his exact steady-precession relation maps to
(39), and his inertia formulas are compatible with the continuous finite-height family.
He is not cited as having published the disk-to-rod sweep or the unchanged
geometry-to-dynamics methodology demonstrated in Appendix~D, and claim~\#22 therefore
remains open and is documented as such. Claim~\#24 is different in kind: it is a bounded
literature-search finding, explicitly limited to the searched accessible record and not
presented as a universal proof of priority.

Registry status: one interpretive claim supported; 23 technical and methodological claims supported; one
literature-positioning claim supported as search-bounded; one external-validation claim
documented open.

\section*{\phantomsection Appendix~F --- Torque Equalization at Dynamic Equilibrium: The Complete
Numerical Demonstration}
\addcontentsline{toc}{section}{Appendix~F --- Torque Equalization at Dynamic
Equilibrium: The Complete Numerical Demonstration}
\label{sec:appF}

This appendix works one body through the whole balance: its purpose (F.1), the body and equation~\eqref{e:35} term by term (F.2 and F.3), the equalization and its tilt sweep (F.4 to F.8), the external checks (F.9 and F.10), and the summary, scope and conclusion (F.11 to F.13).

\subsection*{\phantomsection F.1 Purpose of the numerical demonstration}
\addcontentsline{toc}{subsection}{\protect\numberline{F.1}Purpose of the numerical demonstration}

\S\ref{sec:7.1} defines dynamic equilibrium as the statement that the steady-precession
kinematics satisfy the balance~\eqref{e:32} identically at every instant. That balance sets
two torques about the pivot against each other: the gravity torque $M\vv{R}\times\vv{g}$,
which works to lay the shaft down, and the assembled inertia torque $\Mscr_A$, which works
to bring it up. This appendix works one configuration completely in numbers, so that the
equalization of those two torques --- and the way that equalization breaks when the tilt is
displaced --- can be read directly off a table in N$\cdot$m rather than inferred from the
algebra.

The organizing idea throughout is torque equalization. It is not approximate and not a
modeling assumption: it is equation~\eqref{e:32}, and at the steady tilt the two torques
are equal in magnitude and diametrically opposed, so that their resultant is the zero
vector to the last digit that can be printed. Neither torque disappears there and neither
cancels the other; both are present, each with its own effect on the tilt, and it is the
arithmetic of adding two opposed vectors of equal length that reaches zero. Displace the
tilt by a single degree and the equalization breaks --- but it breaks in a specific
direction on each side, and that directionality is the numerical signature of a restoring
mechanism.

Every number below is computed from the unified vector equation~\eqref{e:35} and its
closure~\eqref{e:37}. Nothing is imported from the classical description except in \S F.9
and \S F.10, where the classical language is used, as in \S\ref{sec:8.3}, solely as the
vocabulary of verification.

\subsection*{\phantomsection F.2 The body, and its two mass integrals}
\addcontentsline{toc}{subsection}{\protect\numberline{F.2}The body, and its two mass
integrals}

The worked body is a homogeneous solid sphere on a light shaft pivoted at a fixed
frictionless point $A$. Its data, with units:

\[ M = 1\ \mathrm{kg} \quad D = 0.2\ \mathrm{m} \quad R = 0.2\ \mathrm{m} \quad
\omega_0 = 1500\ \mathrm{rpm} = 157.0796327\ \mathrm{rad/s} \quad \theta = 45^\circ
\quad g = 9.8\ \mathrm{m/s^{2}} \]

From Table~2 the solid sphere is the hollow-sphere entry at $E = 0$, so the polar second
moment of mass is $J_0 = (3/20)\,M D^{2}$, and full isotropy supplies $J_R = J_0/3$:

\[ J_0 = 0.006000\ \mathrm{kg}\cdot\mathrm{m}^{2} \qquad
J_R = 0.002000\ \mathrm{kg}\cdot\mathrm{m}^{2} \qquad
J_0 - J_R = 0.004000\ \mathrm{kg}\cdot\mathrm{m}^{2} \]

Because the sphere is isotropic, the coefficient $\tfrac{1}{2}(J_0 - 3J_R)$ of the
closure~\eqref{e:37} vanishes identically and the mass-vector operator degenerates to the
single-scalar law $\vv{F}(\vv{v}) = J_R\,\vv{v}$ --- equation~\eqref{e:41}. This is what
makes every term of~\eqref{e:35} an elementary cross product for this body, and it is why
the whole evaluation below can be carried out in closed form.

\subsection*{\phantomsection F.3 Equation~(35), term by term}
\addcontentsline{toc}{subsection}{\protect\numberline{F.3}Equation~(35), term by term}

Freeze the motion at the instant the shaft crosses the vertical plane containing the $Y$
and $Z$ axes, with $Z$ vertical and the center of mass leaning toward positive $Y$. The
four vectors entering~\eqref{e:35} are then $\vv{R} = R(0, \sin\theta, \cos\theta)$ in m,
$\vv{\omega}_0 = \omega_0(0, \sin\theta, \cos\theta)$ in rad/s,
$\vv{\Omega} = \Omega(0, 0, 1)$ in rad/s and $\vv{g} = (0, 0, -g)$ in m/s$^{2}$. All four
lie in or along the $YZ$ plane, so every cross product formed from them points along $X$,
and each term of~\eqref{e:35} has a single non-zero component, in N$\cdot$m, as set out in
Table~F.1:

\begin{table}[H]
\centering
\TABLEFONE
\caption{Table~F.1 --- The six terms of the unified vector equation~(35) for the isotropic
sphere. The fourth vanishes because $\vv{F}(\vv{\Omega})$ is parallel to $\vv{\Omega}$;
the three $J_R$ terms combine with signs $(+1, -1, +1)$ to a single surviving
contribution.}
\end{table}

The negative of the sum of the six entries gives the inertia torque, and
$M\vv{R}\times\vv{g}$ gives the gravity torque, with $\hat{x}$ the unit vector of the
frozen frame's $x$-axis, normal to the plane containing the shaft and the vertical:

\begin{equation}\tag{F1}\label{e:F1} \Mscr_A = \sin\theta\cdot\Omega\left[(J_0 - J_R)\,\omega_0 - MR^{2}\,\Omega\cos\theta\right]\hat{x} \qquad [\mathrm{N}\cdot\mathrm{m}] \end{equation}

\begin{equation}\tag{F2}\label{e:F2} M\vv{R}\times\vv{g} = -MgR\sin\theta\ \hat{x} \qquad [\mathrm{N}\cdot\mathrm{m}] \end{equation}

Both are horizontal and both perpendicular to the vertical plane holding the shaft, and
they point opposite ways along it: at the equalization the inertia torque lies along
positive $X$ and the gravity torque along negative $X$. Equalization is therefore a
statement about two collinear vectors, and reduces to a comparison of two numbers.

\subsection*{\phantomsection F.4 The equalization condition and the precession rate}
\addcontentsline{toc}{subsection}{\protect\numberline{F.4}The equalization condition and
the precession rate}

Imposing~\eqref{e:32}, that is adding~\eqref{e:F2} to~\eqref{e:F1} and requiring the
resultant to be the zero vector, the common
factor $\sin\theta$ divides out --- the reduction of \S\ref{sec:7.5} --- leaving a
quadratic in $\Omega$, every term in N$\cdot$m:

\begin{equation}\tag{F3}\label{e:F3} MR^{2}\cos\theta\cdot\Omega^{2} \ -\ (J_0 - J_R)\,\omega_0\cdot\Omega \ +\ MgR \ =\ 0 \end{equation}

With the values of \S F.2 the coefficients are $0.02828427$ kg$\cdot$m$^{2}$ for
$\Omega^{2}$, $-0.62831853$ kg$\cdot$m$^{2}$/s for $\Omega$, and $1.9600$ N$\cdot$m for
the constant. The discriminant is $+0.17303549$ kg$^{2}\cdot$m$^{4}$/s$^{2}$, positive,
so two real equalization rates exist:

\[ \Omega(\text{slow}) = 3.7537326\ \mathrm{rad/s} = 35.84551\ \mathrm{rpm}
\qquad\qquad
\Omega(\text{fast}) = 18.4606821\ \mathrm{rad/s} = 176.28653\ \mathrm{rpm} \]

The slow root is the precession ordinarily observed. Its existence is not automatic: at this tilt
no equalization exists for this body unless $\omega_0$ is at least $117.7255$ rad/s
$= 1124.2$ rpm, so the chosen 1500 rpm clears that condition, whereas at 1000 rpm the
quantity under the square root of~\eqref{e:D12} would be negative and no equalization at
$45^\circ$ would be possible.

\subsection*{\phantomsection F.5 The equalization itself}
\addcontentsline{toc}{subsection}{\protect\numberline{F.5}The equalization itself}

Substituting the slow root back into~\eqref{e:F1} and~\eqref{e:F2} separately --- that is,
evaluating the two torques independently rather than assuming their equalization --- gives

\[ M\vv{R}\times\vv{g} = -\,1.385929291\ \hat{x}\ \ \mathrm{N}\cdot\mathrm{m}
\qquad\qquad
\Mscr_A = +\,1.385929291\ \hat{x}\ \ \mathrm{N}\cdot\mathrm{m} \]

The two agree in magnitude to every digit shown and carry opposite signs. This is the
equalization, exhibited numerically: the torque gravity exerts about the pivot and the
torque the body's own inertia produces are equal in size, $1.385929291$ N$\cdot$m each,
both directed horizontally at right angles to the shaft's vertical plane and pointing
opposite ways along it --- gravity's negative, working to increase the tilt, and the
inertia torque's positive, working to reduce it. Their resultant is the zero vector,
$0.000000000$ N$\cdot$m, with a residual of order $10^{-16}$ N$\cdot$m, and the tilt is
therefore held at $45^\circ$ while the shaft sweeps steadily around the vertical. Both
torques are present the whole time; nothing has disappeared.

\subsection*{\phantomsection F.6 The tilt sweep that breaks the equalization}
\addcontentsline{toc}{subsection}{\protect\numberline{F.6}The tilt sweep that breaks the equalization}

Now hold the spin at $\omega_0 = 1500$ rpm and the precession at the slow root
of~\eqref{e:F3} carried at full precision, $\Omega = 3.7537326254$ rad/s
$= 35.8455061428$ rpm --- not at the value rounded for display in \S F.4 --- and move the
tilt alone, one degree at a time either side of $45^\circ$. At each tilt the
two torques are recomputed independently from~\eqref{e:F1} and~\eqref{e:F2}. The
equalization, exact at $45^\circ$, fails elsewhere --- and the sign of the failure is what
matters:

\begin{table}[H]
\centering
\TABLEF
\caption{Table~F.2 --- Both torques recomputed from~(35) at five tilts, with $\omega_0$
and $\Omega$ held fixed. ``DOWN'' denotes a net down torque about $A$, which increases the
tilt back toward $45^\circ$; ``UP'' denotes a net up torque about $A$, which decreases it,
again back toward $45^\circ$.}
\end{table}

Below the equalization tilt the gravity torque is the larger of the two, and the surplus
torques the shaft further down about $A$, which increases the tilt and carries it back
toward $45^\circ$. Above the equalization tilt the inertia torque is the larger, and that
surplus torques the shaft upward about $A$, decreasing the tilt and again carrying it back
toward $45^\circ$. On both sides the failure of equalization acts to restore it (see
Figure~9). Throughout this behavior nothing pushes or pulls the shaft: the entire
restoring action is a torque about $A$, as \S\ref{sec:6} established.

Because the two torque entries carry opposite signs and signed numbers can mislead the
eye, the same result is also restated in magnitudes. The ratio in the final column is a pure number, the
N$\cdot$m canceling, and it passes through unity exactly at the equalization tilt (Table~F.3):

\begin{table}[H]
\centering
\TABLEFTHREE
\caption{Table~F.3 --- The five configurations of Table~F.2 in magnitudes. The ratio
$|\Mscr_A| / |M\vv{R}\times\vv{g}|$ crosses unity exactly where the torques equalize.}
\end{table}

Figure~9 shows both torques about the pivot --- gravity and the inertia torque --- as the shaft tilt alone
is varied
from $40^\circ$ to $50^\circ$, with the magnitude of the precession $\Omega$ and the magnitude of the shaft spin $\omega_0$ held constant throughout at their equalization
values. The $45^\circ$ tilt is the exact dynamic-equilibrium tilt by design: there the two
torques are equal, at $1.385929291$ N$\cdot$m, and the shaded band between the curves is
the torque imbalance that appears on either side of it --- gravity torque larger for tilt
angles below $45^\circ$, inertia torque larger for tilt angles above $45^\circ$, so
that on both sides of $45^\circ$ tilt the excess torque acts to restore the torque
equalization at exactly $45^\circ$ tilt angle.

\begin{figure}[H]
\centering
\includegraphics[width=6.5in]{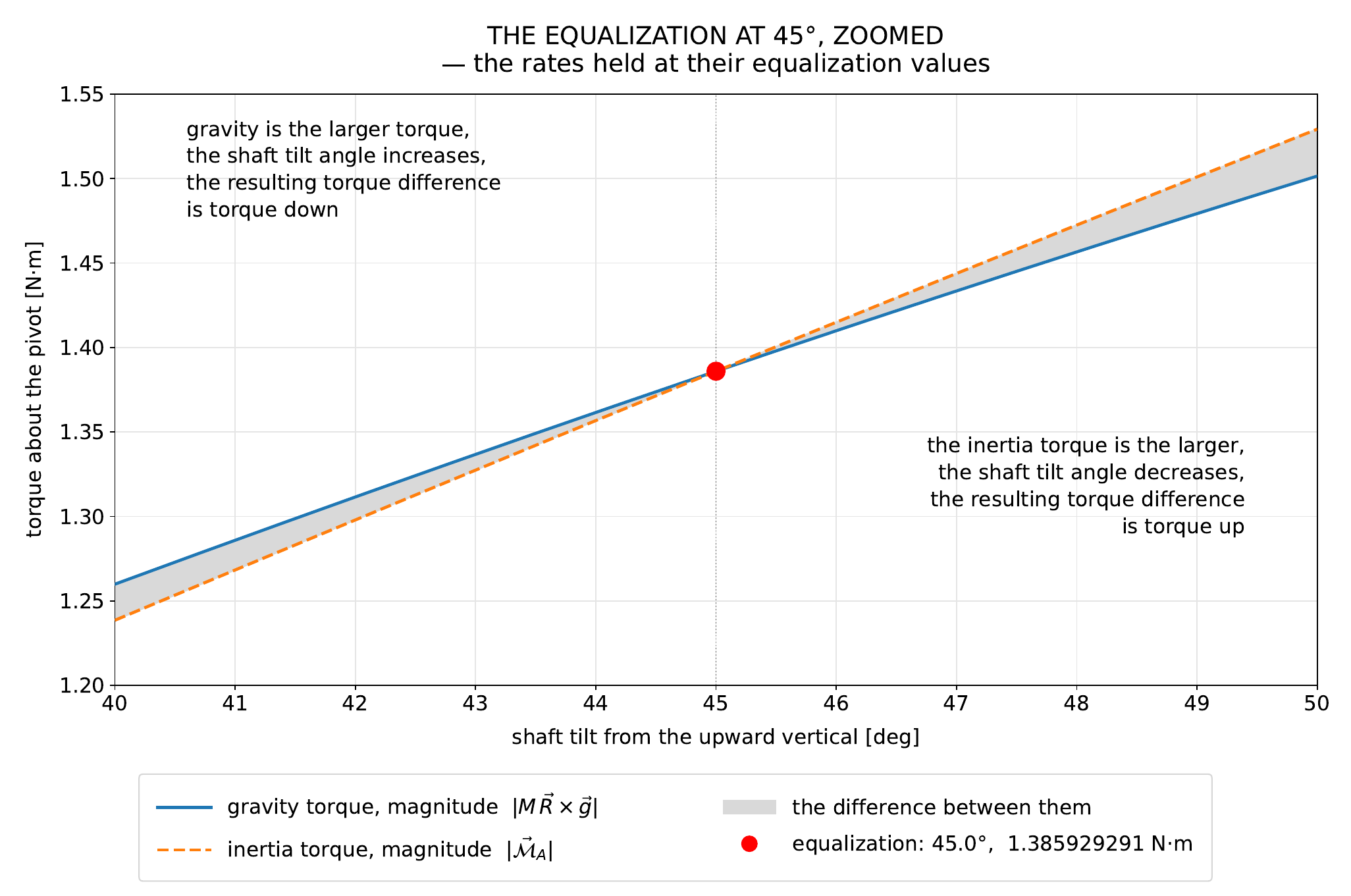}
\caption{Figure~9 --- Torque equalization at $45^\circ$, zoomed: the magnitudes $|M\vv{R}\times\vv{g}|$ and $|\Mscr_A|$ against tilt; the two torque vectors are opposite in direction (Table~F.2).}
\end{figure}

\subsection*{\phantomsection F.7 The rate at which equalization is restored}
\addcontentsline{toc}{subsection}{\protect\numberline{F.7}The rate at which equalization
is restored}

Differentiating~\eqref{e:F1} and~\eqref{e:F2} with respect to the tilt and evaluating at
the equalization gives the rate at which the resultant grows with displacement:

\[ \frac{d}{d\theta}\left[M\vv{R}\times\vv{g} + \Mscr_A\right]_X \ =\ +\,0.281810172\
\mathrm{N}\cdot\mathrm{m/rad} \ =\ +\,0.004919\ \mathrm{N}\cdot\mathrm{m}\ \text{per
degree} \qquad (\text{at } \theta = 45^\circ) \]

The per-degree figure lies between the two one-degree steps of Table~F.2, $0.004789516$ and $0.005047016$~N$\cdot$m, and matches their mean to within $3\times10^{-7}$~N$\cdot$m, as it should.
The derivative is non-zero, so the resultant passes cleanly through zero rather than
grazing it: the equalization tilt is isolated, not one of a range of tilts at which the
torques happen to match. Two further features of the numbers are worth noting. The
imbalance grows very nearly in proportion to the displacement --- about $0.005$ N$\cdot$m
at one degree and about $0.010$ N$\cdot$m at two, on both sides --- which is the behavior
of a linear restoring mechanism near equilibrium. And its scale is small: $0.005$
N$\cdot$m against torques of about $1.386$ N$\cdot$m is roughly a third of one percent, so the restoring surplus is genuine but gentle.

\subsection*{\phantomsection F.8 The different response of the two torques to a change of tilt}
\addcontentsline{toc}{subsection}{\protect\numberline{F.8}The different response of the two torques to a change of tilt}

The mechanism behind the whole table is visible in~\eqref{e:F1} and~\eqref{e:F2}. Both
torques carry the same dimensionless factor $\sin\theta$. If that were the whole of their
tilt dependence they would rise and fall in lockstep, no imbalance could ever arise, and
the equalization would be indifferent to the tilt. But the inertia torque carries an
additional dependence through $\cos\theta$, contributed by the transport term
$M(\vv{\Omega}\cdot\vv{R})(\vv{R}\times\vv{\Omega})$, while the gravity torque does not.
Once the common $\sin\theta$ is divided out, the gravity side is the flat constant
$MgR = 1.9600$ N$\cdot$m at every tilt while the inertia-torque side still varies with
$\cos\theta$, and it is that asymmetry which produces the restoring surplus.

It follows that the effect is invisible to any fast-spin treatment. The fast-top
approximation discards precisely the $\cos\theta$ terms, leaving both torques proportional
to $\sin\theta$ alone; a calculation made that way returns a constant ratio at every tilt
and shows no imbalance. The exact operator expression is what makes the
equalization mechanism visible, and this is a concrete instance of the paper's general
claim that the familiar approximations are recovered as special cases while the exact
structure carries more.

\subsection*{\phantomsection F.9 External verification of the equalization rate}
\addcontentsline{toc}{subsection}{\protect\numberline{F.9}External verification of the
equalization rate}

The rate at which the torques equalize was tested against five independently published
results for the heavy symmetric top, drawn from six publications: the canonical
minimum-spin condition is stated independently by Goldstein~\cite{ref2} and by Thornton
and Marion~\cite{ref4}, and the two statements are one result and are counted here once.
In each case the source's own relation is written exactly as published, the notation is
mapped through the correspondence of \S\ref{sec:8.3}, and the same physical inputs are
substituted. No source is asked to endorse the operator formalism; only the physical
number is compared.

Brand's exact two-branch relation~\cite{ref37} and Bolina's~\cite{ref27} are the two tested here that give both rates in closed form at every non-vertical tilt except the horizontal shaft; the MIT exact balance~\cite{ref24} and Agashe's condition~\cite{ref26}, equally exact, are evaluated here as balances at the worked tilt. Bolina's returns 3.753732625419 rad/s for the
slow branch and 18.460682065373 rad/s for the fast, agreeing with the values of Appendix~F.4 to
$1.8\times10^{-15}$ rad/s on both branches. Brand's returns the same two rates with a
residual of exactly zero, and the reason is worth stating: once his constants are
substituted through $C = J_0 - J_R$ and $A = \tfrac{1}{2}(J_0 + J_R) + MR^2$, his
relation is not merely equivalent to the balance used here but the same quadratic
written in different letters, and his spin variable is this paper's own $\omega_0$
rather than the axial rate, so no intermediate quantity is introduced to make the
comparison. The MIT OpenCourseWare exact balance~\cite{ref24} and Agashe's
uniform-precession condition~\cite{ref26} are each satisfied to machine zero, the latter
with a residual of $6.7\times10^{-16}$ N$\cdot$m and a discriminant of $+0.164315261$ kg$^{2}\cdot$m$^{4}$/s$^{2}$.
The canonical minimum-spin condition of Goldstein~\cite{ref2} and of Thornton and
Marion~\cite{ref4} is satisfied with margin at the slow root, whose $\omega_0 + \Omega\cos\theta = 159.73392$ rad/s stands against a threshold of $123.47154$ rad/s; the two equalization rates of F.4 at the fixed $\omega_0$ are established separately, by the positive discriminant of~\eqref{e:F3}.

\subsection*{\phantomsection F.10 External verification of the torque values themselves}
\addcontentsline{toc}{subsection}{\protect\numberline{F.10}External verification of the
torque values themselves}

The equalization rate being confirmed, the torque columns of Table~F.2 were checked
separately. The gravity column follows from the definition of the cross product,
$|M\vv{R}\times\vv{g}| = MgR\sin\theta$, and reproduces the tabulated values at all five
tilts with residuals at the level of the ninth decimal, that is at the level of the
printed precision.

The inertia-torque column was reproduced by a deliberately independent route: assembling the
angular momentum about the pivot in the classical manner,
$\vv{L} = I_3\,(\omega_0 + \Omega\cos\theta)\,\hat{e}_3 + I_1\,\vv{\omega}_\perp$, with $\hat{e}_3$ the unit vector along the shaft and $\vv{\omega}_\perp$ the part of $\vv{\omega}$ across it, in kg$\cdot$m$^{2}$/s, and
taking the negative of the rate at which the precession sweeps it round,
$-\,d\vv{L}/dt = -\,\vv{\Omega}\times\vv{L}$ in N$\cdot$m, the inertia torque being the
negative of that rate. That construction uses the projected
moments of inertia, the axial-and-transverse decomposition and the component resolution
which the present formulation is built to avoid --- so agreement between the two cannot be
an artifact of shared machinery. All five entries agree, and when the comparison is run at
full internal precision rather than against the printed nine-decimal figures the residuals
fall to between $0$ and $1.3\times10^{-15}$ N$\cdot$m, the floor of double-precision
arithmetic.

\subsection*{\phantomsection F.11 Summary of checks}
\addcontentsline{toc}{subsection}{\protect\numberline{F.11}Summary of checks}

Table~F.4 collects every check performed for this appendix and its outcome.

\begin{table}[H]
\centering
\TABLEFFOUR
\caption{Table~F.4 --- Every check performed for this appendix. No check produced a
discrepancy beyond decimal rounding.}
\end{table}

\subsection*{\phantomsection F.12 Scope and limitations}
\addcontentsline{toc}{subsection}{\protect\numberline{F.12}Scope and limitations}

Four qualifications, stated plainly in the manner of Appendix~D.8.

First, on which external sources apply. The OpenStax~\cite{ref29} and
Jensen-and-Poling~\cite{ref30} examples used in Appendix~D are thin-disk problems compared at $\theta$ = 90$^\circ$ (a horizontal shaft in Jensen and Poling; OpenStax states no tilt) with $J_R\approx$ 0; they are correct
reduction checks in their own regime but cannot test a solid sphere at 45$^\circ$, and
they are therefore not invoked here. The five results used here are those whose
published relations are exact at the worked configuration: Brand's~\cite{ref37} and
Bolina's~\cite{ref27} are exact and tilt-dependent throughout, matching the generality
of this calculation, while the remaining three are exact balances or conditions
evaluated at the same tilt.

Second, on the status of the independent route of \S F.10. The classical
angular-momentum construction is an independent method, not an independently published
numerical datum. Its value is that it crosses formalisms; it is not a citation.

Third, on precision. Residuals of order $10^{-10}$ N$\cdot$m against the printed table
values are limited by the nine decimals shown, not by any physical disagreement. At full
internal precision they fall to the $10^{-15}$ to $10^{-16}$ level throughout.

Fourth, on what the sweep of Table~F.2 establishes, and in what sense. The spin and the
precession rate are held at their equalization values while the tilt alone is moved, and
the two torques are recomputed independently at each tilt. The treatment is geometric and
positional throughout: every quantity in this appendix is a function of the tilt and of
nothing else. It is not a function of time. No transient is followed, no disturbed motion
is integrated, and no rate of change is taken. What the table shows is the equalization
itself --- exact at the steady tilt, and unequal on either side of it, with gravity the
larger below and the inertia torque the larger above --- and that equalization is what
this paper means when it speaks of the dynamic equilibrium constant tilt angle.

It is worth stating plainly what that does and does not claim, because the word stability carries a settled technical meaning elsewhere. In classical mechanics the word has a definite meaning:
Lyapunov stability asks whether every sufficiently small displacement from the
equilibrium stays small for all later time --- a question about the motion that follows
the displacement, and not a requirement that the disturbance decay. That is what a
reader trained in dynamics expects the word to promise --- a time-domain analysis yielding a nutation frequency, a decay
rate, or an eigenvalue with a sign. Nothing of that kind is undertaken here, and nothing
of that kind is claimed.

What this appendix establishes is a different object, and deliberately so. The spin and
the precession rate are held at their equalization values, the tilt alone is displaced,
and both torques are recomputed independently at every tilt. Nothing is released, nothing
is integrated, and no rate of change with respect to time is taken. What emerges is that
the two torques are exactly equal at the equilibrium tilt angle and unequal on either side
of it, and that the inequality reverses direction as the tilt angle passes through that
value. For a reduced tilt angle --- smaller than the equilibrium tilt angle --- the
gravity torque is the larger, and the difference turns the shaft back down about the
pivot: a corrective torque down, which increases the tilt angle toward the equilibrium
tilt angle. For an increased tilt angle --- larger than the equilibrium tilt angle --- the
inertia torque is the larger, and the difference turns the shaft back up about the
pivot: a corrective torque up, which decreases the tilt angle toward the equilibrium tilt
angle. Both corrective torques therefore act to restore the shaft tilt angle toward the
same equilibrium tilt angle, from either side of it. The magnitude of each corrective
torque grows as the tilt-angle displacement increases --- $0.009320229$ N$\cdot$m at a
tilt angle two degrees smaller than the equilibrium tilt angle, and $0.010349837$
N$\cdot$m at a tilt angle two degrees larger, the difference crossing zero transversally
at $0.281810172$ N$\cdot$m per radian. That is a restoring signature under a constrained
displacement: a genuine and independently checkable result about the torques, which this paper verifies numerically; the published sources of \S F.9 confirm the equalization rate and the minimum spin at the steady tilt, not the imbalance on either side of it. It is not the same object as a
Lyapunov analysis.

\subsection*{\phantomsection F.13 Conclusion}
\addcontentsline{toc}{subsection}{\protect\numberline{F.13}Conclusion}

For a 1 kg solid sphere of $0.2$ m diameter mounted at $R = 0.2$ m from a fixed pivot,
spinning at 1500 rpm and tilted $45^\circ$ from the upward vertical in a field of $9.8$ m/s$^{2}$,
the gravity torque and the inertia torque about the pivot equalize at a precession rate
of $\Omega = 3.7537326$ rad/s $= 35.84551$ rpm, both torques then taking the value
$1.385929291$ N$\cdot$m directed horizontally at right angles to the shaft's vertical
plane. A second, fast equalization branch exists at $176.28653$ rpm, and equalization at
$45^\circ$ is possible at all only because $\omega_0$ exceeds $1124.2$ rpm for this
geometry.

Displacing the tilt breaks the equalization in a restoring direction on both sides:
gravity is the larger below $45^\circ$, by $0.009320229$ N$\cdot$m at $43^\circ$, and the
inertia torque is the larger above it, by $0.010349837$ N$\cdot$m at $47^\circ$, with the
difference crossing zero transversally at a rate of $0.281810172$ N$\cdot$m/rad. The
mechanism is the additional $\cos\theta$ dependence carried by the inertia torque and not
by gravity --- a dependence that any fast-spin treatment discards, and which the exact
operator expression retains.

The equalization rate has been confirmed against the five independently published results of the heavy symmetric top collected in Table~F.4; the gravity column has been checked against the elementary closed form, and the inertia-torque column against an independent classical construction that uses precisely the machinery this
formulation avoids. No check produced a discrepancy at any level beyond decimal
rounding.

% Forward equation labels for displays not yet transcribed (appendices only).
% This file is EMPTY when all sixty are placed.
% AUTO-GENERATED by mkfwd.py. One line per equation NOT YET transcribed into a
% section, so that forward \eqref references resolve to the correct printed
% number. Regenerate after every section. The file is EMPTY when the paper is
% complete, and an empty file is the proof that nothing is outstanding.

\end{document}